\documentclass[preprint]{aastex}
\usepackage{amsthm}
\usepackage{amsmath}

\theoremstyle{plain} 
\newcommand{\ath}{{\it Athena }}
\newcommand{\athnosp}{{\it Athena}}
\newcommand{\deldot}{\nabla \cdot}
\newcommand{\delcross}{\nabla \times}
\newcommand{\vct}[1]{\boldsymbol{#1}}
\newcommand{\mat}[1]{\mathbf{#1}}
\newcommand{\tns}[1]{\mathbf{#1}}
\newcommand{\rinv}{\case{1}{R}}
\newcommand{\rhoinv}{\case{1}{\rho}}
\newcommand{\Emf}{{\mathbf{\mathcal{E}}}}
\newcommand{\emf}{{\mathcal{E}}}
\defcitealias{col84}{CW}
\defcitealias{col90}{Colella}
\defcitealias{blo93}{BL}
\defcitealias{gar05}{GS05}
\defcitealias{gar08}{GS08}

\shorttitle{\emph{Athena} in Cylindrical Geometry}
\shortauthors{M. A. Skinner and E. C. Ostriker}

\begin{document}


\title{The \emph{Athena} Astrophysical MHD Code \\ in Cylindrical Geometry}


\author{M. Aaron Skinner\altaffilmark{1} and Eve C. Ostriker\altaffilmark{2}}
\affil{Astronomy Department, University of Maryland,
    College Park, MD 20742}
\altaffiltext{1}{askinner@astro.umd.edu}
\altaffiltext{2}{ostriker@astro.umd.edu}


\begin{abstract}
A method for implementing cylindrical coordinates in the \ath magnetohydrodynamics (MHD) code is described.  The extension follows the approach of \athnosp's original developers and has been designed to alter the existing Cartesian-coordinates code \citep{sto08} as minimally and transparently as possible.  The numerical equations in cylindrical coordinates are formulated to maintain consistency with constrained transport, a central feature of the \ath algorithm, while making use of previously implemented code modules such as the Riemann solvers.  Angular-momentum transport, which is critical in astrophysical disk systems dominated by rotation, is treated carefully.  We describe modifications for cylindrical coordinates of the higher-order spatial reconstruction and characteristic evolution steps as well as the finite-volume and constrained transport updates.  Finally, we present a test suite of standard and novel problems in one-, two-, and three-dimensions designed to validate our algorithms and implementation and to be of use to other code developers.  The code is suitable for use in a wide variety of astrophysical applications and is freely available for download on the web.
\end{abstract}


\section{Introduction}

The \ath code (\citealt{gar05}, hereafter \citetalias{gar05}; \citealt{gar08}, hereafter \citetalias{gar08}; \citealt{sto08}) is a new, second-order Godunov code for solving the equations of ideal magnetohydrodynamics (MHD).  Among its salient features are that it preserves the divergence-free constraint, $\nabla \cdot \mathbf{B} = 0$, to within machine round-off error via unsplit evolution of the magnetic field, and that it employs fully conservative updates of the MHD equations.  This last feature distinguishes \ath from its predecessor, {\it Zeus} \citep{sto92a,sto92b}, which also preserves the divergence-free constraint, but employs operator-split finite-difference methods.  \ath has been extensively tested via both comparison to analytic solutions, and comparison to the results of other numerical MHD codes.  The code package is freely available to the community, and is highly portable and easily configurable, as it is self-contained and does not rely on outside libraries other than MPI for computation on multi-processor distributed memory platforms.

The equations of ideal MHD consist of eight coupled partial differential equations, which are not analytically solvable in general, and fully three-dimensional numerical solutions can be quite costly.  For many astrophysical systems of interest, however, the computational cost for certain problems can be reduced by exploiting geometric symmetry.  For example, the high angular velocity of the plasma in accreting systems implies that most of the mass is confined within a disk.  If the properties are statistically independent of azimuthal angle, $\phi$, these disks can be studied using radial-vertical ($R$-$z$) models, and if vertical variations are of lesser importance, these disks can be studied using radial-azimuthal ($R$-$\phi$) models.  The dynamical properties of winds and jets from astrophysical systems can also be analyzed using axisymmetric models.  Exploiting symmetry in this way to reduce the effective dimension of the problem can greatly simplify the calculations involved and allow finer resolution when and where needed.  In addition, for either reduced-dimensional or fully three-dimensional problems, using a curvilinear coordinate system for rotating, grid-aligned flow is superior for preservation of total angular momentum, and renders imposition of boundary conditions much simpler compared to the Cartesian-grid case.

There are several other publicly available high-resolution shock-capturing codes for astrophysical MHD in wide use, including {\it VAC} \citep{tot96}, {\it BATS-R-US} \citep{pow99}, {\it FLASH} \citep{fry00}, {\it RAMSES} \citep{tey02}, {\it NIRVANA} \citep{zie04}, and {\it PLUTO} \citep{mig07}, to name a few.  Although these and other codes enjoy increasing popularity within the community, as of this writing only {\it VAC} and {\it PLUTO} have the capability for MHD in curvilinear coordinates.

In this paper, we describe our adaptation of \ath to support cylindrical geometry, and present a suite of tests designed to validate our algorithms and implementation.  These tests include standard as well as novel problems, and may be of use to other code developers.  A guiding principal of our approach is to alter the existing \ath code as minimally and as transparently as possible.  This will involve a careful formulation of the MHD equations so that the finite-volume algorithm remains consistent with constrained transport, and so that the built-in Riemann solvers (as well as computation of wavespeeds and eigenfunctions) need not be changed.  Finally, we pay particular attention to angular-momentum transport, which is critical in systems dominated by rotation.

The plan of this paper is as follows:  In \S\ref{eqns}, we describe the conservative system of mathematical equations that we shall solve, and in \S\ref{algover}, we briefly outline the main steps used in \ath to evolve the system numerically.  In \S\ref{lin}, we describe the projected primitive variable system used in the reconstruction step.  In \S\S\ref{recon} and~\ref{charevo}, we describe the modifications needed for cylindrical coordinates in the higher-order spatial reconstruction and characteristic evolution steps, respectively.  In \S\S\ref{fvm} and~\ref{ct}, we describe the implementation in cylindrical coordinates of the finite volume and constrained transport updates, respectively, and then in \S\ref{alg}, we summarize the steps of the whole algorithm in detail.  In \S\ref{tests}, we present code verification tests and results, and we conclude in \S\ref{conclusion}.

Our version of the code, including the suite of test problems we have developed, is freely available for download on the Web.

{\section{The Equations of MHD} \label{eqns}}

The coordinate-free conservative form of the equations of ideal MHD are:
{\begin{mathletters} \label{eqns:all}
\begin{eqnarray}
\partial_t \rho + \deldot (\rho \vct{v}) &=& 0, \label{eqns:cont} \\
\partial_t (\rho \vct{v}) + \deldot (\rho \vct{v} \vct{v} - \vct{B} \vct{B} + P^* \tns{I}) &=& -\rho \,\vct{\nabla} \Phi, \label{eqns:mom} \\
\partial_t E + \deldot \left[(E + P^*) \vct{v} - \vct{B} (\vct{B} \cdot \vct{v})\right] &=& -\rho \,\vct{v}\cdot\nabla\Phi, \label{eqns:energy} \\
\partial_t \vct{B} + \deldot (\vct{v} \vct{B} - \vct{B} \vct{v}) &=& 0. \label{eqns:ind} 
\end{eqnarray}
\end{mathletters}}
Here, $\rho$ is the mass density, $\rho \vct{v}$ is the momentum density, $\vct{B}$ is the magnetic field vector, and $\tns{I}$ is the identity tensor.  The total pressure is defined as $P^* \equiv P + (\vct{B} \cdot \vct{B})/2$, where $P=nkT$ is the thermal pressure, and $E$ is the total energy density defined as $E \equiv \epsilon + \rho(\vct{v} \cdot \vct{v})/2 + (\vct{B} \cdot \vct{B})/2$, where $\epsilon$ is the internal energy density.  An ideal gas equation of state $P=(\gamma - 1)\epsilon$ is assumed, where $\gamma$ is the ratio of specific heats.  As written, the equations have been scaled in such a way that the magnetic permeability is $\mu = 1$ (for cgs units, $\vct{B}$ is replaced by $\vct{B}/\sqrt{4\pi}$).  Optionally, we can include a static gravitational potential $\Phi=\Phi(\vct{x})$ in equations~(\ref{eqns:mom}) and (\ref{eqns:energy}); the energy equation~(\ref{eqns:energy}) can also be generalized by including radiative heating and cooling terms.

Ignoring terms on the right-hand sides, equations~(\ref{eqns:all}) can be summarized by the single evolution equation in ``conservative'' form:
\begin{equation}
\partial_t \vct{Q} + \deldot \vct{F} = 0, \label{eqns:compact}
\end{equation}
where $\vct{Q} = \vct{Q}(\vct{x},t)$ is the set of conserved quantities
\begin{equation}
\vct{Q} \equiv \left(
\begin{array}{c}
\rho \\
\rho \vct{v} \\
E \\
\vct{B}
\end{array} \right),
\end{equation}
and
\begin{equation}
\tns{F} \equiv \left[ \begin{array}{c} 
\rho\vct{v} \\ 
\tns{M} \\ 
(E+P^*)\vct{v} - \vct{B}(\vct{v} \cdot \vct{B}) \\
\tns{J} \end{array} \right]
\end{equation}
is a structure whose components represent the (nonlinear) fluxes associated with the various components of $\vct{Q}$.  For added simplicity, we have defined the momentum and induction tensors:
\begin{equation}
\tns{M} \equiv \rho\vct{v}\vct{v} - \vct{B}\vct{B} + P^* \tns{I}, \label{eqns:mdef}
\end{equation}
and
\begin{equation}
\tns{J} \equiv \vct{v}\vct{B} - \vct{B}\vct{v}. \label{eqns:jdef}
\end{equation}
Note that additional terms such as gravitational forces appearing on the right-hand sides of equations~(\ref{eqns:all}) are treated separately as source terms, hence are not part of the conservative system in equation~(\ref{eqns:compact}).

In Cartesian coordinates, equation~(\ref{eqns:compact}) can be expanded in a straightforward manner:
\begin{equation}
\partial_t \vct{Q} + \partial_x \vct{F}_x + \partial_y \vct{F}_y + \partial_z \vct{F}_z = 0,  \label{eqns:cartform}
\end{equation}
where $\vct{F}_x$, $\vct{F}_y$, and $\vct{F}_z$ are one-dimensional vectors representing the fluxes of various components of $\vct{Q}$ in each orthogonal coordinate direction.  

In order to extend the existing algorithm in \ath to curvilinear coordinates, we introduce geometric scale factors and source terms arising from the covariant derivatives in the curved metric.  Following this approach, the existing code can be made to support cylindrical geometry with only moderate adjustments, which is what we describe in this paper.  

We expand equation~(\ref{eqns:compact}) according to the form of the divergence operator in cylindrical coordinates \citep[see e.g.][]{mih84} when acting on a vector,
\begin{equation}
\deldot \vct{v} = \rinv \partial_R \left( R \,v_R \right) + \rinv \partial_\phi v_\phi + \partial_z v_z,
\end{equation}
and when acting on a tensor,
\begin{mathletters}
\begin{eqnarray}
(\deldot \tns{T})_R &=& \rinv \partial_R \left( R \,T_{RR} \right) + \rinv \partial_\phi T_{\phi R} + \partial_z T_{zR} - \rinv T_{\phi\phi} \label{eqns:cyltensordiva} \\
(\deldot \tns{T})_\phi &=& \rinv \partial_R \left( R \,T_{R\phi} \right) + \rinv \partial_\phi T_{\phi \phi} + \partial_z T_{z\phi} + \rinv T_{\phi R} \label{eqns:cyltensordivb} \\
(\deldot \tns{T})_z &=& \rinv \partial_R \left( R \,T_{Rz} \right) + \rinv \partial_\phi T_{\phi z} + \partial_z T_{zz}. \label{eqns:cyltensordivc}
\end{eqnarray}
\end{mathletters}
The extra non-derivative terms in equations~(\ref{eqns:cyltensordiva}) and~(\ref{eqns:cyltensordivb}) are the so-called geometric source terms, and represent ``fictitious'' forces, e.g. the centrifugal and Coriolis forces.  Once source terms are introduced, the finite-volume updates are no longer fully conservative.  As we shall next show, however, all but one of the geometric source terms can be eliminated from the equations.  This remaining geometric source term, appearing in the radial momentum equation, is often balanced by gravity in realistic astrophysical problems.

\subsection{Continuity Equation}
Expanding the derivative operators in cylindrical coordinates in equation~(\ref{eqns:cont}), we have the continuity equation in conservative variable form:
\begin{equation}
\partial_t \rho + \rinv \partial_R (R \,\rho v_R) + \rinv \partial_\phi (\rho v_\phi) + \partial_z (\rho v_z) = 0. \label{eqns:contcylcons}
\end{equation}

\subsection{Momentum Equation}
For the momentum equation~(\ref{eqns:mom}), we have, in terms of the symmetric momentum tensor $\tns{M}$ (eq. \ref{eqns:mdef}):
\begin{mathletters}
\begin{eqnarray}
\partial_t(\rho v_R) + \rinv \partial_R (R \,M_{RR}) + \rinv \partial_\phi M_{\phi R} + \partial_z M_{zR} &=& \rinv M_{\phi\phi} \label{eqns:momcyla} \\
\partial_t(\rho v_\phi) + \rinv \partial_R (R \,M_{R\phi}) + \rinv \partial_\phi M_{\phi \phi} + \partial_z M_{z\phi} &=& -\rinv M_{\phi R} \label{eqns:momcylb} \\
\partial_t(\rho v_z) + \rinv \partial_R (R \,M_{Rz}) + \rinv \partial_\phi M_{\phi z} + \partial_z M_{zz} &=& 0.  \label{eqns:momcylc}
\end{eqnarray}
\end{mathletters}
\citet{mig07} note that the symmetric character of $\tns{M}$ allows further simplification of equation~(\ref{eqns:momcylb}) in cylindrical coordinates, since 
\begin{equation}
(\deldot \tns{M})_\phi = \case{1}{R^2} \partial_R (R^2 \,M_{R\phi}) + \rinv \partial_\phi M_{\phi \phi} + \partial_z M_{z\phi}.
\end{equation} 
This leads to the so-called ``angular momentum-conserving form'' of the $\phi$-momentum equation:
\begin{equation}
\partial_t(\rho \,R \,v_\phi) + \rinv \partial_R (R^2 \,M_{R\phi}) + \rinv \partial_\phi (R \,M_{\phi \phi}) + \partial_z (R \,M_{z\phi}) = 0 \label{eqns:angmomform}
\end{equation} 
Note that in this form, the conserved quantity is \emph{angular} momentum and there is no source term.  However, since the original Cartesian version of \ath makes use of linear momenta in the flux calculations and since we do not wish to alter those calculations, we rewrite this equation once more to obtain the $\phi$-momentum equation that we shall use:
\begin{equation}
\partial_t(\rho \,v_\phi) + \case{1}{R^2} \partial_R (R^2 \,M_{R\phi}) + \rinv \partial_\phi M_{\phi \phi} + \partial_z M_{z\phi} = 0.  \label{eqns:redangmomform}
\end{equation} 
This leaves the term $M_{\phi\phi}/R = (\rho v_\phi^2 - B_\phi^2 + P^*)/R$ appearing in the $R$-momentum equation~(\ref{eqns:momcyla}) as the only geometric source term in the cylindrical coordinate expansion of equation~(\ref{eqns:mom}).  In practice, this source term is often (partially) balanced by the radial component of the gravitational source term
\begin{equation}
-\rho \,\vct{\nabla} \Phi  \label{eqns:gravsource_mom}
\end{equation}
for many astrophysical applications.

\subsection{Energy Equation}
For the total energy equation~(\ref{eqns:energy}) in conservative form, we have
\begin{eqnarray}
\partial_t E &+& \rinv \partial_R \left[ R \,((E + P^*) v_R - B_R \,(\vct{B} \cdot \vct{v})) \right] + \rinv \partial_\phi ((E + P^*) v_\phi - B_\phi (\vct{B} \cdot \vct{v})) \nonumber \\
&+& \partial_z ((E + P^*) v_z - B_z (\vct{B} \cdot \vct{v})) = 0.
\end{eqnarray}
The gravitational source term for the energy equation is 
\begin{equation}
-\rho \,\vct{v} \cdot \vct{\nabla} \Phi.  \label{eqns:gravsource_en}
\end{equation}

\subsection{Induction Equation}
Finally, for the induction equation~(\ref{eqns:ind}), we have in terms of the antisymmetric induction tensor, $\tns{J}$ (eq. \ref{eqns:jdef}):
\begin{mathletters}
\begin{eqnarray}
\partial_t B_R + \rinv \partial_\phi J_{\phi R} + \partial_z J_{zR} &=& 0 \label{eqns:indcyla} \\
\partial_t B_\phi + \rinv \partial_R (R \,J_{R\phi}) + \partial_z J_{z\phi} &=& \case{J_{R \phi}}{R} \label{eqns:indcylb} \\
\partial_t B_z + \rinv \partial_R (R \,J_{Rz}) + \rinv \partial_\phi J_{\phi z} &=& 0. \label{eqns:indcylc}
\end{eqnarray}
\end{mathletters}
It is important for the preservation of the divergence constraint that we avoid source terms in the magnetic fluxes.  Thus, we rewrite the $\phi$-induction equation~(\ref{eqns:indcylb}) as
\begin{equation}
\partial_t \left(\case{B_\phi}{R}\right) + \rinv \partial_R \left( R \,\case{J_{R\phi}}{R} \right) + \partial_z \left(\case{J_{z\phi}}{R}\right) = 0. \label{eqns:indcylb2}
\end{equation} 
Note that in this form, the conserved quantity is $B_\phi/R$ and there is no source term.  Equation~(\ref{eqns:indcylb2}) has also modified the fluxes of $B_\phi$, but since \ath does not actually use flux differences to evolve the magnetic fields (see \S\ref{ct}), this is not a serious problem.  We will use a reduced form equivalent to equation~(\ref{eqns:indcylb2}) for $B_\phi$ in the reconstruction step (see \S\ref{lin}):
\begin{equation}
\partial_t B_\phi + \partial_R J_{R\phi} + \partial_z J_{z\phi} = 0. \label{eqns:indcylb3}
\end{equation}
In this form, we use the fluxes $J_{R\phi}$ and $J_{z\phi}$ originally appearing in equation~(\ref{eqns:indcylb}), not the modified fluxes $J_{R\phi}/R$ and $J_{z\phi}/R$ appearing in equation~(\ref{eqns:indcylb2}).

{\section{Overview of the Numerical Algorithm} \label{algover}}

The algorithm used in \ath to evolve the system in equation~(\ref{eqns:compact}) uses a Godunov-type finite volume (FV) scheme.  A simplified version of the algorithm presented in \citet{sto08} is as follows:
\begin{enumerate}
\item Using cell-centered volume averages at time $t^n$, compute left and right (L/R) interface states with the reconstruct-evolve-average (REA) method based on the linearized one-dimensional evolution equations. \label{algover:lrstates}
\item Add the parallel components of source terms to the L/R states. \label{algover:lrsource}
\item Compute the first-order interface fluxes from the L/R states using an exact or approximate Riemann solver. \label{algover:fluxes1}
\item Update the L/R states' magnetic fields using constrained transport (CT) \citep{eva88}.
\item Correct the L/R states' remaining non-magnetic variables with transverse flux gradients and the transverse components of source terms. \label{algover:transverse}
\item Compute the second-order interface fluxes from the corrected L/R states using the Riemann solver. \label{algover:fluxes2}
\item Using the second-order fluxes, advance the interface magnetic fields to time $t^{n+1} = t + \Delta t$ with CT.
\item Using the second-order fluxes, advance the remaining cell-centered quantities to time $t^{n+1}$ with the FV method. \label{algover:fv}
\item Add the time- and volume-averaged source terms to the cell-centered quantities. \label{algover:fvsource}
\item Average the interface magnetic field components to obtain the cell-centered field components at time $t^{n+1}$.  \label{algover:ccmag}
\item Compute a new timestep $\Delta t$ based on the CFL condition and repeat steps~(\ref{algover:lrstates})-(\ref{algover:newdt}) until $t^{n+1} \ge t_f$. \label{algover:newdt}
\end{enumerate}

Currently, \ath includes a wide variety of non-linear Riemann solvers \citep[see][for a complete list]{sto08}.  In our tests, we use the solvers based on the HLL flux \citep{har83} as well as Roe's 
linearized method \citep{roe81}.  Among the solvers based on the HLL flux are the HLLE solver \citep{ein91}, which uses a single intermediate state, the HLLC solver \citep{tor99}, which extends the original HLLE solver by including a contact wave, and the HLLD solver \citep{miy05}, which extends the original HLLE solver by including both contact and Alfv\'en waves.  The HLLE solver has the advantage that it is simple and therefore faster than more accurate solvers such as Roe's method, and like all solvers based on the HLL fluxes, it is positive-definite for 1D problems.  However, since it neglects the contact wave, and additionally the Alfv\'en waves for MHD, it is overly-diffusive for these waves.  On the other hand, for hydrodynamics, the HLLC solver produces fluxes that are as accurate, if not better, than those produced by Roe's method, but at a considerably lower computational cost.  For MHD, it has been shown that the HLLD solver is of comparable accuracy to the MHD extension of Roe's method for several tests using \athnosp, although it is much faster \citep{sto08}.  The advantage of Roe's linearized method is that it includes all waves in a given problem, yielding less diffusive and, hence, more accurate results for intermediate waves that are neglected by the methods based on HLL fluxes, although for some values of the left and right states, Roe's method will fail to return positive density and/or pressure in the intermediate state(s).  Finally, we reiterate that with the approach we have adopted, it is not necessary to make any changes to the computation of wavespeeds, eigenfunctions, or fluxes in any of these methods.

Although no changes are required for the solution of the Riemann problem at interfaces, several changes are required in other parts of the \ath algorithm in order to accommodate non-Cartesian coordinates.  In the next sections, we discuss the geometry-specific details of computing the L/R states (steps~\ref{algover:lrstates}-\ref{algover:lrsource}; see \S\S\ref{lin}-\ref{charevo}), the FV method (steps~\ref{algover:fv}-\ref{algover:fvsource}; see \S\ref{fvm}), and the incorporation of CT into the corner transport upwind (CTU) method of \citet{col90} (see \S\ref{ct}).  Finally, we will recapitulate the steps of the algorithm in greater detail and explain the computation of the new timestep (step~\ref{algover:newdt}; see \S\ref{alg}).

{\section{The Linearized Evolution Equations} \label{lin}}
In \athnosp, the left and right (L/R) interface states (the inputs to the Riemann solver) are computed using a modified form of the system in equation~(\ref{eqns:compact}).  The equations, written in primitive variable form, are projected in a single coordinate direction, and the resulting system is linearized and then evolved.  The projection in the $\phi$-direction yields a system that can be obtained from the corresponding Cartesian projection \citepalias[see][\S3.1]{gar05} by making the substitution $\partial_y \mapsto R^{-1} \partial_\phi$.  However, the projection in the $R$-direction differs more significantly as a result of geometric scale factors.

For the projection in the $R$-direction, we begin with the primitive variable form of equation~(\ref{eqns:compact}), take $\partial_\phi \equiv 0$ and $\partial_z \equiv 0$, expand the remaining $R$-partials, and move the non-derivative terms to the right-hand side to obtain the system:
\begin{equation}
\partial_t \vct{w} + \mat{A} \partial_R \vct{w} = \vct{s}, \label{lin:linform}
\end{equation}
where
\begin{equation}
\vct{w} = \left[ \begin{array}{c}
\rho \\
v_R \\
v_\phi \\
v_z \\
P \\
B_\phi \\
B_z
\end{array} \right]
\end{equation}
is the vector of primitive variables, omitting the parallel component of the magnetic field,
\begin{equation} 
\mat{A} = \left[ \begin{array}{ccccccc}
v_R & \rho & 0 & 0 & 0 & 0 & 0 \\
0 & v_R & 0 & 0 & 1/\rho & B_\phi/\rho & B_z/\rho \\
0 & 0 & v_R & 0 & 0 & -B_R/\rho & 0 \\
0 & 0 & 0 & v_R & 0 & 0 & -B_R/\rho \\
0 & \gamma P & 0 & 0 & v_R & 0 & 0 \\
0 & B_\phi & -B_R & 0 & 0 & v_R & 0 \\
0 & B_z & 0 & -B_R & 0 & 0 & v_R 
\end{array} \right] \label{lin:wavemat}
\end{equation}
is the wave matrix, and $\vct{s} = \vct{s}_{\rm MHD} + \vct{s}_{\rm grav} + \vct{s}_{\rm geom}$ is the source term vector, a combination of the MHD source terms arising from the $\deldot \vct{B}$ constraint, gravity source terms from a static potential, and the geometric source terms inherent in the cylindrical coordinate system.  As in the Cartesian version of \athnosp, the form of the MHD source terms differs slightly in the 2D and 3D cases (see \S\ref{linind} below), but the forms of the gravity and geometric source terms are independent of dimension.

The hyperbolic wave matrix, $\mat{A}$, given in equation~(\ref{lin:wavemat}), is linearized by taking it to be a constant function of the primitive variable state $\vct{w}$ at time $t^n$.  However, it is only indirectly accessed through the system of eigenvectors and eigenvalues of $\mat{A}$ (see \S\ref{charevo} below).  We write the projected equations in cylindrical coordinates using this specific form in order to make use of the eigensystem solution previously implemented in \athnosp.

In the remainder of this section, we derive the cylindrical coordinate form of the primitive variable system given in equation~(\ref{lin:linform}), and in the process obtain the geometric source terms.

\subsection{Continuity Equation}
Expanding the derivative operators in cylindrical coordinates in equation~(\ref{eqns:cont}) and projecting in the $R$-direction, we have for the continuity equation in primitive variable form:
\begin{equation}
\partial_t \rho + \rho \partial_R v_R + v_R \partial_R \rho = -\rinv \rho v_R. \label{lin:cont}
\end{equation}
The left-hand side of equation~(\ref{lin:cont}) contains all the terms from equation~(\ref{lin:linform}), and the term on the right-hand side is the first component of the geometric source term vector, $\vct{s}_{\rm geom}$.  Furthermore, if we make the substitution $R \mapsto x$ and ignore the source term, we recover the $x$-projection of the continuity equation in Cartesian coordinates.

\subsection{Momentum Equation}
For the momentum equation, we begin with the conservative form of equation~(\ref{eqns:mom}) and use the continuity equation and divergence-free constraint to eliminate terms and obtain:
\begin{equation}
\rho \,\partial_t \vct{v} + \rho(\vct{v} \cdot \nabla)\vct{v} - (\vct{B} \cdot \nabla)\vct{B} + \nabla P^* = 0. \label{lin:mom}
\end{equation}
By explicitly enforcing $\deldot \vct{B}=0$ here, we ensure that any numerical error in the divergence of the magnetic field can not influence the evolution of momentum during the reconstruction step.

Next, we divide through by $\rho$, substitute $P^* = P + B^2/2$, project in the $R$-direction, expand the partials, and move the source terms to the right-hand side to obtain:
\begin{mathletters}
\begin{eqnarray}
\partial_t v_R + v_R \partial_R v_R + \case{1}{\rho} \partial_R P + \case{1}{\rho} B_\phi \partial_R B_\phi + \case{1}{\rho} B_z \partial_R B_z &=& \rinv (v_\phi^2 - \case{1}{\rho} B_\phi^2), \label{lin:moma} \\
\partial_t v_\phi + v_R \partial_R v_\phi - \case{1}{\rho} B_R \partial_R B_\phi &=& -\rinv (v_\phi v_R - \case{1}{\rho} B_\phi B_R), \label{lin:momb} \\
\partial_t v_z + v_R \partial_R v_z - \case{1}{\rho} B_R \partial_R B_z &=& 0. \label{lin:momc}
\end{eqnarray}
\end{mathletters}
Recall that the $\phi$-momentum equation~(\ref{eqns:angmomform}) can be expressed in angular-momentum conserving form and thus avoid a geometric source term.  However, we must include the source term on the right-hand side of equation~(\ref{lin:momb}) in primitive variable form in order to preserve the specific structure of the coefficient matrix, $\mat{A}$, on the left-hand side of equation~(\ref{lin:linform}).  Finally, the gravity source terms in the momentum equation are given by the components of $-\vct{\nabla} \Phi$ in cylindrical coordinates.

\subsection{Energy Equation}
We begin with the internal energy equation in coordinate-free form:
\begin{equation}
\partial_t P + \vct{v} \cdot \nabla P + \gamma P \,\deldot \vct{v} = 0.
\end{equation}
Then, projecting the equations in the $R$-direction, expanding the partials, and moving the source term to the right-hand side, we obtain:
\begin{equation}
\partial_t P + v_R \partial_R P + \gamma P \partial_R v_R = -\rinv \gamma P v_R.
\end{equation}
In primitive form, there is no gravity source term in the energy equation.

{\subsection{Induction Equation} \label{linind}}
For the induction equation~(\ref{eqns:ind}), also written as $\partial_t \vct{B} - \delcross (\vct{v} \times \vct{B}) = 0$, we begin with the components given in equations~(\ref{eqns:indcyla}), (\ref{eqns:indcylc}), and (\ref{eqns:indcylb3}).  Moving terms proportional to $R^{-1} \partial_R (R B_R)$, $R^{-1} \partial_\phi B_\phi$, and $\partial_z B_z$ to the right-hand side, we obtain:
\begin{mathletters}
\begin{eqnarray}
\partial_t B_R + \rinv \partial_\phi (v_\phi B_R) - B_\phi \rinv \partial_\phi v_R && \nonumber \\
+ \partial_z (v_z B_R) - B_z \partial_z v_R &=& v_R [ \rinv \partial_\phi B_\phi + \partial_z B_z ], \label{lin:inda} \\
\partial_t B_\phi + \partial_R (v_R B_\phi) - B_R \partial_R v_\phi && \nonumber \\
+ \partial_z (v_z B_\phi) - B_z \partial_z v_\phi &=& v_\phi [ \rinv \partial_R (R B_R) + \partial_z B_z ] - \rinv v_\phi B_R, \label{lin:indb} \\
\partial_t B_z + \rinv \partial_R [ R (v_R B_z)] - B_R \partial_R v_z && \nonumber \\ 
+ \rinv \partial_\phi (v_\phi B_z) - B_\phi \rinv \partial_\phi v_z &=& v_z [\rinv \partial_R (R B_R) + \rinv \partial_\phi B_\phi] \label{lin:indc}.
\end{eqnarray}
\end{mathletters}
In 2D, the divergence-free constraint in cylindrical coordinates with $\partial_z \equiv 0$ implies that the right-hand side of equation~(\ref{lin:indc}) is identically zero.  Applying the divergence constraint in equation~(\ref{lin:indc}), projecting in the $R$-direction, and expanding the $R$-partials on the left-hand side, we obtain:
{\begin{mathletters} \label{lin:indproj}
\begin{eqnarray}
\partial_t B_R &=& 0, \label{lin:indproja} \\
\partial_t B_\phi + B_\phi \partial_R v_R - B_R \partial_R v_\phi + v_R \partial_R B_\phi &=& v_\phi \rinv \partial_R (R B_R) - \rinv v_\phi B_R, \label{lin:indprojb} \\
\partial_t B_z + B_z \partial_R v_R - B_R \partial_R v_z + v_R \partial_R B_z &=& -\rinv v_R B_z \label{lin:indprojc}.
\end{eqnarray}
\end{mathletters}}
Note that the left-hand sides of equations~(\ref{lin:indprojb}) and~(\ref{lin:indprojc}) exactly match the Cartesian form \citepalias[][eq. 30]{gar05} if we make the substitution $R \mapsto x$.  Also, the divergence term on the right-hand side of equation~(\ref{lin:indprojb}) matches the divergence term on the right-hand side of the Cartesian form, except that it appears in cylindrical coordinate form.  However, the additional source terms $- v_\phi B_R / R$ in equation~(\ref{lin:indprojb}) and $-v_R B_z/R$ in equation~(\ref{lin:indprojc}), which vanish as $R \to \infty$, are curvature-related terms that are unique to the system in cylindrical coordinates.

In 3D, the cancellation of the $\deldot \vct{B}$ terms on the right-hand side of equation~(\ref{lin:indc}) is no longer possible.  Instead, \citetalias{gar08} introduce an algorithm that adds a limited amount of the MHD source terms from the transverse directions to the source terms in each splitting direction.  This is done in such a way that the overall induction equation is not altered and so that the sum of the MHD source terms is minimized.  These constraints take the form of minmod limiter functions that reduce to the underlying 2D algorithm in the limit of 2D, grid-aligned problems \citepalias[see][\S3]{gar08}.  The 3D algorithm in cylindrical coordinates yields the system
{\begin{mathletters} \label{lin:ind3d}
\begin{eqnarray}
\partial_t B_R + \{ \rinv \partial_\phi ( v_\phi B_R - B_\phi v_R ) - v_R L_{R\phi} (\partial_z B_z) \} && \nonumber \\
+ \{ \partial_z ( v_z B_R - B_z v_R ) - v_R L_{Rz} (\rinv \partial_\phi B_\phi) \} &=& 0, \label{lin:ind3da} \\
\partial_t B_\phi + \{ \partial_R ( v_R B_\phi - B_R v_\phi ) - v_\phi L_{\phi R} (\partial_z B_z) \} && \nonumber \\ 
+ \{ \partial_z ( v_z B_\phi - B_z v_\phi ) - v_\phi L_{\phi z} [\rinv \partial_R (R B_R)] \}  &=& 0, \label{lin:ind3db} \\
\partial_t B_z + \{ \rinv \partial_R [ R ( v_R B_z - B_R v_z ) ] - v_z L_{zR} (\rinv \partial_\phi B_\phi) \} && \nonumber \\
+ \{ \rinv \partial_\phi ( v_\phi B_z - B_\phi v_z ) - v_z L_{z \phi} [\rinv \partial_R (R B_R)] \} &=& 0, \label{lin:ind3dc} 
\end{eqnarray}
\end{mathletters}}
where
{\begin{mathletters} \label{lin:limiter1}
\begin{eqnarray}
L_{R\phi} (\partial_z B_z) &\equiv& {\rm minmod} (-\rinv \partial_\phi B_\phi, \;\partial_z B_z), \label{lin:limitera} \\
L_{\phi R} (\partial_z B_z) &\equiv& {\rm minmod} (-\rinv \partial_R (R B_R), \;\partial_z B_z), \label{lin:limiterb} \\
L_{zR} (\rinv \partial_\phi B_\phi) &\equiv& {\rm minmod} (-\rinv \partial_R (R B_R), \;\rinv \partial_\phi B_\phi). \label{lin:limiterc}
\end{eqnarray}
\end{mathletters}}
Note that the limiter $L_{ij}$ is only applied to the equation for $B_i$ projected in the $j$-direction.  We require that
{\begin{mathletters} \label{lin:limiter2}
\begin{eqnarray}
L_{Rz} (\rinv \partial_\phi B_\phi) &=& -L_{R\phi} (\partial_z B_z), \label{lin:limiterd} \\
L_{\phi z} [\rinv \partial_R (R B_R)] &=& -L_{\phi R} (\partial_z B_z), \label{lin:limitere} \\
L_{z \phi} [\rinv \partial_R (R B_R)] &=& -L_{zR} (\rinv \partial_\phi B_\phi), \label{lin:limiterf} 
\end{eqnarray}
\end{mathletters}}
so that the limiters cancel pairwise when summed over all projections.  The limiters defined in equations~(\ref{lin:limiter1}) and (\ref{lin:limiter2}) are the same as in the 3D Cartesian formulae \citepalias[][eqs. 11,15]{gar08}, with $\partial_x B_x \mapsto R^{-1} \partial_R (R B_R)$ and $\partial_y B_y \mapsto R^{-1} \partial_\phi B_\phi$.  

Projecting equations~(\ref{lin:ind3d}) in the $R$-direction, expanding the partials (except for those appearing in the limiter functions), moving all of the source terms to the right-hand side, and using the properties of the minmod function together with the $\deldot \vct{B}=0$ constraint, we obtain:
\begin{mathletters}
\begin{eqnarray}
\partial_t B_R &=& 0, \label{lin:ind3dproja} \\ 
\partial_t B_\phi + B_\phi \partial_R v_R + v_R \partial_R B_\phi - B_R \partial_R v_\phi &=& v_\phi \;{\rm minmod} [\rinv \partial_R (R B_R), \;-\rinv \partial_\phi B_\phi] \nonumber \\
&&- \rinv v_\phi B_R, \label{lin:ind3dprojb} \\
\partial_t B_z + B_z \partial_R v_R + v_R \partial_R B_z - B_R \partial_R v_z &=& v_z \;{\rm minmod} [\rinv \partial_R (R B_R), \;-\partial_z B_z] \nonumber \\
&&- \rinv v_R B_z. \label{lin:ind3dprojc}  
\end{eqnarray}
\end{mathletters}
Note that for the 2D case with $\partial_z \equiv 0$, the $\deldot \vct{B}$ constraint implies that the arguments of the minmod function in equation~(\ref{lin:ind3dprojb}) are equal and that the minmod function in equation~(\ref{lin:ind3dprojc}) evaluates to zero, so that we recover the 2D system in equations~(\ref{lin:indproj}).  The minmod terms on the right-hand side of equations~(\ref{lin:ind3dprojb}) and (\ref{lin:ind3dprojc}) are analogous to the corresponding terms in Cartesian coordinates derived in \citetalias{gar08}, and the remaining terms are geometric.

\subsection{Source Terms}
In summary, for the primitive variable equations in cylindrical coordinates, the MHD source term vectors are given by equations (18) and (19) of \citetalias{gar08} via the substitutions $\partial_x B_x \mapsto R^{-1} \partial_R (R B_R)$ and $\partial_y B_y \mapsto R^{-1} \partial_\phi B_\phi$, and the geometric source term vector is given by
\begin{equation}
\vct{s}_{\rm geom} \equiv \left[ \begin{array}{c}
-\rinv \rho v_R \\
\rinv (v_\phi^2 - \rhoinv B_\phi^2 ) \\
-\rinv (v_\phi v_R - \rhoinv B_\phi B_R ) \\
0 \\
-\rinv \gamma P v_R \\
-\rinv v_\phi B_R \\
-\rinv v_R B_z
\end{array} \right].  \label{lin:geomsource}
\end{equation}
Since the geometric source terms arise directly from the scale factors in the $R$-partials, we associate the geometric source term $\vct{s}_{\rm geom}$ exclusively with the $R$-direction.  Note that $\| \vct{s}_{\rm geom} \| \to 0$ in the limit of vanishing curvature, i.e. as $R \to \infty$.

We emphasize that the geometric source terms in equation~(\ref{lin:geomsource}) are used \emph{only} in obtaining the L/R states, \emph{not} for the final FV update.  Finally, the gravity source terms for the L/R states are given by the cylindrical coordinate components of $-\vct{\nabla} \Phi$ in the momentum equation, and there is no gravity source term in the energy equation.

{\section{Spatial Reconstruction} \label{recon}}
In \athnosp, spatial reconstruction is performed in a directionally-split fashion using piecewise polynomial approximations as outlined in \citealt{col84} (hereafter \citetalias{col84}), and \citealt{col90} (hereafter \citetalias{col90}).  Here, we focus on piecewise linear and quadratic reconstructions, which yield second- and third-order approximations to smooth profiles, respectively.  For a given coordinate direction, $\xi$, we form the piecewise linear or quadratic reconstruction of each primitive variable, $a(\xi)$, from the set $\{a_i\}$ of cell-centered volume-averages (including ghost-zones) at time $t^n$, holding indices $j$ and $k$ fixed.  In each case, we require for consistency that the volume-average of the reconstruction equal the volume-averaged data in the $i$th cell, i.e.
\begin{equation}
a_i = \left< a(\xi) \right>_i \equiv \frac{1}{V_{ijk}} \int_{V_{ijk}} a(\xi) \, dV. \label{recon:consist}
\end{equation}
Instead of defining $a(\xi)$ in the $i$th zone explicitly, i.e. for $\xi \in [\xi_{i-1/2},\xi_{i+1/2}]$, we find it more convenient to define the auxiliary parameter $s \in [0,1]$ by 
\begin{equation}
s \equiv \frac{\xi-\xi_{i-1/2}}{\Delta \xi}, \label{recon:fracvar}
\end{equation}
where $\Delta \xi \equiv \xi_{i+1/2} - \xi_{i-1/2}$ is the width of the interval, so that $\xi = \xi_{i-1/2} + s \,\Delta \xi$.

We also employ slope-limiting and monotonization procedures to ensure that the resulting reconstructions are total-variation-diminishing (TVD) while providing somewhat steeper slopes at discontinuities.  Of course, this can destroy the local formal order of the reconstruction, especially at extrema, but we pay this price for stability.  Note, however, that while monotonicity is a sufficient condition for a reconstruction to be TVD, it is not always necessary \citep{lev02}.  Recently, \citet{col08} have described a slope-limiting method that, when combined with piecewise quadratic reconstruction, preserves the local order of convergence of the reconstruction at extrema.  This has been implemented for Cartesian coordinates in the latest versions of \athnosp, but not for cylindrical coordinates, hence will not be described further here.

For reconstructions in Cartesian coordinates, the procedures for the $y$- and $z$-directions are identical to the procedure for the $x$-direction.  For the reconstructions in cylindrical coordinates, the only non-trivial difference from the Cartesian procedure occurs in the $R$-direction since the discrete cell-volumes change with $R$, but not with $\phi$ or $z$.  Thus, we take $\xi = x$ for the Cartesian cases and $\xi = R$ for the cylindrical cases.  The Cartesian formulae apply, with suitable relabeling of coordinates, for $\phi$- and $z$-reconstructions.

{\subsection{Piecewise Linear (2nd-order) Reconstruction} \label{plm}}
Piecewise linear reconstruction approximates each primitive variable by defining in the $i$th zone,
\begin{equation}
a(s) \equiv a_{L,i} + s \,\Delta a_i \equiv a_{R,i} - (1-s) \,\Delta a_i, \label{recon:plm3}
\end{equation}
where $\Delta a_i \equiv a_{R,i} - a_{L,i}$ represents the difference of some quantity $a$ over the zone, and $a_{L,i}$ and $a_{R,i}$ are the values of $a$ at the left and right interfaces of the zone, respectively.  Thus, to specify $a(s)$ completely, we need only to define $\Delta a_i$ and $a_{L,i}$ for each zone as functions of the volume-averages, $a_i$.

{\subsubsection{PLM in Cartesian Coordinates} \label{cartplm}}
From the consistency requirement in equation~(\ref{recon:consist}) with Cartesian coordinates,
\begin{equation}
a_i = \frac{1}{\Delta x} \int_{x_{i-1/2}}^{x_{i+1/2}} a(x) \,dx = \int_0^1 a(s) \,ds. \label{recon:cartplm2}
\end{equation}
Substituting equation~(\ref{recon:plm3}) into equation~(\ref{recon:cartplm2}) and integrating, we obtain
\begin{equation}
a_i = a_{L,i} + \onehalf \Delta a_i = a_{R,i} - \onehalf \Delta a_i, \label{recon:cartplm3}
\end{equation}
from which
{\begin{mathletters} \label{recon:cartplm4}
\begin{eqnarray}
a_{L,i} &=& a_i - \onehalf \Delta a_i, \label{recon:cartplm4a} \\
a_{R,i} &=& a_i + \onehalf \Delta a_i. \label{recon:cartplm4b}
\end{eqnarray}
\end{mathletters}}
The usual Cartesian formulae for the differences over the zone are
\begin{equation}
\Delta a_i \equiv \left\{ \begin{array}{ll}
\onehalf \left( a_{i+1} - a_{i-1} \right), &{\rm centered} \\
\left( a_{i+1} - a_i \right), &{\rm forward} \\
\left( a_i - a_{i-1} \right), &{\rm backward} \end{array} \right.. \label{recon:cartplmslopes}
\end{equation}
It is clear from equation~(\ref{recon:cartplmslopes}) that constant and linear profiles are reconstructed exactly.  In order to make the reconstruction TVD, these differences (or ``slopes,'' as they are commonly called) are limited using a slight variant of the monotonized central-difference (MC) limiter as described by \citet{lev02}, and flattened to avoid the introduction of new extrema.  Leveque argues that limiting should be performed in characteristic variables so that the accuracy of the reconstruction for smooth wave families is not adversely affected by limiting in other non-smooth wave families.  This requires a special description for systems of conservation laws including a bounded linear transformation from primitive to characteristic variables, with inverse transformation from characteristic variables back to primitive.  We do not go into detail here, but note that the inverse transformation is not guaranteed to be monotonicity-preserving, hence an additional monotonization is performed on the resulting primitive variable differences.  In the previous implementation, this was done in a non-conservative manner, and we have since implemented a related scheme which preserves the consistency requirement of equation~(\ref{recon:consist}).

Finally, by applying the monotonized, limited differences $\Delta a_i$ in equations~(\ref{recon:cartplm3}) and~(\ref{recon:cartplm4}), for smooth profiles we obtain second-order accurate approximations to the values of $a$ across each zone at time $t^n$, except possibly at local extrema.  

{\subsubsection{PLM in Cylindrical Coordinates} \label{cylplm}}
From the consistency requirement of equation~(\ref{recon:consist}) with cylindrical coordinates,
\begin{equation}
a_i = \frac{1}{R_i \,\Delta R} \int_{R_{i-1/2}}^{R_{i+1/2}} a(R) \,R\,dR = \frac{1}{R_i} \int_0^1 a(s) (R_{i-1/2} + s \,\Delta R) \,ds. \label{recon:cylplm2}
\end{equation}
Substituting equation~(\ref{recon:plm3}) into equation~(\ref{recon:cylplm2}) and integrating, we obtain
\begin{equation}
a_i = a_{L,i} + \onehalf \Delta a_i \left( 1 + \gamma_i \right) = a_{R,i} - \onehalf \Delta a_i \left( 1 - \gamma_i \right), \label{recon:cylplm3}
\end{equation}
where 
\begin{equation}
\gamma_i \equiv \frac{\Delta R}{6 \,R_i} \label{recon:gamma}
\end{equation} 
is a correction factor for curvature.  Note that $\gamma_i \to 0$ for fixed $\Delta R$ as $R \to \infty$, or for fixed $R$ as $\Delta R \to 0$, and from the formulae in equation~(\ref{recon:cylplm3}), we recover the Cartesian formulae in equation~(\ref{recon:cartplm3}).  Solving for $a_{L,i}$ and $a_{R,i}$, we obtain
{\begin{mathletters} \label{recon:cylplm4}
\begin{eqnarray}
a_{L,i} &=& a_i - \onehalf \Delta a_i \left( 1 + \gamma_i \right), \label{recon:cylplm4a} \\
a_{R,i} &=& a_i + \onehalf \Delta a_i \left( 1 - \gamma_i \right). \label{recon:cylplm4b}
\end{eqnarray}
\end{mathletters}}
Next, we wish to define the differences $\Delta a_i$ such that constant and linear profiles are reconstructed exactly.  Assuming a linear profile, say $a(R) = CR$ with some constant slope $C$, we enforce the consistency relation of equation~(\ref{recon:cylplm2}) to obtain 
\begin{equation}
a_i \equiv \left< a \right>_i = \left< C R \right>_i = C \left< R \right>_i, \label{recon:cylplm6}
\end{equation}
where
\begin{equation}
\left< R \right>_i = R_i + \frac{(\Delta R)^2}{12\,R_i} \label{recon:ravgcyl}
\end{equation}
is the $R$-coordinate of the volume centroid of the $i$th zone.  To obtain an exact reconstruction for a linear profile, we require that $\Delta a_i = C \,\Delta R$.  First, we consider the Cartesian formula for a centered-difference to obtain
\begin{equation}
\onehalf (a_{i+1} - a_{i-1}) = \onehalf \left( C \left<R_{i+1}\right> - C\left<R_{i-1}\right> \right) = C \,\Delta R \,\left( 1 - \frac{(\Delta R)^2}{12 \,R_{i+1} \,R_{i-1}} \right). \label{recon:cyllinear} 
\end{equation} 
If we divide the Cartesian centered-difference formula in equation~(\ref{recon:cartplmslopes}) by the term in parentheses from equation~(\ref{recon:cyllinear}), we obtain the desired difference, $C \,\Delta R$, exactly.  This---along with similar calculations for the forward- and backward-difference slopes---suggests the following definitions:
\begin{equation}
\Delta a_i \equiv \left\{ \begin{array}{ll}
{\onehalf \left( a_{i+1} - a_{i-1} \right)}/{\left( 1 - \frac{(\Delta R)^2}{12 \,R_{i+1} \,R_{i-1}} \right)}, &{\rm centered} \\
{\left( a_{i+1} - a_i \right)}/{\left( 1 - \frac{(\Delta R)^2}{12 \,R_{i+1} \,R_i} \right)}, &{\rm forward} \\
{\left( a_i - a_{i-1} \right)}/{\left( 1 - \frac{(\Delta R)^2}{12 \,R_i \,R_{i-1}} \right)}, &{\rm backward} \end{array} \right. . \label{recon:cylplmslopes}
\end{equation}
It is clear from equation~(\ref{recon:cylplmslopes}) that constant profiles yield $\Delta a_i = 0$, hence these are also reconstructed exactly.  In order to make the reconstruction TVD, these differences are limited and monotonized as in the Cartesian case.  Finally, by applying the resulting differences to equations~(\ref{recon:plm3}) and~(\ref{recon:cylplm4}), for smooth profiles we obtain second-order accurate approximations to the values of $a$ across each zone at time $t^n$, except possibly at local extrema.  

{\subsection{Piecewise Parabolic (3rd-order) Reconstruction} \label{ppm}}
Piecewise parabolic reconstruction approximates the profile of each primitive variable in the $i$th zone as 
\begin{equation}
a(s) \equiv a_{L,i} + s \,\Delta a_i + s(1-s)a_{6,i} \equiv a_{R,i} - (1-s) \,\Delta a_i + s(1-s)a_{6,i}, \label{recon:ppm3}
\end{equation}
where, as for linear reconstruction, $\Delta a_i \equiv a_{R,i} - a_{L,i}$ represents the average difference of some quantity $a$ over the zone, and $a_{L,i}$ and $a_{R,i}$ are the values of $a$ at the left and right interfaces of the zone, respectively.  The term $a_{6,i}$ is the so-called parabolic coefficient \citepalias[see][eq. 1.5]{col84}.  To specify $a(s)$ completely, we need only to define $\Delta a_i$, $a_{L,i}$, and $a_{6,i}$ for each zone as functions of the set of volume-averages, $\{a_i\}$.

{\subsubsection{PPM Reconstruction in Cylindrical Geometry} \label{cylppm}}
To satisfy the consistency requirement of equation~(\ref{recon:consist}) in cylindrical coordinates, we substitute equation~(\ref{recon:ppm3}) into equation~(\ref{recon:cylplm2}) and integrate, then solve for $a_{6,i}$ in terms of the quantities $a_i$ and $\Delta a_i$:
\begin{equation}
a_{6,i} \equiv 6 \left[ a_i - a_{L,i} - \onehalf \Delta a_i \left( 1 + \gamma_i \right) \right]. \label{recon:cylppm4} \\
\end{equation}
Equation~(\ref{recon:cylppm4}) is equivalent to the definition of the parabolic coefficient appearing in equation (18) of \citet{blo93} (hereafter \citetalias{blo93}).  Recall that $\gamma_i \to 0$ for fixed $\Delta R$ as $R \to \infty$ or for fixed $R$ as $\Delta R \to 0$ (see eq. \ref{recon:gamma}), and in these limits we recover the Cartesian version of the parabolic coefficient, which is the same as equation~(\ref{recon:cylppm4}) except with $\gamma_i=0$ \citepalias[see][eq. 1.5]{col84}.

We wish to define the values of $a_{L,i}$, $a_{R,i}$, and $\Delta a_i \equiv a_{R,i} - a_{L,i}$ such that constant, linear, and parabolic profiles are reconstructed exactly, or at least to second-order.  By constructing a quartic polynomial from the $\{a_i\}$, one can show \citepalias[see][\S3]{blo93} that the cylindrical formulae can be obtained from the corresponding Cartesian formulae \citepalias[see][eq. 1.6]{col84} by making the canonical substitution $a_l \mapsto a_l \,R_l$.  This yields 
{\begin{mathletters} \label{recon:cylppm5}
\begin{eqnarray}
a_{L,i} \,R_{i-1/2} &=& \onehalf \left( a_i \,R_i + a_{i-1} \,R_{i-1} \right) - \case{1}{6} \left( \delta a_i \,R_i - \delta a_{i-1} \,R_{i-1} \right), \label{recon:cylppm5a} \\ 
a_{R,i} \,R_{i+1/2} &=& \onehalf \left( a_{i+1} \,R_{i+1} + a_i \,R_i \right) - \case{1}{6} \left( \delta a_{i+1} \,R_{i+1} - \delta a_i \,R_i \right). \label{recon:cylppm5b} 
\end{eqnarray}
\end{mathletters}}
Here, the centered-, forward-, and backward-differences in zone $i$ are
{\begin{mathletters} \label{recon:cylppm6}
\begin{equation}
\delta a_i \equiv \left\{ \begin{array}{ll}
\onehalf \left( a_{i+1} \,R_{i+1} - a_{i-1} \,R_{i-1} \right)/R_i, &{\rm centered} \\
\left( a_{i+1} \,R_{i+1} - a_{i} \,R_{i} \right)/R_{i+1/2}, &{\rm forward} \\
\left( a_{i} \,R_{i} - a_{i-1} \,R_{i-1} \right)/R_{i-1/2}, &{\rm backward} \end{array} \right., \label{recon:cylppmslopes}
\end{equation}
\end{mathletters}}
and similarly for the $i+1$ and $i-1$ zones.  Note that the corresponding Cartesian formulae are given by taking $R_i \to 1$ in equations~(\ref{recon:cylppm5}) and~(\ref{recon:cylppm6}).

First, assuming a constant profile, $a(R) = C$, and taking the volume average across the $i$th zone, we have that $a_i = C$.  Using centered-differences, it follows that $\delta a_i = C\,\Delta R/R_i$, hence from equations~(\ref{recon:cylppm5}), we see that $a_{L,i} = C$ as desired.  For forward- and backward-differences, a constant profile yields $a_{L,i}=C \left[ 1 + O \left( (\Delta R)^3 \right) \right]$.

Next, assuming a linear profile, $a(R) = CR$, and volume-averaging, we have $a_i$ given using equations~(\ref{recon:cylplm6}) and~(\ref{recon:ravgcyl}) above.  Using either centered-, forward-, or backward-differences, it follows that $\delta a_i = 2 C \,\Delta R$, hence from equations~(\ref{recon:cylppm5}), we see that $a_{L,i} = C R_{i-1/2}$ and $a_{R,i} = C R_{i+1/2}$, as desired.  

Finally, assuming a parabolic profile, $a(R) = C R^2$, it is straightforward to show that the volume average of $R$ across the $i$th zone is 
\begin{equation}
a_i = C \left< R^2 \right>_i = C \left( R_i^2 + \frac{(\Delta R)^2}{4} \right).  
\end{equation}
Using centered-differences, it follows that $\delta a_i = C \,\Delta R \left[ 3R_i + 5 (\Delta R)^2/(4 R_i) \right]$, hence from equations~(\ref{recon:cylppm5}), we see that $a_{L,i} = C R_{i-1/2}^2$, as desired.  For forward- or backward-differences, it can be shown that $a_{L,i} = C R_{i-1/2}^2 \left[ 1 + O \left( \Delta R/R_{i-1/2} \right)^3 \right]$.  Thus, we conclude that parabolic profiles can be recovered up to the required order.  

As in \S\ref{plm}, the slopes (meaning the centered-, forward- and backward-difference $\delta a_i$'s) are monotonized in characteristic form using a slight variant of the MC limiter, as in \citet{lev02}.  Then, after computing the parabolic interpolant, the slopes are re-monotonized to ensure that the interpolation introduces no new extrema.  Following \citetalias{blo93}, the values of $a_{L,i}$ and $a_{R,i}$ are reset to
\begin{equation}
a^*_{L,i} = a^*_{R,i} = a_i
\end{equation}
whenever $a_i$ is a local extremum with respect to $a_{L,i}$ and $a_{R,i}$, or to
{\begin{mathletters} \label{recon:cylppm7} 
\begin{eqnarray}
a^*_{L,i} &=& \frac{6a_i - a_{R,i}\,(4 + 3\gamma_i)}{2 - 3\gamma_i}, \label{recon:cylppm7a} \\
a^*_{R,i} &=& \frac{6a_i - a_{L,i}\,(4 - 3\gamma_i)}{2 + 3\gamma_i}, \label{recon:cylppm7b}
\end{eqnarray}
\end{mathletters}}
whenever they are close enough to $a_i$ so that the parabolic interpolation function introduces new extrema.  The test for this case, 
\begin{equation}
| a_{R,i} - a_{L,i} | \ge | a_{6,i} |, 
\end{equation}
is geometry-independent.  As a result, for smooth profiles we obtain second-order accurate approximations to the values of $a$ across each zone at time $t^n$, except possibly at local extrema.  By taking $\gamma_i = 0$, we recover the Cartesian versions of equations~(\ref{recon:cylppm7}) \citepalias[see][eq. 1.10]{col84}.

{\section{Characteristic Evolution and Averaging} \label{charevo}}
The final step in the calculation of the one-dimensional L/R states is a characteristic time-evolution from $t^n$ to $t^{n+1/2}$ following the methods of \citetalias{col84} and \citetalias{col90}.  This is accomplished by computing the time-averages of the solutions to the linearized primitive variable systems described in \S\ref{lin} at zone interfaces over this half-timestep.  The particular form of the averages depends on the direction, the order of the reconstruction, the coordinate system, etc.  

First, recall the modified primitive variable system described in \S\ref{lin} by equation~(\ref{lin:linform}), where $\mat{A}$ is the linearized hyperbolic wave matrix for the 1D equations projected in the $R$-direction, which is given by equation~(\ref{lin:wavemat}).  Recall further that (weakly) hyperbolic systems of conservation laws have square wave matrices with real eigenvalues.  Thus, let $\lambda^1 \leq \cdots \leq \lambda^M$ represent the $M$ real (but not necessarily distinct) eigenvalues of $\mat{A}$ corresponding to the $M$ linearly independent left- and right-eigenvectors, $\{\vct{l}^\nu,\vct{r}^\nu\}$, where $\nu = 1,\dots,M$.  These eigenvectors are orthonormalized so that $\vct{l}^\mu \cdot \vct{r}^\nu = \delta_{\mu \nu}$.  Thus, any vector $\vct{w} \in \mathbb{R}^M$ (in particular the centered-, forward-, or backward-differences across the $i$th zone, $\vct{\Delta w}_i$) has the right-eigenvector expansion
\begin{equation}
\vct{w} = \sum_{\nu = 1}^M a^\nu \vct{r}^\nu \label{charevo:expand}
\end{equation}
with the coefficients $a^\nu = \vct{l}^\nu \cdot \vct{w}$ representing the components of the projection of $\vct{w}$ onto the left-eigenspace of $\mat{A}$.  

Next, we note that the characteristic form of the primitive variable system is obtained by multiplication of equation~(\ref{lin:linform}) on the left by $\mat{L}$, the matrix whose rows are the left-eigenvectors of $\mat{A}$, i.e. $\mat{L} = \{\vct{l}^1,\ldots,\vct{l}^M\}^T$, hence $\mat{L} \mat{A} = \mat{\Lambda} \mat{L}$ where $\mat{\Lambda}$ is the diagonal matrix consisting of the eigenvalues of $\mat{A}$.  Neglecting source terms for the moment, we obtain the homogeneous linear system
\begin{equation}
\partial_t \vct{a} + \mat{\Lambda} \,\partial_R \vct{a} = 0, \label{charevo:charform}
\end{equation}
where $\vct{a} \equiv \mat{L} \vct{w}$ is the vector of characteristic variables.  From the form of equation (\ref{charevo:charform}), these eigenvalues $\{\lambda^\nu\}$ evidently represent the signal speeds of wave families along characteristics.  Furthermore, the system in equation~(\ref{charevo:charform}) decouples into $M$ constant-coefficient linear advection equations of the form
\begin{equation}
\partial_t a + \lambda \,\partial_R a = 0, \label{charevo:chardecouple} \\
\end{equation}
which have the solution
\begin{equation}
a(R,t^n+\tau) = a^n(\xi = R-\lambda \tau), \label{charevo:chardecouplesoln} \\
\end{equation}
where $a^n(\xi)$ is the reconstructed solution at time $t^n$.  Since the solution in equation (\ref{charevo:chardecouplesoln}) depends only on $a(\xi)$, for each characteristic wave impinging on the interface, the contribution to the time-averaged interface state is given by the volume average of this reconstruction over the domain of dependence defined by the wave's characteristic speed, $\lambda$, and the time interval $(t^n,t^n+\Delta t)$.  

In a time $\Delta t$, a right-moving wave travels a distance $\lambda \,\Delta t$ in the $R$-direction.  The volume in cylindrical coordinates of the domain of dependence of the left interface state at $R_{i+1/2}$ upon this wave is given by $V_{\rm DOD} = (R_{i+1/2}- \lambda \,\Delta t/2) \,\lambda \,\Delta t \,\Delta\phi \,\Delta z$.  Thus, with $\chi_R \equiv \lambda \,\Delta t / \Delta R$ and $\chi_L \equiv -\lambda \,\Delta t / \Delta R$, we have the average of $a$ over $V_{\rm DOD}$ equal to
{\begin{mathletters} \label{charevo:cylplmavg6}
\begin{equation}
f^a_{L,i+1/2}(\chi_R) = \frac{1}{(R_{i+1/2} - \onehalf \chi_R \,\Delta R) \,\chi_R} \int_{1-\chi_R}^{1} a(s) \,(R_{i-1/2} + s \,\Delta R) \,ds. \label{charevo:cylplmavg6a}
\end{equation}
Similarly, over the domain of dependence of the right interface state at $R_{i-1/2}$ upon a left-moving wave, we have
\begin{equation}
f^a_{R,i-1/2}(\chi_L) = \frac{1}{(R_{i-1/2} + \onehalf \chi_L \,\Delta R) \,\chi_L} \int_{0}^{\chi_L} a(s) \,(R_{i-1/2} + s \,\Delta R) \,ds. \label{charevo:cylplmavg6b}
\end{equation}
\end{mathletters}}

{\subsection{PLM Evolution in Cylindrical Geometry} \label{plmevolve}}
Here, we describe the evaluation of the L/R states at time $t^{n+1/2}$ based on a piecewise linear reconstruction of the underlying profile at time $t^n$, defined by equations~(\ref{recon:plm3}),~(\ref{recon:cylplm4}) and~(\ref{recon:cylplmslopes}) of \S\ref{plm}.  Substituting $a(s)$ into equations~(\ref{charevo:cylplmavg6}) and integrating, we obtain
{\begin{mathletters} \label{charevo:cylplmavg7}
\begin{equation}
f^a_{L,i+1/2}(\chi_R) = a_{R,i} - \onehalf \chi_R \,\Delta a_i \,\left(1-\beta_{R,i}(\chi_R) \right) \label{charevo:cylplmavg7a}
\end{equation}
on the right side of the zone (left of the interface) and
\begin{equation}
f^a_{R,i-1/2}(\chi_L) = a_{L,i} + \onehalf \chi_L \,\Delta a_i \,\left(1+\beta_{L,i}(\chi_L) \right) \label{charevo:cylplmavg7b}
\end{equation}
\end{mathletters}}
at the left of the zone (right of the interface).  Here, $a_{L,i}$ and $a_{R,i}$ are the values of $a$ at the left and right interfaces of the $i$th zone, respectively, $\Delta a_i$ is the monotonized difference of $a$ across the zone from equation~(\ref{recon:cylplmslopes}), and we have defined the functions
{\begin{mathletters} \label{charevo:beta}
\begin{eqnarray}
\beta_{R,i}(\chi_R) &\equiv& \frac{\chi_R \,\Delta R}{6 ( R_{i+1/2}-\onehalf \chi_R \,\Delta R ) }, \label{charevo:betar} \\
\beta_{L,i}(\chi_L) &\equiv& \frac{\chi_L \,\Delta R}{6 ( R_{i-1/2}+\onehalf \chi_L \,\Delta R ) }, \label{charevo:betal}
\end{eqnarray}
\end{mathletters}}
as additional correction factors due to the curvature of the zone.  Note that \mbox{$\beta_{R,i}(\chi_R),\beta_{L,i}(\chi_L) \to 0$} as $R \to \infty$, i.e. in the limit of vanishing curvature, in which case equations~(\ref{charevo:cylplmavg7}) reduce to the Cartesian formulae.  Note further that in averaging $a(s)$ over the whole zone, i.e. taking $\chi_{L/R}=1$, we have $\beta_{R,i}(1) = \beta_{L,i}(1) = \gamma_i$ (see eq. \ref{recon:gamma}), and from equations~(\ref{charevo:cylplmavg7}), we recover the averages of equation~(\ref{recon:cylplm3}).

{\subsection{PPM Evolution in Cylindrical Geometry} \label{ppmevolve}}
Here, we describe the evaluation of the L/R states at time $t^{n+1/2}$ based on a piecewise parabolic reconstruction of the underlying profile at time $t^n$, defined by equations~(\ref{recon:ppm3}),~(\ref{recon:cylppm4}),~(\ref{recon:cylppm5}), and~(\ref{recon:cylppm6}) of \S\ref{ppm}.  Substituting $a(s)$ into equations~(\ref{charevo:cylplmavg6}) and integrating, we obtain expressions analogous to equations~(\ref{charevo:cylplmavg7}) for the time-average of the right(left)-moving waves over the domain of dependence of the left(right) interface state at $R_{i+1/2}$($R_{i-1/2}$) upon these waves:
{\begin{mathletters} \label{charevo:cylppmavg7}
\begin{eqnarray}
f^a_{L,i+1/2}(\chi_R) &=& a_{R,i} - \onehalf \chi_R \left[ \Delta a_i - \left( 1 - \twothirds \chi_R \right) a_{6,i} \right] \nonumber \\ 
&& + \onehalf \chi_R \left[ \Delta a_i - \left( 1 - \chi_R \right) a_{6,i} \right] \beta_{R,i}(\chi_R), \label{charevo:cylppmavg7a} \\ 
f^a_{R,i-1/2}(\chi_L) &=& a_{L,i} + \onehalf \chi_L \left[ \Delta a_i + \left( 1 - \twothirds \chi_L \right) a_{6,i} \right] \nonumber \\ 
&& + \onehalf \chi_L \left[ \Delta a_i + \left( 1 - \chi_L \right) a_{6,i} \right] \beta_{L,i}(\chi_L), \label{charevo:cylppmavg7b}.
\end{eqnarray}
\end{mathletters}}
The functions $\beta_{R,i}(\chi_R)$ and $\beta_{L,i}(\chi_L)$ defined in equations~(\ref{charevo:beta}) are the correction factors due to the curvature of the zone.  The PLM result in equations~(\ref{charevo:cylplmavg7}) corresponds to setting $a_{6,i}=0$.  Note that equations~(\ref{charevo:cylppmavg7}) also reduce to the Cartesian formulae \citepalias[see][eq. 1.12]{col84} when $\beta \to 0$ as $R \to \infty$.  Note further that in averaging $a(s)$ over the whole zone, i.e. taking $\chi_{L/R}=1$, we have $\beta_{R,i}(1) = \beta_{L,i}(1) = \gamma_i$, and the results in equations~(\ref{charevo:cylppmavg7}) are consistent with equation~(\ref{recon:cylppm4}).

\subsection{Sum Over Characteristics}
Once time-averaged L/R states have been obtained in the characteristic variables (as in eqs. \ref{charevo:cylplmavg7} or \ref{charevo:cylppmavg7}), we convert back to the primitive variables using $\mat{R}$, the matrix consisting of the right-eigenvectors of $\mat{A}$.  Since $\mat{L}$ transforms from primitive to characteristic variables via $\vct{a} = \mat{L}\vct{w}$, and since $\mat{R}\mat{L} = \mat{I}$, where $\mat{I}$ is the standard identity matrix, $\mat{R}\vct{a}=\vct{w}$ accomplishes the inverse transformation.

Defining $a^{\nu,n+1/2}_{L,i+1/2} \equiv f^a_{L,i+1/2}(\chi_R^\nu)$ and $a^{\nu,n+1/2}_{R,i-1/2} \equiv f^a_{R,i-1/2}(\chi_L^\nu)$ for each characteristic variable, we obtain the total time-averaged L/R states in primitive variable form by summing the projections of these contributions onto the right-eigenspace:
\begin{equation}
\vct{w}^{n+1/2}_{L/R,i \pm 1/2} = \mat{R} \,\vct{a}^{n+1/2}_{L/R,i \pm 1/2} = \sum_{\nu = 1}^{M} a^{\nu,n+1/2}_{L/R,i \pm 1/2} \vct{r}^\nu. \label{charevo:wavesum}
\end{equation}
For example, in the Cartesian case \citep[][eq. 42]{sto08}, using equation~(\ref{charevo:cylplmavg7a}) or (\ref{charevo:cylppmavg7a}) with $\beta=0$, we have
\begin{equation}
\vct{w}^{n+1/2}_{L,i+1/2} = \vct{w}_{i}^n + \left( \frac{1}{2} - \frac{\lambda^M \,\Delta t}{2\Delta x} \right) \left( \Delta \vct{w} \right)_i - \frac{\Delta t}{2\Delta x} \sum_\nu \left( \lambda^\nu - \lambda^M \right) [ \vct{l}^\nu \cdot (\Delta \vct{w})_i] \vct{r}^\nu. \label{charevo:42sto08}
\end{equation}

As written in equation~(\ref{charevo:wavesum}), the inverse transformation includes a sum over all waves.  However, following \citetalias{col84} and \citetalias{col90}, contributions to $\vct{w}$ on a given interface from waves that propagate in the opposite direction may be discarded to yield a more robust solution for strongly nonlinear problems, i.e. the waves are upwinded in the appropriate direction.  Thus, for left-moving waves with $\lambda < 0$, we may set $\chi_{L/R}=0$ in equation~(\ref{charevo:cylplmavg7a}) to obtain $f^a_{L,i+1/2}(0) = a_{R,i}$.  Similarly for right-moving waves with $\lambda > 0$, we may set $\chi_{L/R}=0$ in equation~(\ref{charevo:cylplmavg7b}) to obtain $f^a_{R,i-1/2}(0) = a_{L,i}$.

\citet{sto08} have noted that this upwinding destroys the formal second-order convergence for smooth flows.  In this case, the 1D L/R states are accurate to the desired order if and only if we account for \emph{all} waves, including those propagating toward the interfaces from the outside of the zone.  With this approach, the sum in equation~(\ref{charevo:42sto08}) includes all $\lambda^\nu$.  This means that for a given zone we integrate an extrapolation of the local reconstruction profile over a domain of dependence that lies \emph{outside} the zone.  However, for non-smooth flows, we have found that this can lead to significant errors near discontinuities, since we may end up extrapolating the local reconstruction beyond a point of discontinuity into a region where it is no longer a good approximation of the profile.  Thus, for flows that may contain discontinuities, we follow \citetalias{col84} and restrict to integration only over characteristics that propagate toward the interfaces from the interior of the zone, as previously described.  However, we do not use the reference states for waves propagating away from the interface.  Thus, we compute the states using
{\begin{mathletters} \label{upwindonly}
\begin{eqnarray}
\vct{w}^{n+1/2}_{L,i+1/2} = \vct{w}_{i}^n + \frac{1}{2} \left( \Delta \vct{w} \right)_i - \frac{\Delta t}{2\Delta x} \sum_{\nu:\;\lambda^\nu > 0} \lambda^\nu \,[ \vct{l}^\nu \cdot (\Delta \vct{w})_i] \vct{r}^\nu, \label{charevo:upwindonlyl} \\
\vct{w}^{n+1/2}_{R,i-1/2} = \vct{w}_{i}^n - \frac{1}{2} \left( \Delta \vct{w} \right)_i - \frac{\Delta t}{2\Delta x} \sum_{\nu:\;\lambda^\nu < 0} \lambda^\nu \,[ \vct{l}^\nu \cdot (\Delta \vct{w})_i] \vct{r}^\nu, \label{charevo:upwindonlyr}
\end{eqnarray}
\end{mathletters}}
for Cartesian PLM, which we refer to as ``upwind-only'' integration.  For PLM in cylindrical coordinates, factors $(1-\beta_{R,i}(\xi_R^\nu))$ and $(1+\beta_{L,i}(\xi_L^\nu))$ are included in
the sums in equations~(\ref{charevo:upwindonlyl}) and~(\ref{charevo:upwindonlyr}), respectively, using equations~(\ref{charevo:cylplmavg7a}) and~(\ref{charevo:cylplmavg7b}).  Equations~(\ref{charevo:cylppmavg7a}) and~(\ref{charevo:cylppmavg7b}) are used to obtain analogous expressions for PPM in cylindrical coordinates.

{\section{Finite Volume Method} \label{fvm}}
The FV method uses approximate time- and area-averaged interface fluxes to update volume-averaged quantities.  In cylindrical coordinates $(R,\phi,z)$, the differential volume element is
\begin{equation}
dV = R \,dR \,d\phi \,dz, \label{fvm:volelement}
\end{equation}
and the finite grid cell volume and interface area are:
\begin{mathletters}
\begin{eqnarray}
V_{ijk} &=& R_i \,\Delta R \,\Delta \phi \,\Delta z, \label{fvm:dv} \\
A_{R; \;i \pm 1/2,j,k} &=& R_{i \pm 1/2} \,\Delta \phi \,\Delta z, \label{fvm:da1} \\
A_{\phi; \;i,j \pm 1/2,k} &=& \Delta R \,\Delta z, \label{fvm:da2} \\
A_{z; \;i,j,k \pm 1/2} &=& R_i \,\Delta R \,\Delta \phi. \label{fvm:da3} 
\end{eqnarray}
\end{mathletters}
To derive the FV method, we integrate the system in equations~(\ref{eqns:compact}) over the volume of a given grid cell, apply Gauss's Divergence Theorem, and integrate in time from $t^n$ to $t^{n+1}$ to obtain
\begin{eqnarray}
\vct{Q}_{ijk}^{n+1} = \vct{Q}_{ijk}^n &-& \frac{\Delta t}{R_i \,\Delta R} \left( R_{i+1/2} \,\vct{F}^{n+1/2}_{R; \;i+1/2,j,k} - R_{i-1/2} \,\vct{F}^{n+1/2}_{R; \;i-1/2,j,k} \right) \nonumber \\
&-& \frac{\Delta t}{R_i \,\Delta \phi} \left( \vct{F}^{n+1/2}_{\phi; \;i,j+1/2,k} - \vct{F}^{n+1/2}_{\phi; \;i,j-1/2,k} \right) \nonumber \\
&-& \frac{\Delta t}{\Delta z} \left( \vct{F}^{n+1/2}_{z; \;i,j,k+1/2} - \vct{F}^{n+1/2}_{z; \;i,j,k-1/2} \right) \nonumber \\
&+& \Delta t \,\vct{S}^{n+1/2}_{ijk}, \label{fvm:update1}
\end{eqnarray}
where $\vct{Q}$ represents the volume-averaged conserved quantities, $\vct{F}$ represents the time- and area-averaged fluxes, and $\vct{S}$ represents the time- and volume-averaged source terms.

For cylindrical coordinates, in \S\ref{eqns} we rewrote the $\phi$-momentum equation in a modified angular-momentum preserving form in order to reduce the number of source terms on the right-hand side of the system.  Therefore, when we apply the procedure in equation~(\ref{fvm:update1}) to equation~(\ref{eqns:angmomform}), it yields the FV update for the \emph{angular} momentum, $R\rho v_\phi$, not the linear momentum, $\rho v_\phi$.  However, it can be shown that the radial contribution from a quasi-FV update of the $\rho v_\phi$ equation~(\ref{eqns:redangmomform}), 
\begin{equation}
- \frac{\Delta t}{R_i^2 \,\Delta R} \left( R_{i+1/2}^2 F_{R; \;i+1/2,j,k}^{n+1/2} - R_{i-1/2}^2 F_{R; \;i-1/2,j,k}^{n+1/2} \right) \label{fvm:update2}
\end{equation}
is equivalent to the corresponding terms in the true FV update of equation~(\ref{eqns:momcylb}),
\begin{equation}
- \frac{\Delta t}{R_i \,\Delta R} \left( R_{i+1/2} F_{R; \;i+1/2,j,k}^{n+1/2} - R_{i-1/2} F_{R; \;i-1/2,j,k}^{n+1/2} \right) - \Delta t \,\left\langle \frac{M_{R \phi}}{R} \right\rangle_{ijk}^{n+1/2}, \label{fvm:update3}
\end{equation}
to second-order away from the origin for smooth flows.  Note that equation~(\ref{fvm:update3}) contains the volume- and time-averaged geometric source term, $-M_{R \phi}/R = -(\rho v_R v_\phi - B_R B_\phi)/R$, whereas equation~(\ref{fvm:update2}) has no source term.  

The only nonzero component of the geometric source term, $\vct{S}^{n+1/2}_{{\rm geom},ijk}$, required for the FV update is in the radial momentum equation~(\ref{eqns:momcyla}).  For this term, we must compute
\begin{equation}
\left\langle \frac{M_{\phi\phi}}{R} \right\rangle_{ijk}^{n+1/2} = \frac{1}{\Delta t \,\Delta V_{ijk}} \int_{t^n}^{t^{n+1}} \int_{V_{ijk}} \frac{\rho v_\phi^2 - B_\phi^2 + P^*}{R} \,dV \,dt. \label{fvm:geomsource}
\end{equation}
Na\"ively, one might compute this source term from volume averaged quantities at time $t^n$.  However, to achieve second-order accuracy, it is necessary to use time-centered estimates of these quantities, i.e. advanced to the half-timestep $t^{n+1/2}$.  One can think of this as a sort of trapezoid rule applied to the time domain.  This half-timestep advance is performed using a combination of FV updates on the volume-centered variables $\rho$ and $\rho v_\phi$ and CT updates on the interface-centered $B_\phi$.  For the $P^*$ contribution, the FV and CT updates to $t^{n+1/2}$ are too costly, since they would be required for every variable and they, in turn, would require source term calculations.  Instead, we compute the total pressure contribution directly from the fluxes at $R$-interfaces.  The appropriate second-order average is
\begin{equation}
\left\langle \frac{P^*}{R} \right\rangle_{ijk} \approx \frac{R_{i+1/2} P^*_{i+1/2} + R_{i-1/2} P^*_{i-1/2}}{2 R_i^2}, \label{fvm:ptotal}
\end{equation}
where $P^*_{i\pm 1/2}$ are the time-averaged pressure fluxes returned directly from the Riemann solver.  

Our application of the geometric source term is similar to what is done for the gravity source terms in the existing \ath code.  However, we have found it easier to maintain centrifugal balance numerically by using the analytic gravitational acceleration function $\vct{g}(\vct{x}) \equiv -\vct{\nabla} \Phi(\vct{x})$ in the momentum equation, rather than approximations of the gradient using finite-differences of the static potential, $\Phi$.  We approximate the gravitational source term for the momentum equation by
\begin{equation}
\vct{S}^{n+1/2}_{{\rm grav},ijk} \equiv -\langle \rho \,\vct{\nabla} \Phi \rangle_{ijk}^{n+1/2} \simeq \langle \rho \rangle_{ijk}^{n+1/2} \,\vct{g}(\left\langle \vct{x} \right\rangle_{ijk}), \label{fvm:gravsource}
\end{equation}
where $\langle \vct{x} \rangle_{ijk}$ is the volume-centroid of cell $(i,j,k)$, the radial component of which is given by equation~(\ref{recon:ravgcyl}).  Note that for the case of solid-body rotation with uniform density, $\vct{g} \propto R$, so that the gravitational source term given by equation~(\ref{fvm:gravsource}) is exact.  For the energy equation, we rely on the previously implemented FV update based on the potential function $\Phi(\vct{x})$, which allows the energy equation to be written conservatively.  

To compute the gravitational source terms appearing in the calculation of the L/R states, which only appear in the momentum equation in primitive variable form, we evaluate $\vct{g}(\vct{x})$ at the area-centroid of each interface.  Note that the $R$-coordinate of the area-centroid of $\phi$- and $z$-interfaces coincides with the $R$-coordinate of the volume-centroid of each adjacent grid cell.

{\section{Constrained Transport} \label{ct}}
In this section, we discuss modifications in cylindrical coordinates to the constrained transport (CT) algorithm described in \citetalias{gar05} and \citet{eva88}.  As argued in those papers, the integral form of the induction equation~(\ref{eqns:ind}) is most naturally expressed in terms of finite area-averages rather than volume-averages.  In this way, the equation becomes a statement of the conservation of total magnetic flux through a given grid cell and as such automatically preserves the $\deldot \vct{B}$ constraint.  

{\subsection{Integral Form and Consistency Relations} \label{consist}}
To see this, we rewrite the induction equation  as
\begin{equation}
\partial_t \vct{B} + \delcross \Emf = 0, \label{ct:ind}
\end{equation}
where $\Emf = -\vct{v} \times \vct{B}$ is the electric field in ideal MHD (the electromotive force [EMF]).  Then, integrating over the oriented bounding surface of grid cell $(i,j,k)$ and applying Stokes' Theorem, we find that
{\begin{mathletters} \label{ct:indall}
\begin{eqnarray}
B^{n+1}_{R; \;i\pm 1/2,j,k} &=& B^{n}_{R; \;i\pm 1/2,j,k} - \frac{\Delta t}{R_{i\pm 1/2} \,\Delta \phi} \left( \emf^{n+ 1/2}_{z; \;i\pm 1/2,j+ 1/2,k} - \emf^{n+ 1/2}_{z; \;i\pm 1/2,j- 1/2,k} \right), \nonumber \\
&&+ \frac{\Delta t}{\Delta z} \left( \emf^{n+ 1/2}_{\phi; \;i\pm 1/2,j,k+ 1/2} - \emf^{n+ 1/2}_{\phi; \;i\pm 1/2,j,k- 1/2} \right) \label{ct:inda} \\
B^{n+1}_{\phi; \;i,j\pm 1/2,k} &=& B^{n}_{\phi; \;i,j\pm 1/2,k} + \frac{\Delta t}{\Delta R} \left( \emf^{n+ 1/2}_{z; \;i+ 1/2,j\pm 1/2,k} - \emf^{n+ 1/2}_{z; \;i- 1/2,j\pm 1/2.k} \right) \nonumber \\
&&- \frac{\Delta t}{\Delta z} \left( \emf^{n+ 1/2}_{R; \;i,j\pm 1/2,k+ 1/2} - \emf^{n+ 1/2}_{R; \;i,j\pm 1/2,k- 1/2} \right), \label{ct:indb} \\
B^{n+1}_{z; \;i,j,k\pm 1/2} &=& B^{n}_{z; \;i,j,k\pm 1/2} - \frac{\Delta t}{R_i \,\Delta R} \left( \emf^{n+ 1/2}_{\phi; \;i+ 1/2,j,k\pm 1/2} - \emf^{n+ 1/2}_{\phi; \;i- 1/2,j,k\pm 1/2} \right) \nonumber \\
&&+ \frac{\Delta t}{R_i \,\Delta \phi} \left( \emf^{n+ 1/2}_{R; \;i,j+ 1/2,k\pm 1/2} - \emf^{n+ 1/2}_{R; \;i,j- 1/2,k\pm 1/2}\right), \label{ct:indc}
\end{eqnarray}
\end{mathletters}}
where 
{\begin{mathletters} \label{ct:flux}
\begin{eqnarray}
B^{n}_{R; \;i\pm 1/2,j,k} &\equiv& \frac{1}{R_{i\pm 1/2} \,\Delta \phi \,\Delta z} \int_{z_{k- 1/2}}^{z_{k+ 1/2}} \int_{\phi_{j- 1/2}}^{\phi_{j+ 1/2}} B_R(R_{i\pm 1/2},\phi,z,t^n) \,R_{i\pm 1/2} \,d\phi \,dz, \label{ct:fluxa} \\
B^{n}_{\phi; \;i,j\pm 1/2,k} &\equiv& \frac{1}{\Delta R \,\Delta z} \int_{z_{k- 1/2}}^{z_{k+ 1/2}} \int_{R_{i- 1/2}}^{R_{i+ 1/2}} B_\phi(R,\phi_{j\pm 1/2},z,t^n) \,dR \,dz, \label{ct:fluxb} \\
B^{n}_{z; \;i,j,k\pm 1/2} &\equiv& \frac{1}{R_i \,\Delta R \,\Delta \phi} \int_{\phi_{j- 1/2}}^{\phi_{j+ 1/2}} \int_{R_{i- 1/2}}^{R_{i+ 1/2}} B_z(R,\phi,z_{k\pm 1/2},t^n) \,R \,dR \,d\phi, \label{ct:fluxc}
\end{eqnarray}
\end{mathletters}}
are the interface area-averaged components of the magnetic field normal to each surface (i.e. the magnetic flux per unit area) and
{\begin{mathletters} \label{ct:emf3}
\begin{eqnarray}
\emf^{n+ 1/2}_{R; \;i,j\pm 1/2,k\pm 1/2} &\equiv& \frac{1}{\Delta t \,\Delta R} \int_{t^n}^{t^{n+1}} \int_{R_{i- 1/2}}^{R_{i+ 1/2}}  \emf_R(R,\phi_{j\pm 1/2},z_{k\pm 1/2},t) \,dR \,dt, \label{ct:emf3a} \\
\emf^{n+ 1/2}_{\phi; \;i\pm 1/2,j,k\pm 1/2} &\equiv& \frac{1}{\Delta t \,R_{i\pm 1/2} \,\Delta \phi} \int_{t^n}^{t^{n+1}} \int_{\phi_{i- 1/2}}^{\phi_{i+ 1/2}} \emf_\phi(R_{i\pm 1/2},\phi,z_{k\pm 1/2},t) \,R_{i\pm 1/2} \,d\phi \,dt, \label{ct:emf3b} \\
\emf^{n+ 1/2}_{z; \;i\pm 1/2,j\pm 1/2,k} &\equiv& \frac{1}{\Delta t \,\Delta z} \int_{t^n}^{t^{n+1}} \int_{z_{i- 1/2}}^{z_{i+ 1/2}} \emf_z(R_{i\pm 1/2},\phi_{j\pm 1/2},z,t) \,dz \,dt \label{ct:emf3c}
\end{eqnarray}
\end{mathletters}}
are the corner-centered EMFs averaged over the edges bounding each surface.  The EMFs in equations~(\ref{ct:emf3}) are approximated to some desired order of accuracy and the surface-averaged field components are evolved using equations~(\ref{ct:indall}).  

The interface-centered, area-averaged magnetic field components in equations~(\ref{ct:flux}) comprise the fundamental representation of the magnetic field in \athnosp.  However, one often needs to refer to the cell-centered, volume-averaged magnetic field components as well.  Therefore, we adopt the averages
{\begin{mathletters} \label{ct:avg}
\begin{eqnarray}
B^{R,n}_{i,j,k} &\equiv& \frac{1}{2R_i} \left( R_{i- 1/2} B^{n}_{R; \;i- 1/2,j,k} + R_{i+ 1/2} B^{n}_{R; \;i+ 1/2,j,k} \right), \label{ct:avga} \\
B^{\phi,n}_{i,j,k} &\equiv& \frac{1}{2} \left( B^{n}_{\phi; \;i,j- 1/2,k} + B^{n}_{\phi; \;i,j+ 1/2,k} \right), \label{ct:avgb} \\
B^{z,n}_{i,j,k} &\equiv& \frac{1}{2} \left( B^{n}_{z; \;i,j,k- 1/2} + B^{n}_{z; \;i,j,k+ 1/2} \right). \label{ct:avgc}
\end{eqnarray}
\end{mathletters}}
Note the use of an $R$-weighted average of the $B_R$ interface values in equation~(\ref{ct:avga}); it is straightforward to show that this is the appropriate second-order accurate average in cylindrical coordinates.  Equations~(\ref{ct:avg}) imply consistency relations between the Godunov fluxes computed by the Riemann solver (the fluxes of the volume-averaged magnetic field components) and the corner-centered EMFs (the fluxes of the area-averaged magnetic field components).  These relations define how they are computed from each other \citepalias[see][for details]{gar05}.

\subsection{Calculating the EMFs}
The primary modification to the CTU+CT algorithm described in \citetalias{gar05} for cylindrical coordinates concerns the calculation of the upwinded, corner-centered EMF component $\emf_z$.  As we shall demonstrate, we must combine spatial gradients of different curvature to form $\emf_z$.  However, to form $\emf_R$ or $\emf_\phi$, we combine spatial gradients of the \emph{same} curvature, hence the effect of that curvature is cancelled out and no subsequent modification is necessary.  

For example, to compute $\emf_{z; \;i-1/2,j-1/2}$, we estimate $(\partial_\phi \emf_z)_{i-1/2,j-3/4}$ and use a centered-difference scheme to calculate one estimate:
\begin{equation}
\emf_{z; \;i-1/2,j-1/2} = \emf_{z; \;i-1/2,j-1} + \case{R_{i-1/2} \Delta \phi}{2} \left( \partial_\phi \emf_z \right)_{i-1/2,j-3/4}.
\end{equation}
In the same manner, we integrate $\emf_z$ to the corner from each of the remaining adjacent interface centers and take the arithmetic average:
\begin{eqnarray}
\emf_{z; \;i-1/2,j-1/2} &=& \onequarter \left( \emf_{z; \;i-1/2,j-1} + \emf_{z; \;i-1/2,j} + \emf_{z; \;i-1,j-1/2} + \emf_{z; \;i,j-1/2} \right) \nonumber \\
&&+ \case{R_{i-1/2} \,\Delta \phi}{8} \left[ \left( \partial_\phi \emf_z \right)_{i-1/2,j-3/4} - \left( \partial_\phi \emf_z \right)_{i-1/2,j-1/4} \right] \nonumber \\
&&+ \case{\Delta R}{8} \left[ \left( \partial_R \emf_z \right)_{i-3/4,j-1/2} - \left( \partial_R \emf_z \right)_{i-1/4,j-1/2} \right]. \label{ct:emfavg}
\end{eqnarray}

To ensure stability, for $(\partial_\phi \emf_z)_{i-1/2,j-3/4}$, we use the upwinding scheme \citepalias[][eq. 50]{gar05} based on the sign of the mass flux at the center of each interface:
\begin{equation}
(\partial_\phi \emf_z)_{i-1/2,j-3/4} = \left\{
\begin{array}{ll}
(\partial_\phi \emf_z)_{i-1,j-3/4}, & v_{R; \;i-1/2,j-1}>0 \\
(\partial_\phi \emf_z)_{i,j-3/4}, & v_{R; \;i-1/2,j-1}<0 \\
\onehalf \left[ (\partial_\phi \emf_z)_{i-1,j-3/4} + (\partial_\phi \emf_z)_{i,j-3/4} \right]
, & {\rm otherwise}
\end{array} \right. .  \label{ct:emfupwind}
\end{equation}
The formulae for the remaining gradients are analogous.  

To obtain the estimates of $\partial_\phi \emf_z$ needed in equation~(\ref{ct:emfupwind}), we use a centered-difference scheme based on the cell-centered EMFs, computed using volume-averages of $\rho$, $\rho \vct{v}$, and $\vct{B}$, and on the interface-centered EMFs, which come directly from the fluxes:
{\begin{mathletters} \label{ct:emf6}
\begin{eqnarray}
\left( \partial_\phi \emf_z \right)_{i,j-3/4} &=& \case{2}{R_{i} \,\Delta \phi} \left( \emf_{z; \;i,j-1/2} - \emf_{z; \;i,j-1} \right), \label{ct:emf6a} \\
\left( \partial_\phi \emf_z \right)_{i-1,j-3/4} &=& \case{2}{R_{i-1} \,\Delta \phi} \left( \emf_{z; \;i-1,j-1/2} - \emf_{z; \;i-1,j-1} \right). \label{ct:emf6b}
\end{eqnarray}
\end{mathletters}}
Note that this scheme has no dependence on the particular type of Riemann solver used to calculate the fluxes.  Furthermore, note the different radial scale factors appearing in equations~(\ref{ct:emf6}) resulting from the combination of $\phi$-gradients at different radii; the factors of $\Delta \phi$ cancel the factor of $\Delta \phi$ from equation~(\ref{ct:emfavg}), but the radial scale factors themselves do not cancel and must be inserted into the algorithm.  On the other hand, when $\phi$-gradients are combined at the \emph{same} radius, as is the case for the corner-integration of $\emf_{R}$, both the $\Delta \phi$ and the corresponding radial scale factors cancel, hence no modification is required.

{\section{The \ath Algorithm} \label{alg}
In this section, we summarize in somewhat greater detail the main steps of the six-solve version of the CTU+CT algorithm adapted from \citet{sto08} for cylindrical coordinates \citepalias[see also][for details]{gar05,gar08}.
\begin{enumerate}
\item Compute the first-order source terms, $\vct{S}^{*}_{i,j,k}$, in both conservative and primitive variable forms using the initial volume-averaged data at time $t^n$.  These include the geometric source terms, $\vct{S}_{\rm geom}$ (eq.~\ref{lin:geomsource} for primitive variables, and eq.~\ref{fvm:geomsource} for conserved variables), the gravitational source terms, $\vct{S}_{\rm grav}$, computed from static accelerations and potentials (eq.~\ref{fvm:gravsource} for conserved variables), and the MHD source terms arising from the $\deldot \vct{B}$ constraint (e.g. eqs.~\ref{lin:limiter1} and \ref{lin:limiter2} for 3D).
\label{alg:step1}
\item Compute the L/R interface states, $\vct{Q}^{L/R,*}_{i-1/2,j,k}$, $\vct{Q}^{L/R,*}_{i,j-1/2,k}$, and $\vct{Q}^{L/R,*}_{i,j,k-1/2}$, by using the desired reconstruction scheme on the initial data in primitive variable form.  This requires reconstruction with characteristic evolution as described in \S\S\ref{recon} and \ref{charevo}, followed by application of the parallel components of the (primitive variable) source terms from step~(\ref{alg:step1}).  \label{alg:step2}
\item Compute the first-order interface fluxes, $\vct{F}^*_{R; \;i-1/2,j,k}$, $\vct{F}^*_{\phi; \;i,j-1/2,k}$, and $\vct{F}^*_{z; \;i,j,k-1/2}$, from the interface states via an exact or approximate Riemann solver. \label{alg:step3}
\item Compute the corner-centered electric field components, $\emf^*_{R; \;i,j-1/2,k-1/2}$, $\emf^*_{\phi; \;i-1/2,j,k-1/2}$, and $\emf^*_{z; \;i-1/2,j-1/2,k}$, from components of the interface-centered fluxes from step~(\ref{alg:step3}) and the cell-centered electric field computed using the initial data at time $t^n$, via equations~(\ref{ct:emfavg}) and~(\ref{ct:emfupwind}) for the $z$-components. \label{alg:step4}
\item Update the interface magnetic field components for a half-timestep using the CT difference equations~(\ref{ct:indall}) and the EMFs from step~(\ref{alg:step4}). \label{alg:step5}
\item Compute the updated L/R interface states, $\vct{Q}^{L/R,n+1/2}_{i-1/2,j,k}$, $\vct{Q}^{L/R,n+1/2}_{i,j-1/2,k}$, and $\vct{Q}^{L/R,n+1/2}_{i,j,k-1/2}$, by applying transverse flux gradients to the non-magnetic variables of the interface states and then adding the transverse components of the source terms from step~(\ref{alg:step1}). \label{alg:step6}
\item Use the fluxes from step~(\ref{alg:step3}) and the source terms from step~(\ref{alg:step1}) to compute the velocities at the half-timestep using conservative FV updates of the cell-centered density and momentum at time $t^n$.  Average the half-timestep interface magnetic field components from step~(\ref{alg:step5}) to obtain the cell-centered magnetic field components at time $t^{n+1/2}$ using equations~(\ref{ct:avg}).  Then, calculate the cell-centered electric field components, $\emf^{n+1/2}_{R; \;i,j,k}$, $\emf^{n+1/2}_{\phi; \;i,j,k}$, and $\emf^{n+1/2}_{z; \;i,j,k}$, using the cell-centered velocities and magnetic fields at time $t^{n+1/2}$.  \label{alg:step7}
\item Compute the second-order interface fluxes, $\vct{F}^{n+1/2}_{R; \;i-1/2,j,k}$, $\vct{F}^{n+1/2}_{\phi; \;i,j-1/2,k}$, and $\vct{F}^{n+1/2}_{z; \;i,j,k-1/2}$, using the updated interface states from steps~(\ref{alg:step5}) and (\ref{alg:step6}) via the Riemann solver. \label{alg:step8}
\item Compute the corner-centered electric field components, $\emf^{n+1/2}_{R; \;i,j-1/2,k-1/2}$, $\emf^{n+1/2}_{\phi; \;i-1/2,j,k-1/2}$, and $\emf^{n+1/2}_{z; \;i-1/2,j-1/2,k}$, from components of the updated fluxes from step~(\ref{alg:step8}) and the cell-centered electric field components computed in step~(\ref{alg:step7}), as in step~(\ref{alg:step4}).  \label{alg:step9}
\item Use the fluxes from step~(\ref{alg:step8}) to obtain half-timestep conservative FV updates of the cell-centered density and $\phi$-momentum at time $t^n$, similar to step~(\ref{alg:step7}).  Compute the cell-centered total pressure, $P^*$, from the interface-centered quantities returned by the Riemann solver in step~(\ref{alg:step8}) using equation~(\ref{fvm:ptotal}).  Combine the cell-centered $B_\phi$ from step~(\ref{alg:step7}) with $\rho$, $\rho v_\phi$, and $P^*$ to construct the second-order geometric source term (eq. \ref{fvm:geomsource}) at time $t^{n+1/2}$.  Combine $\rho$ and the static gravitational acceleration, $\vct{g}$, to construct the components of the gravity source term for the momentum equation at time $t^{n+1/2}$.  Combine the interface-centered $\rho \vct{v}$, obtained directly from the fluxes in step~(\ref{alg:step8}), to construct the gravity source term for the energy equation at time $t^{n+1/2}$.  \label{alg:step10}
\item Using the fluxes from step~(\ref{alg:step8}) and source terms from step~(\ref{alg:step10}), advance the cell-centered quantities from time $t^n$ to $t^{n+1}$ using conservative FV updates on the hydrodynamic variables (mass, momentum, and energy) and using the CT difference equations~(\ref{ct:ind}) with the EMFs from step~(\ref{alg:step9}) to update the interface magnetic field components. \label{alg:step11}
\item Average the updated interface magnetic field components from step~(\ref{alg:step11}) to compute the updated cell-centered values using equations~(\ref{ct:avg}).  \label{alg:step12}
\item Increment the time to $t^{n+1} = t^n + \Delta t$ and then compute a new timestep using the standard CFL condition based on the maximum signal speed at cell centers and on the size of the grid cells.  Here, we must use $R_i\,\Delta\phi$ to estimate the CFL stability criterion, since the Riemann solvers compute linear wavespeeds.  \label{alg:step13}
\end{enumerate}

This algorithm is simplified for the purely hydrodynamic case.  Besides having fewer variables to store, reconstruct, and evolve, there is no need to compute MHD source terms, magnetic components of geometric source terms, or corner- or cell-centered EMFs, or to apply FV or CT updates to magnetic field components.

In adapting the code for cylindrical coordinates, we have altered several steps of the original Cartesian algorithm to varying degrees.  First, we significantly change the computation of the L/R states for the $R$-direction in step~(\ref{alg:step2}); only a minor change is needed for the $\phi$-direction and no change is needed for the $z$-direction.  Second, we add geometric scale factors to the flux differences in the conservative FV updates performed in steps~(\ref{alg:step7}),~(\ref{alg:step10}), and~(\ref{alg:step11}), and we include the geometric source terms computed in steps~(\ref{alg:step1}) and~(\ref{alg:step10}).  As detailed in \S\S\ref{eqns} and~\ref{lin}, the geometric source terms applied in steps~(\ref{alg:step2}) and~(\ref{alg:step6}) differ from those applied in steps~(\ref{alg:step7}),~(\ref{alg:step10}), and~(\ref{alg:step11}) because of the differences in the primitive and conservative forms of the evolution equations, and furthermore, the source term contribution from the total pressure, $P^*$, is computed differently in steps~(\ref{alg:step1}) and~(\ref{alg:step10}), as discussed in \S\ref{fvm}.  Finally, we change the CT calculation in steps~(\ref{alg:step4}) and~(\ref{alg:step10}) to reflect the additional geometric scale factors appearing in the cylindrical coordinate version of the induction equation~(\ref{ct:ind}) and to enforce modified consistency relations, as described in \S\ref{consist}.

{\section{Code Verification Tests} \label{tests}}

In this section, we present a suite of tests of our cylindrical coordinate adaptation of the \ath code.  Some are drawn from tests published by other authors (\citetalias{gar05}, \citetalias{gar08}, \citealt{lon00}, \citealt{sak85}), which were originally written to test Cartesian codes, including \ath itself, while others are new.  Where possible, we have tried to make comparisons with the existing Cartesian tests in order to demonstrate our code's ability to recover their results both qualitatively and quantitatively.  We include tests in one, two, and three spatial dimensions, in both hydrodynamics and MHD, with solutions that are both smooth and non-smooth, and having varying levels of symmetry.

{\subsection{Force Balance} \label{cylforce}}
In this problem, we investigate various steady equilibria in order to evaluate the code's ability to balance forces.  While there are many possible tests to choose from, we give here two representative examples demonstrating simple magnetohydrostatic equilibria.

First, we consider the axisymmetric magnetic field
\begin{equation}
\vct{B} = \frac{B_0}{R} \hat{\phi}, \label{cylforce:cylbphi}
\end{equation}
for which the outward magnetic pressure and inward tension forces sum to zero.  Note that the magnetic field given by equation~(\ref{cylforce:cylbphi}) satisfies the $\deldot \vct{B} = 0$ constraint.  We use $B_0=1$ and set the velocity $\vct{v}=0$, mass density $\rho=1$, and gas pressure $P=1$.  Figure~\ref{cylbphi:l1error} shows the convergence of the $L_2$ norm of the $L_1$ error vector (RMS error) for the solution at time $t=10$, defined as
\begin{equation}
\delta \vct{q} = \frac{1}{N} \sum_i | \vct{q}_i - \vct{q}_i^0 |,
\end{equation}
where $\vct{q}_i^0$ is the initial solution.  For reference, we plot a line of slope $-2$ (dashed) alongside the error to demonstrate that the convergence is second-order in $1/N$.  These data were computed using the HLLD fluxes and third-order reconstruction; the results were similar for all combinations of Roe or HLLD fluxes, second- or third-order reconstruction, and 1D, 2D, or 3D integrators.  However, because the 2D and 3D algorithms differ significantly from the 1D version, especially in their inclusion of transverse flux gradients and CT updates, we also present the results of the same test using these integrators on grids which are essentially one-dimensional, but contain a few grid cells in each transverse direction considered.  By symmetry, it is clear that any number of cells may be used in the transverse directions, but since the grid cell volumes change with $R$ in multidimensions, the CFL condition will determine the timestep based on grid cell volume \emph{as well as} the maximum signal speed, so the absolute errors should \emph{not} be compared between, say, the 1D and 2D algorithms.  Only the order of convergence of each individual algorithm is meaningful.  Additionally, we have performed tests with the outward acceleration $v_\phi^2/R$ from a solid-body rotation profile $v_\phi=\Omega_0 R$ balanced by the gradient of the static gravitational potential $\Phi = (\Omega_0 R)^2/2$, as well as with constant $v_z \ne 0$, and find similar results in all cases.  

Second, we consider the non-axisymmetric magnetic field 
\begin{equation}
\vct{B} = \frac{B_0 \,cos(\psi)}{R}\hat{R}, \label{cylforce:cylbr}
\end{equation}
which when combined with the gas pressure
\begin{equation}
P = P_0 + \frac{B_0^2 \,[1 + \sin^2(\psi)]}{2 R^2}
\end{equation}
yields zero net force in the $\phi$-direction.  The combination of static gravitational potential 
\begin{equation}
\Phi_1 = - \frac{B_0^2}{2\rho_0 R^2}
\end{equation}
and mass density
\begin{equation}
\rho = \rho_0 [1 + \sin^2(\psi)]
\end{equation}
together balance the gradient of the gas pressure in the $R$-direction.  Here, we use the angular coordinate $\psi = 2\pi(\phi-\phi_{\rm min})/(\phi_{\rm max}-\phi_{\rm min})$ so that $B_R$, $P$, and $\rho$ are all periodic in the $\phi$-domain.  Note that the magnetic field given by equation~(\ref{cylforce:cylbr}) satisfies the $\deldot \vct{B}=0$ constraint.  We use $B_0=1$, $P_0=1$, $\rho_0=1$, and once again use the solid-body rotation profile $v_\phi=\Omega_0 R$ balanced by the gradient of the static gravitational potential $\Phi_2= (\Omega_0 R)^2/2$, where $\Omega_0=\pi/4$.  Note that the total potential is given by $\Phi=\Phi_1+\Phi_2$.

Figure~\ref{cylbr:l1error} shows the convergence of the RMS error for the solution at time $t=10$.  For reference, we once again plot a line of slope $-2$ (dashed) alongside the error to demonstrate that the convergence is second-order.  These data were computed using the HLLD fluxes, third-order reconstruction, and the 2D integrator; the results were similar for all combinations of the Roe or HLLD fluxes, second- or third-order reconstruction, and 2D or 3D integrators.  We use a computational domain of size $N$-by-$N$ with $R \in [1,2]$, $\phi \in [0,\pi/4]$, and $z=0$.  We use periodic boundary conditions in the $\phi$- and $z$-directions, and Dirichlet boundary conditions in the $R$-direction.  Similar results were obtained using a Neumann boundary condition in the $R$-direction.  

\begin{figure}
\centering
\epsscale{0.85}
\plotone{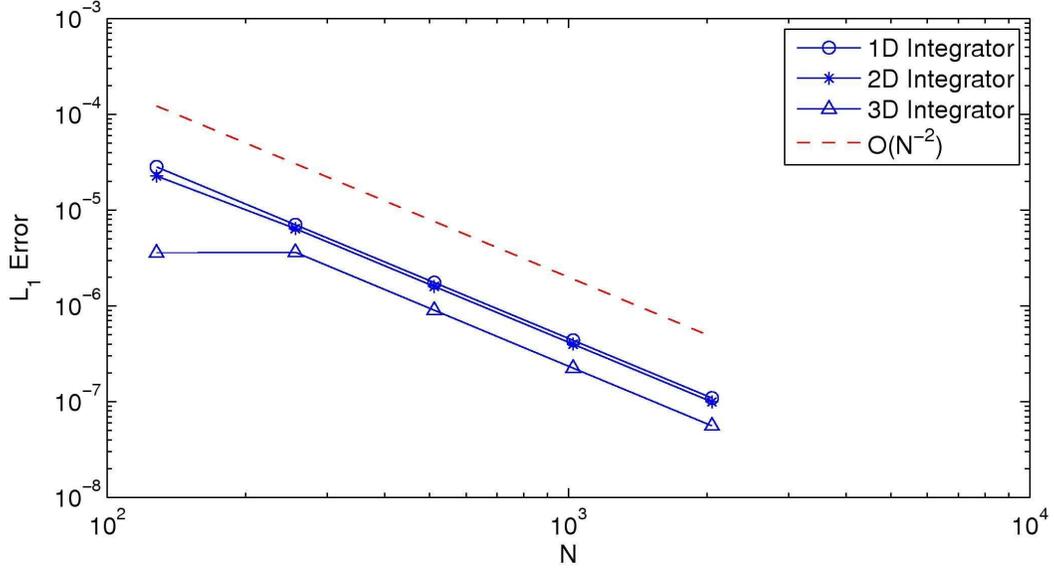}
\caption{Convergence of the RMS error in the $L_1$-norm for the $B_\phi$ force-balance problem in 1D, 2D and 3D.  For reference, we have plotted a line of slope $-2$ (dashed) to show that the convergence is second-order in $1/N$. \label{cylbphi:l1error}}
\end{figure}

\begin{figure}
\centering
\epsscale{0.85}
\plotone{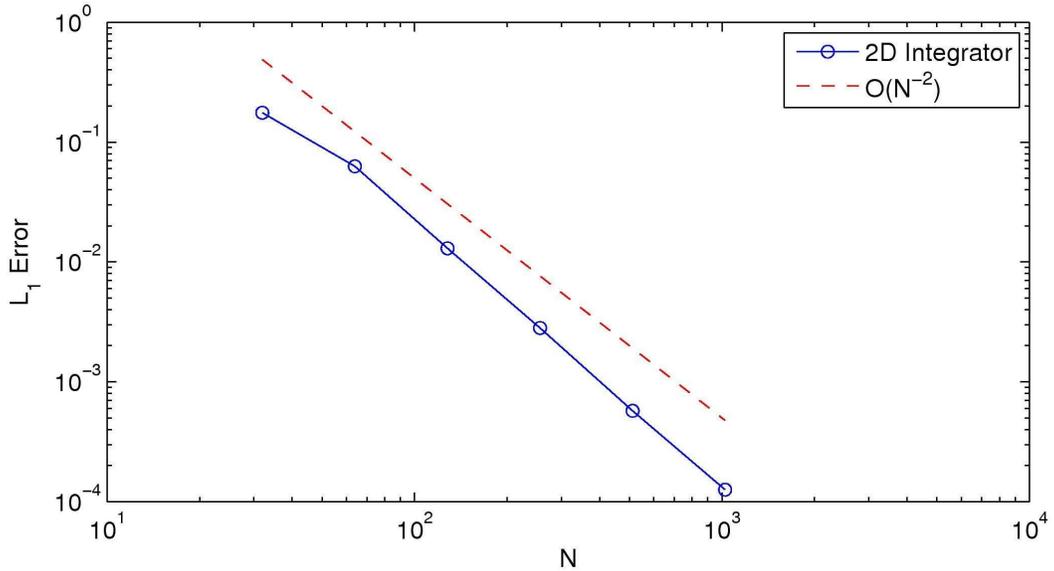}
\caption{Convergence of the $L_1$-error for the 2D $B_R$ force-balance problem in 2D.  For reference, we have plotted a line of slope $-2$ (dashed) to show that the convergence is second-order in $1/N$. \label{cylbr:l1error}}
\end{figure}

{\subsection{Rotational Stability} \label{cylrayleigh}}
In this problem, we investigate the stability of rotating disks evolved with our code, using the 2D integrator.  Given a differential rotation profile, $\Omega(R)$, Rayleigh's criterion for stability to axisymmetric, infinitesimal disturbances is that specific angular momentum increase outward:
\begin{equation}
\partial_R \left[ (R^2 \Omega(R))^2 \right] > 0. \label{cylrayleigh:stabcrit}
\end{equation}
While it is possible that systems satisfying Rayleigh's criterion are still subject to growth of finite-amplitude non-axisymmetric disturbances, laboratory measurements at Reynolds number up to $2 \times 10^6$ have found that Couette flows violating Rayleigh's criterion show large angular momentum transport associated with turbulence, while those satisfying Rayleigh's criterion do not \citep{ji06}.

For our test, we consider power-law rotational profiles of the form $\Omega(R) \propto R^{-q}$, where $q$ is a constant, the so-called ``shear parameter.''  Rayleigh's criterion in a differentially rotating system of this form predicts stability when $q<2$, and instability for $q>2$.  For example, Keplerian rotation with $q=1.5$ is predicted to be stable (for unmagnetized flows).

Using a constant background density and pressure, we set
\begin{equation}
v_\phi(R) = R \,\Omega(R) = \Omega_0 R^{1-q}
\end{equation}
and set the gravitational potential so that rotational equilibrium is achieved.  Next, we perturb $v_\phi$ to
\begin{equation}
\tilde{v}_\phi = v_\phi + \delta v_\phi,
\end{equation}
where $\delta v_\phi$ is a random variable uniformly distributed in $[-\epsilon,\epsilon]$, and $\epsilon$ is small, typically on the order of $10^{-4}$.  The same initial perturbation is used for each value of $q$ considered.  We use a grid of $200 \times 400$ cells over the domain $[3,7] \times [0,\pi/2]$, which is chosen so that $R_{\text{avg}} \,\Delta \phi \sim 2 \Delta R$.  We set $\rho_0 = 200$, $P_0 = 1$, and use an adiabatic index of $\gamma=5/3$ which gives $c_s \approx 0.09$.  With  $\Omega_0 = 2\pi$, $v_{\phi,\text{min}} \approx 0.9$, which puts the Mach numbers in a range of approximately $10$-$20$ over the domain.  Thus, the flow is rotationally dominated.  

As a diagnostic of instability, we compute a scaled mean perturbed angular momentum flux,
\begin{equation}
\frac{\langle R \,\rho v_R \,\delta v_\phi \rangle}{\langle R P \rangle} = \frac{\iint R \,\rho v_R (v_\phi-R\Omega) \,R\,dR\,d\phi}{\iint RP \,R\,dR\,d\phi}.
\end{equation}
For stable flows, this will remain on the order of the initial perturbation, but for unstable flows, it will diverge exponentially.  Figure~\ref{cylrayleigh:angmomtransp} shows the values of the dimensionless angular momentum flux as a function of time for $t \in [0,300]$ for various values of the shear parameter near the marginal stability limit of $q=2$.  Consistent with Rayleigh's criterion, the flows with $q < 2$ remain stable, and those with $q > 2$ go unstable.  For the $q=2.05$ case, the instability reaches saturation more quickly (around $t=90$) and the mass flies off the grid, but the characteristic exponential growth is observed before this point.  This test demonstrates the code's accurate conservation of angular momentum near the boundary of rotational stability.  

Additionally, we investigate the long-term stability of the rotation profiles given by the shear parameters $q=1$, typical of galactic disk systems, and $q=1.5$, typical of Keplerian systems.  For this test we use the unperturbed equilibrium solutions as initial data.  As a diagnostic of the error, we compute the cumulative mean of the dimensionless background angular momentum flux (proportional to the radial accretion rate):
\begin{equation}
\frac{\langle \langle R \,\rho v_R \,(\Omega R) \rangle \rangle}{\langle \langle R P \rangle \rangle} = \frac{\iiint R \,\rho v_R (\Omega R) \,R\,dR\,d\phi\,dt}{\iiint RP \,R\,dR\,d\phi\,dt}.
\end{equation}
Figure~\ref{cylrayleigh:radialaccretion} shows the dimensionless cumulative mean as a function of time for $t \in [0,300]$.  We use the same computational domain and background state as above.  Since $v_\phi=\Omega R \propto R^{1-q}$ is constant for $q=1$, and constant profiles are reconstructed exactly in our code, we observe relatively small errors for this level of discretization, indicating that angular momentum is conserved very well for systems of astrophysical interest.

\begin{figure}
\centering
\epsscale{0.85}
\plotone{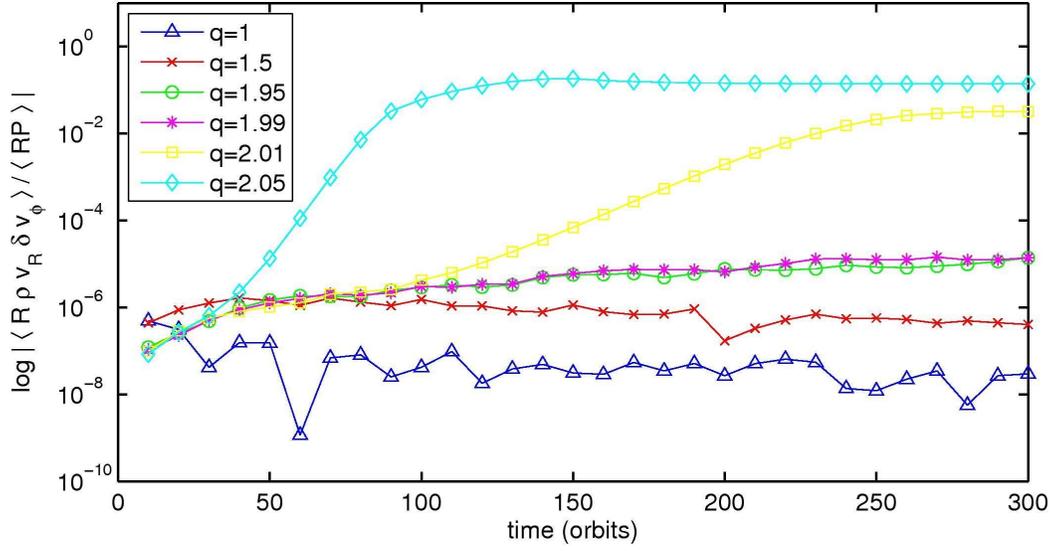}
\caption{Mean dimensionless angular momentum transport as a function of time in the Rayleigh rotational stability test for various values of $q$. \label{cylrayleigh:angmomtransp}}
\end{figure}

\begin{figure}
\centering
\epsscale{0.85}
\plotone{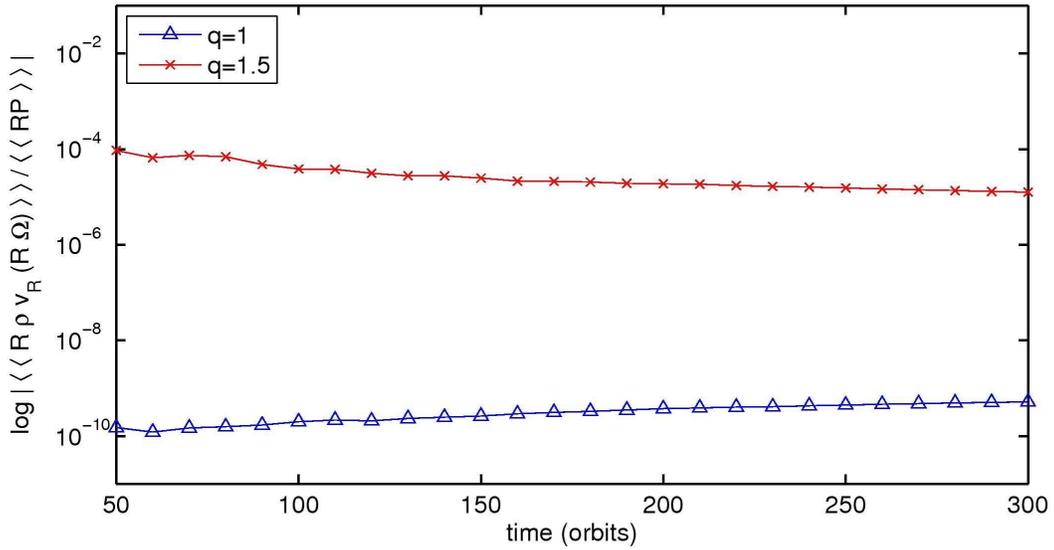}
\caption{The cumulative mean (time- and space-average) of the dimensionless background radial angular momentum flux as a function of time for shear parameters $q=1$ and $q=1.5$ with unperturbed initial data. \label{cylrayleigh:radialaccretion}}
\end{figure}

{\subsection{Adiabatic Blast Wave} \label{cylblast_B0}}
In this problem, we investigate a strong 2D shock using the HLLC solver.  We use the parameter set of \citetalias{gar08} and compare the outputs of our cylindrical code with those from the Cartesian version of \athnosp.  For the Cartesian version we use the domain $(x,y) \in [-0.5,0.5] \times [-0.75,0.75]$, and for the cylindrical version we use the domain $(R,\phi) \in [1,2] \times [-0.5,0.5]$ so that the physical domain spans an arc-length of $R_\text{mid}(\phi_\text{max}-\phi_\text{min})=1.5$ at $R_\text{mid}=1.5$, giving a roughly similar domain sizes.  The initial conditions consist of a circular region of hot gas with radius $R=0.1$ and pressure $P=10$ in an ambient medium of uniform pressure $P_0=0.1$ and density $\rho_0=1$.  We use a computational grid of $200 \times 300$ cells, third-order reconstruction, and the HLLC fluxes with upwind-only integration for the L/R states in each version of the test.  Contour plots of the density, pressure and specific kinetic energy densities of the evolved state at time $t=0.2$ are shown in Figure~\ref{cylblast_B0:2d_contour} using the cylindrical (left column) or Cartesian (right column) version of \athnosp.  For additional comparison, 1D plots of these variables along a horizontal line through the center of the blast are shown in Figure~\ref{cylblast_B0:2d_lineout}, demonstrating excellent agreement between the Cartesian and cylindrical versions of \athnosp.  Notice in the Cartesian version that symmetry is perfectly preserved by the integrator, which is most easily seen in the grid noise in the interior of the shell in Figure~\ref{cylblast_B0:2d_contour}.  Symmetry is also preserved rather well in the cylindrical version although in this case the grid is non-uniform.

\begin{figure}
\centering
\epsscale{0.6}
\plottwo{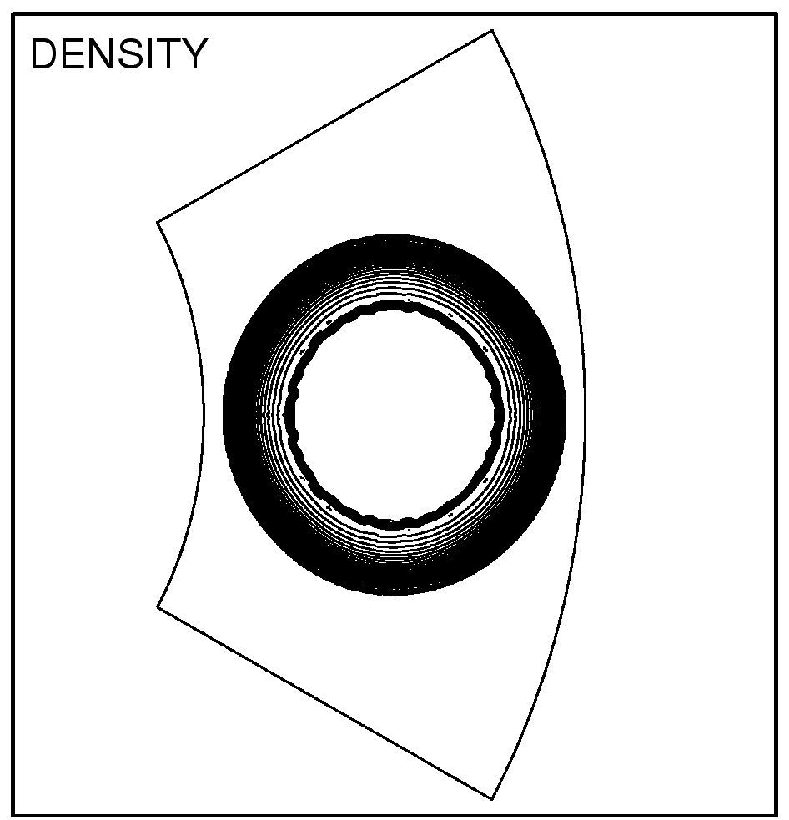}{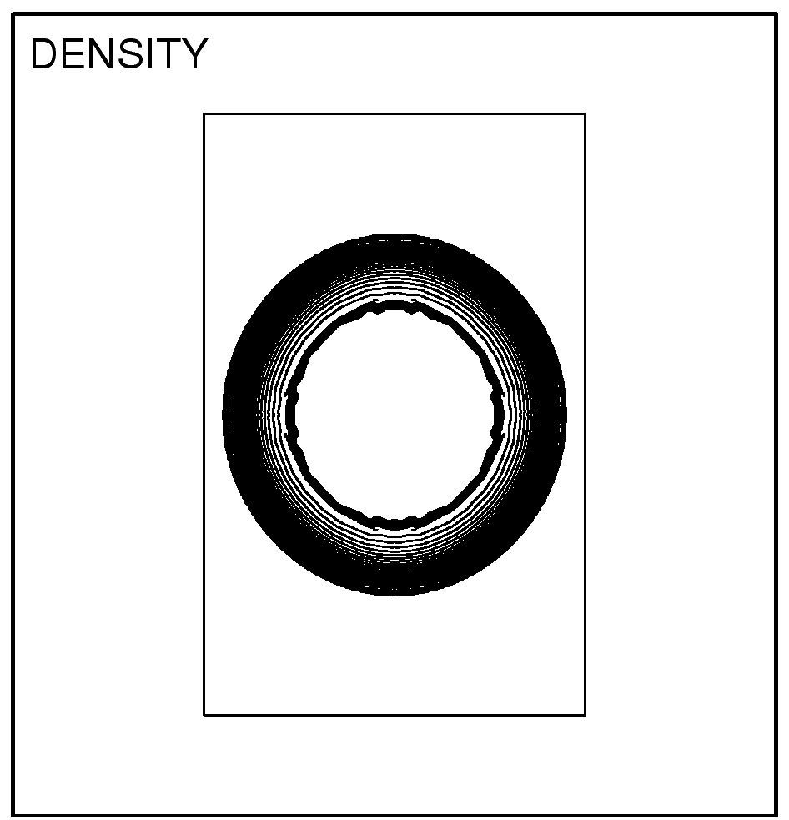} \\
\plottwo{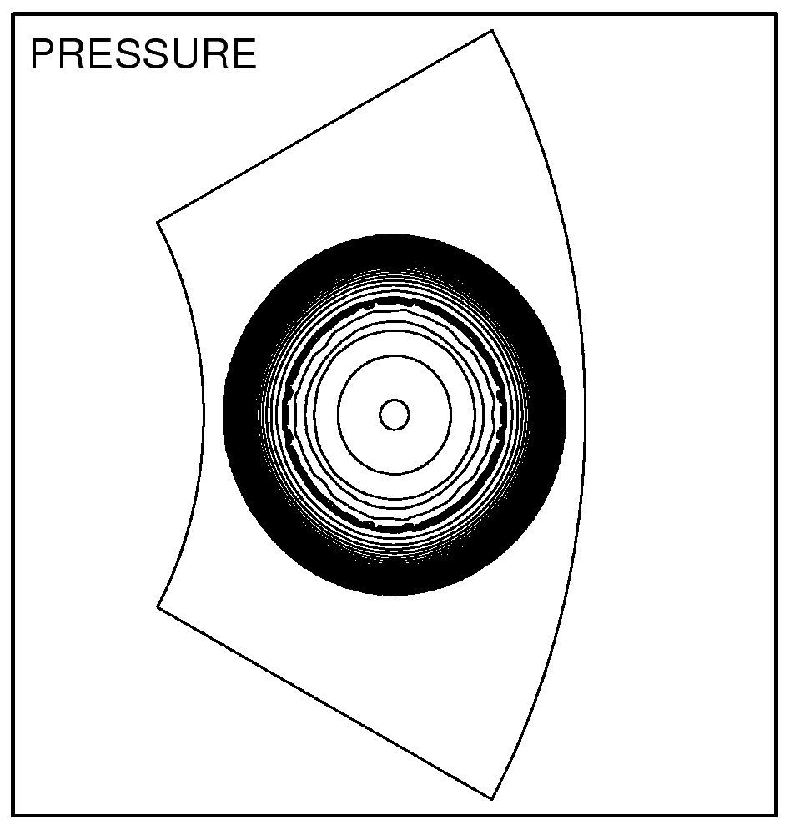}{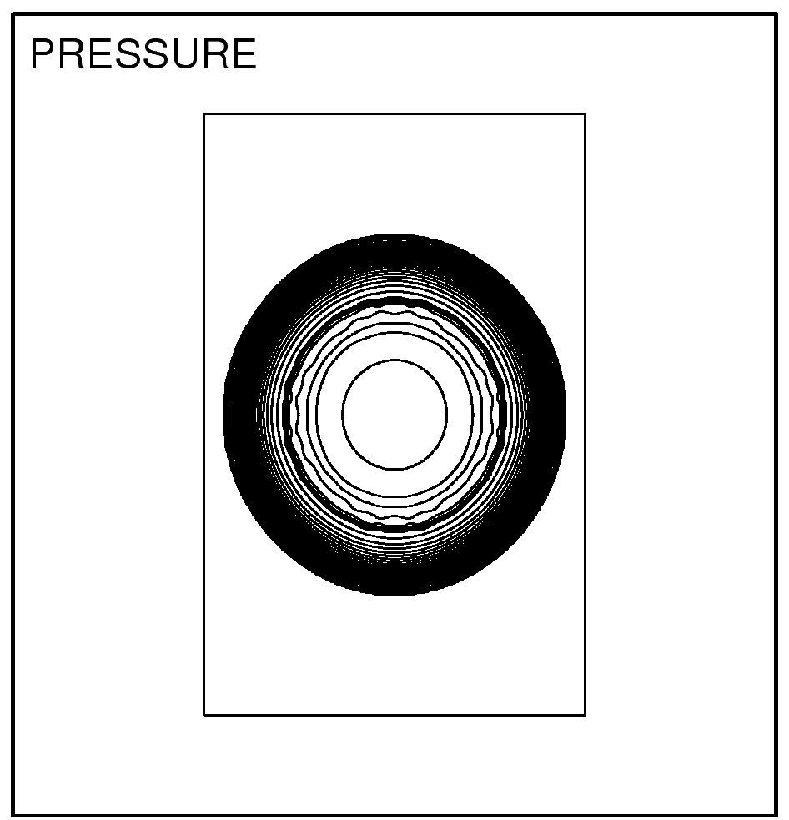} \\
\plottwo{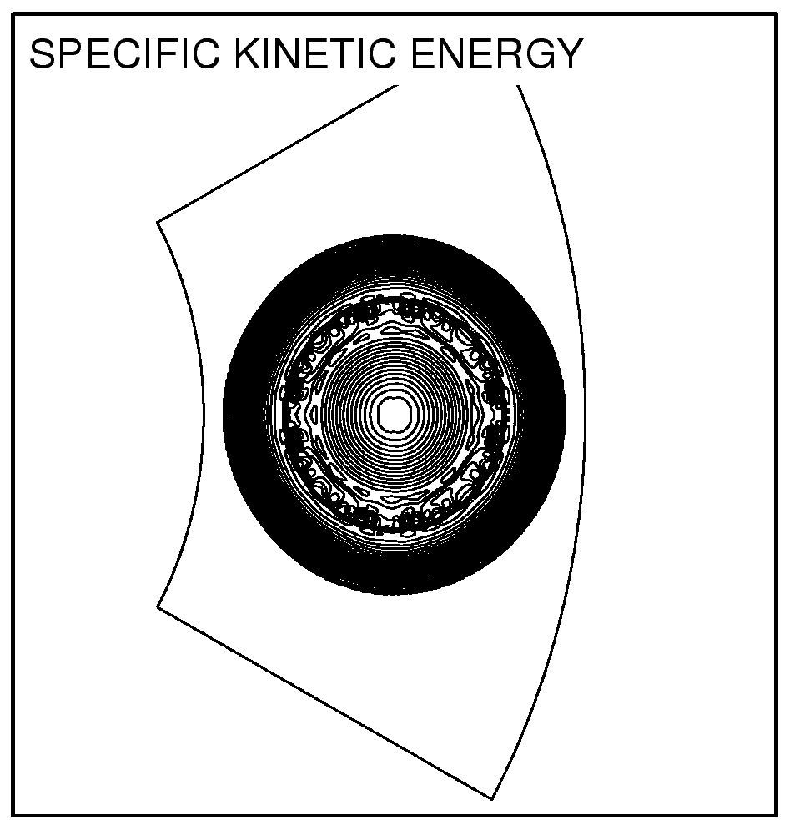}{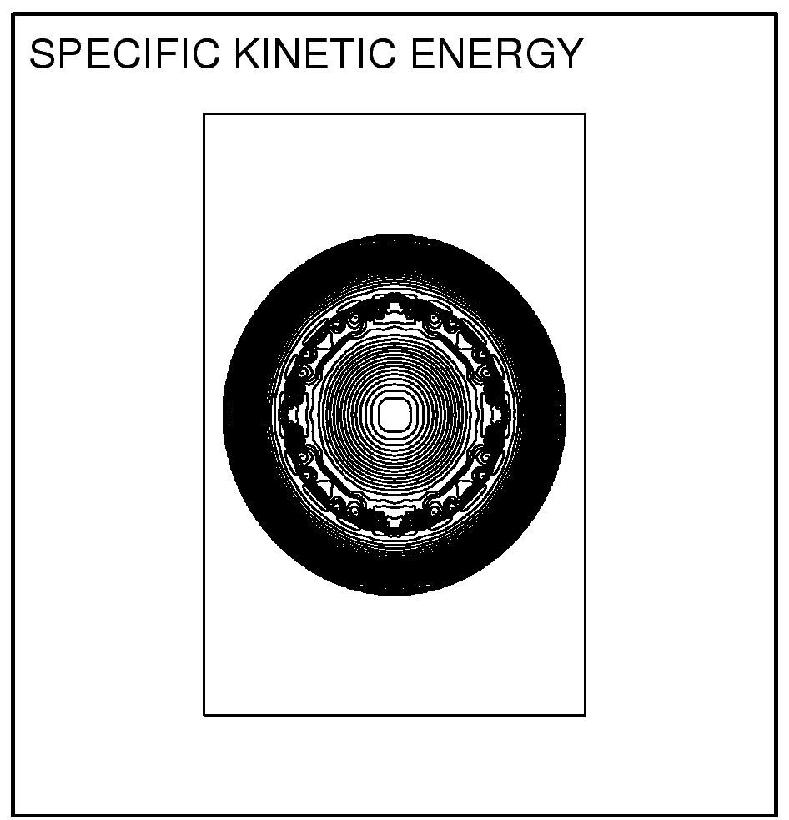}
\caption{Contours of selected variables of the evolved state (at time $t=0.2$) for the 2D hydrodynamic blast wave test using $200 \times 300$ grid cells, third-order reconstruction, HLLC fluxes, and the cylindrical (left column) or Cartesian (right column) versions of \athnosp.  Thirty equally spaced contours between the minimum and maximum are drawn in each plot. \label{cylblast_B0:2d_contour}}
\end{figure}

\begin{figure}
\centering
\epsscale{0.7}
\plotone{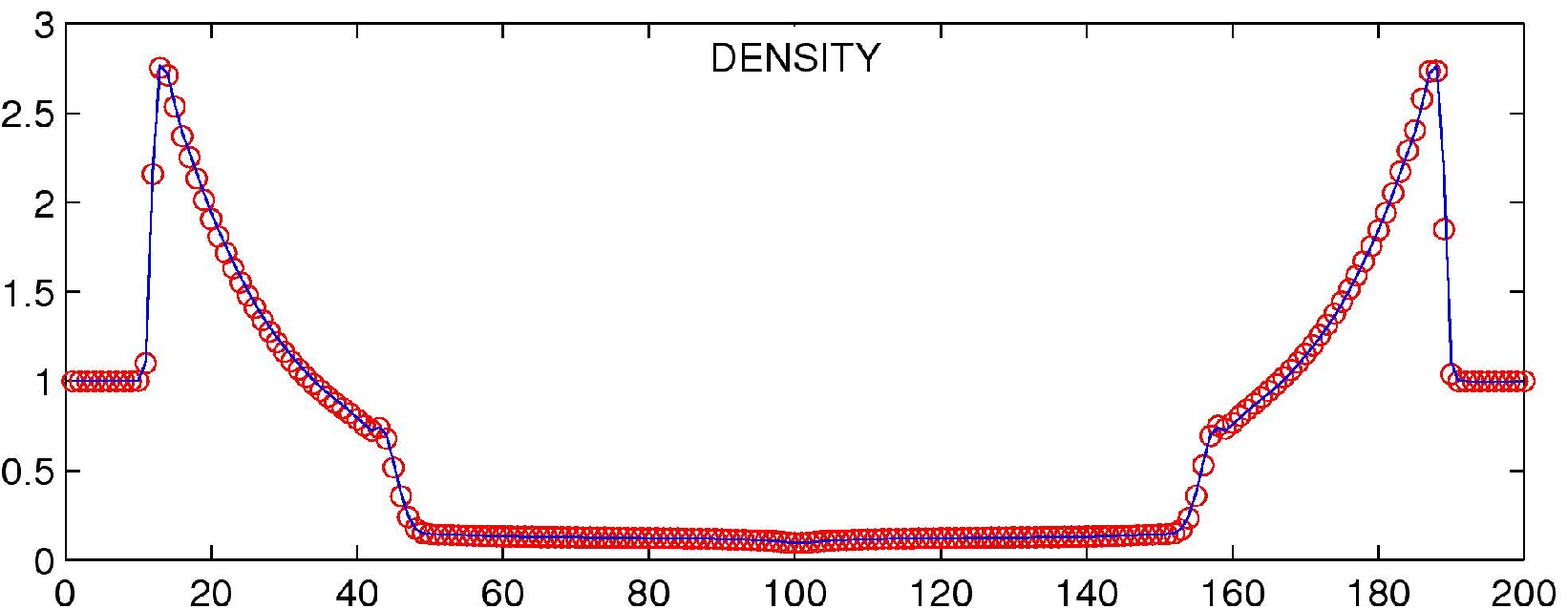} \\
\plotone{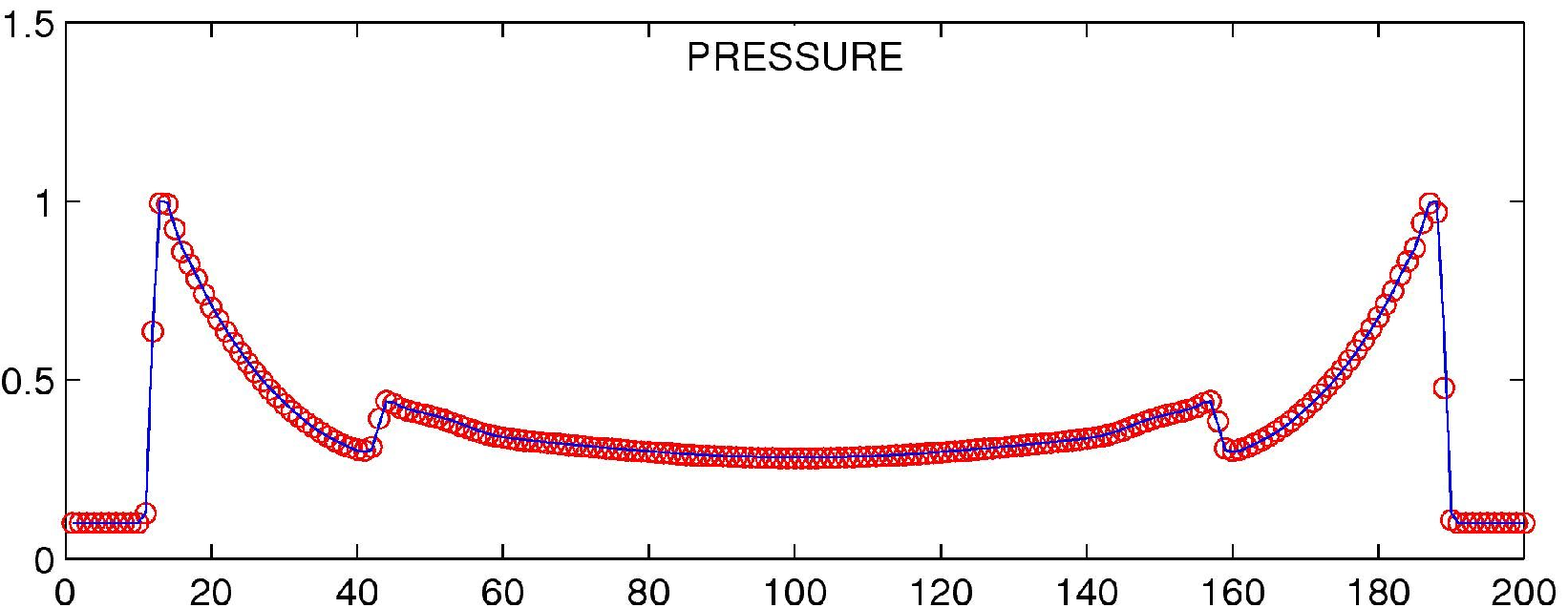} \\
\plotone{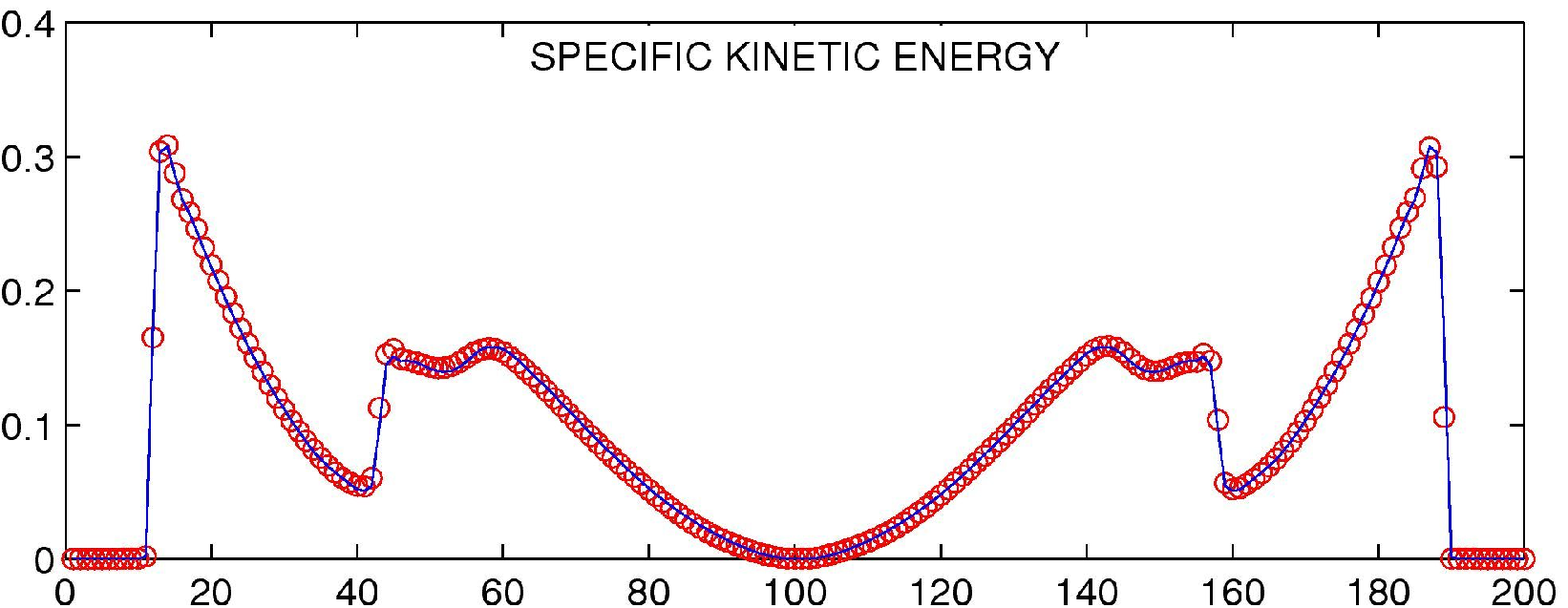}
\caption{Plots of selected variables along a horizontal line through the center of the blast at time \mbox{$t=0.2$} for the 2D hydrodynamic blast wave test (see Fig. \ref{cylblast_B0:2d_contour} legend), using the cylindrical (circles) or Cartesian (solid line) versions of \athnosp. \label{cylblast_B0:2d_lineout}}
\end{figure}

{\subsection{Rotating Wind} \label{cylwindrot}}
In this problem, we investigate a steady, axisymmetric, rotating hydrodynamic wind as a test of angular momentum transport across a sonic transition.  We adopt a Newtonian gravitational potential of the form $\Phi_g = -GM/R$.  The constants of motion are given by
\begin{align}
K &= P \rho^{-\gamma}, \\
\dot{M} &= R \rho v_R, \label{cylwindrot:massflux} \\
J &= R v_\phi.
\end{align}
This flow must satisfy the Bernoulli equation, $\mathcal{B} = {\rm constant}$, along streamlines for
\begin{equation}
\mathcal{B} \equiv \onehalf v^2 + h  + \Phi_g, \label{cylwindrot:bernoulli}
\end{equation}
where
\begin{equation}
h \equiv \int \frac{dP}{\rho} = \frac{c_s^2}{\gamma-1}
\end{equation}
is the specific enthalpy of the gas.

We scale density and pressure to their values at infinity, $\rho_\infty \equiv \rho(\infty)$ and $P_\infty \equiv P(\infty)$, and the radial coordinate to some finite fiducial value, $R_B$.  In terms of $\alpha \equiv \rho / \rho_\infty$ and $\chi \equiv R / R_B$, the constant entropy parameter is $K = c_\infty^2/(\gamma \,\rho_\infty^{\gamma-1})$, where $c_\infty^2 \equiv \gamma P_\infty/\rho_\infty$ is the square of the sound speed at infinity.  At any radius, the local sound speed, radial velocity, and specific enthalpy satisfy
\begin{eqnarray}
c_s^2 &=& c_\infty^2 \alpha^{\gamma-1}, \\
v_R^2 &=& \mathcal{M}_R^2 c_\infty^2 \alpha^{\gamma-1}, \label{cylwindrot:vr} \\
h &=& \frac{1}{\gamma-1} c_\infty^2 \alpha^{\gamma-1},  \label{cylwindrot:h}
\end{eqnarray}
where $\mathcal{M}_R \equiv v_R / c_s$ is the radial Mach number.

We define the dimensionless radial mass flux by
\begin{equation}
\lambda \equiv \frac{\dot{M}}{R_B \rho_\infty c_\infty} = \chi \mathcal{M}_R \alpha^{(\gamma+1)/2} \label{cylwindrot:lambda}
\end{equation}
and the dimensionless angular momentum by
\begin{equation}
\omega \equiv \frac{J}{R_B c_\infty} = \frac{\chi v_\phi}{c_\infty}. \label{cylwindrot:omega}
\end{equation}
Solving equation~(\ref{cylwindrot:lambda}) for $\alpha$ and introducing $\beta \equiv 2(\gamma-1)/(\gamma+1)$, we find that $\alpha^{\gamma-1} = [\lambda/(\chi \mathcal{M}_R)]^\beta$.  Finally, taking $R_B$ to be the Bondi radius, $R_B \equiv GM / c_\infty^2$, we obtain the dimensionless Bernoulli equation
\begin{equation}
\left( \frac{1}{2} \mathcal{M}_R^{2-\beta} + \frac{1}{\gamma-1} \mathcal{M}_R^{-\beta} \right) \lambda^{\beta} = \left[ \frac{\tilde{\mathcal{B}}}{\gamma-1} \chi^{\beta} + \chi^{\beta-1} - \frac{\omega^2}{2} \chi^{\beta-2} \right], \label{cylwindrot:scaledbernoulli}
\end{equation}
where $\tilde{\mathcal{B}} \equiv \mathcal{B}/h_\infty$ is the dimensionless Bernoulli constant, and $h_\infty \equiv c_\infty^2/(\gamma-1)$ is the specific enthalpy at infinity.

Figure~\ref{cylwindrot:lambdacontours} shows the contours of $\lambda$ for various values of $\omega$, using an adiabatic index of $\gamma=5/3$.  For the $\omega = 0$ case, we recover a cylindrical version of Parker's spherically symmetric wind \citep[see, e.g.,][]{spi78}.  The bold lines represent the transonic solutions (wind and accretion) passing through the X-type saddle points.  These solutions can be found by first writing equation~(\ref{cylwindrot:scaledbernoulli}) as $\mathcal{F}(\mathcal{M}_R) f(\lambda) = \mathcal{G}(\chi)$ and requiring that $\mathcal{F}'=\mathcal{G}'=0$.  The first constraint implies that $\mathcal{M}_R = 1$, i.e. the saddle point is the sonic point.  The second constraint yields a quadratic in $\chi$, and for $\omega \in (0,\omega_{\rm max})$, there are two distinct, positive solutions,
\begin{equation}
\chi_\pm = \frac{3-\gamma \pm \sqrt{(3-\gamma)^2-16\tilde{\mathcal{B}}\omega^2}}{4\tilde{\mathcal{B}}},
\end{equation}
where $\omega_{\rm max} \equiv (3-\gamma)/(4\tilde{\mathcal{B}}^{1/2})$.  It can be shown that $\chi_-$ gives an O-type critical point and $\chi_+$ gives the desired transonic critical point.  For $\omega > \omega_{\rm max}$, no transonic solutions exist.  Note further that no transonic solutions exist for $\chi < \chi_{\rm min}$, where $\chi_{\rm min}$ represents the point at which the transonic wind solution for which $\mathcal{M}_R \to \infty$ as $\chi \to \infty$ joins the transonic accretion solution for which $\mathcal{M}_R \to 0$ as $\chi \to \infty$ (see, e.g., the lower-left panel of Figure~\ref{cylwindrot:lambdacontours}).  We define $\chi_{\rm min}$ to be the smallest value of $\chi \ge 0$ for which $\mathcal{G}(\chi) \ge 0$, which is given by
\begin{equation}
\chi_{\rm min} = \frac{-(\gamma-1)+\sqrt{(\gamma-1)^2+2\tilde{\mathcal{B}}\omega^2}}{2\tilde{\mathcal{B}}}.
\end{equation}
Finally, the critical value of the radial mass flux, $\lambda_{\rm c}$, is defined by equation~(\ref{cylwindrot:scaledbernoulli}) with $\chi = \chi_+$ and $\mathcal{M}_R = 1$, which is given by 
\begin{equation}
\lambda_{\rm c} = \left[ \chi_+^{\beta-2} (\chi_+ - \omega^2) \right]^{1/\beta}.
\end{equation}

For our code test, we use $\gamma = 5/3$ (i.e. $\beta = 1/2$), $\omega=0.3$, and $\tilde{\mathcal{B}}=1$, which give the transonic solution shown in Figure~\ref{cylwindrot:lambdacontours}c.  The critical point occurs at $\chi_+ \approx 0.479$, with $\lambda_c \approx 1.377$.  We solve the problem on the domain $\chi \in [\chi_-,2]$, where $\chi_- \approx 0.188$, using bisection with a tolerance of $\epsilon = 10^{-10}$ to evaluate $\mathcal{M}_R$ at each $\chi$ from equation~(\ref{cylwindrot:scaledbernoulli}).  Once $\mathcal{M}_R$ is known, $v_R$, $\rho$, and $P$ follow algebraically.  We choose units such that $GM = c_\infty = 1$, which yields $R_B = 1$.  We fix the solution at the inner and outer boundaries and evolve the initial solution long enough for it to settle into equilibrium.  

Figure~\ref{cylwindrot:l1error} shows the convergence of the $L_2$ norm of the $L_1$ error vector for the solution at time $t=5.0$.  These data were computed using the Roe fluxes, second-order reconstruction, and the 1D; the results were similar for all combinations of Roe or HLLC fluxes and second- or third-order reconstruction.  The test was also performed using the 2D and 3D integrator with a few grid cells in the transverse directions.  

This test clearly demonstrates the code's ability to maintain smooth, steady hydrodynamic flows in both subsonic and supersonic regimes, as well as its ability to conserve angular momentum to second-order in cylindrical geometry.

\begin{figure}
\centering
\epsscale{0.85}
\plottwo{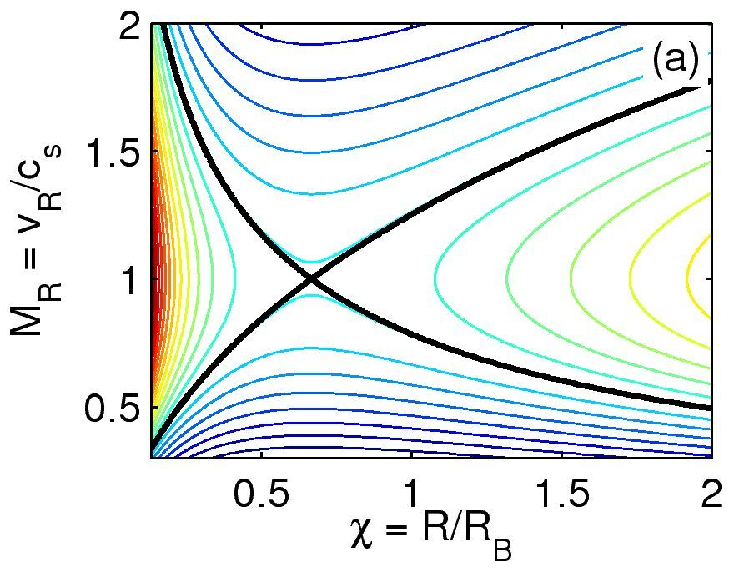}{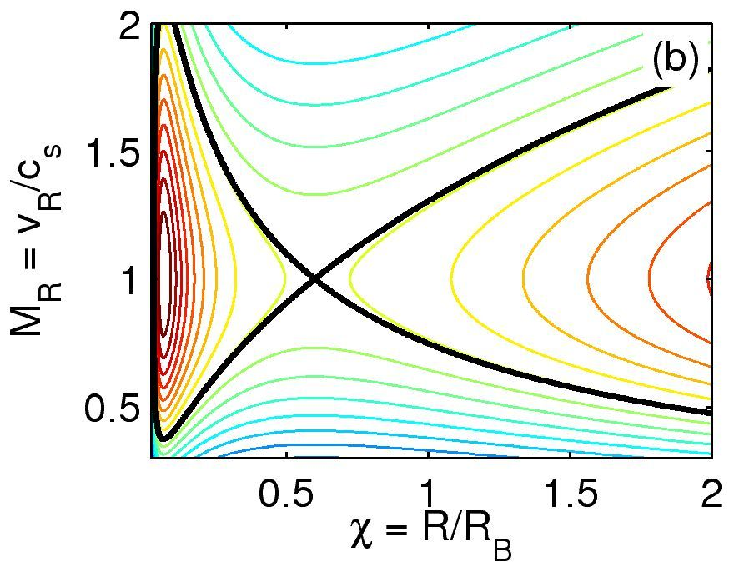} \\
\plottwo{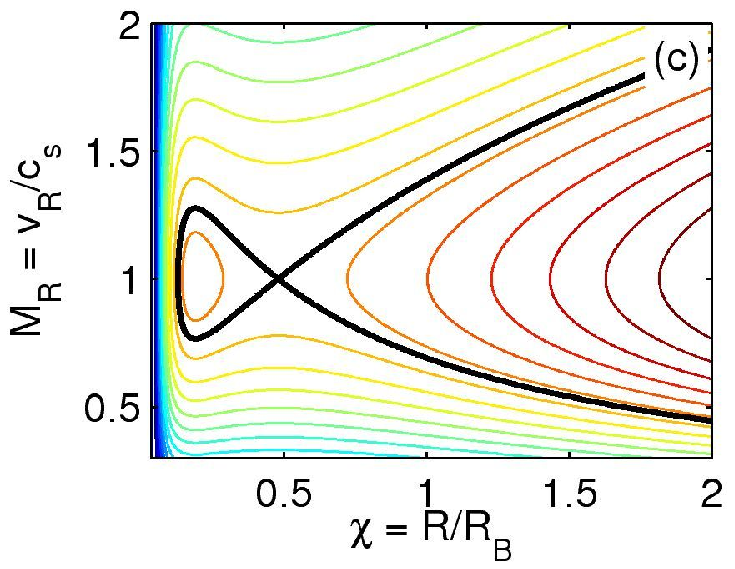}{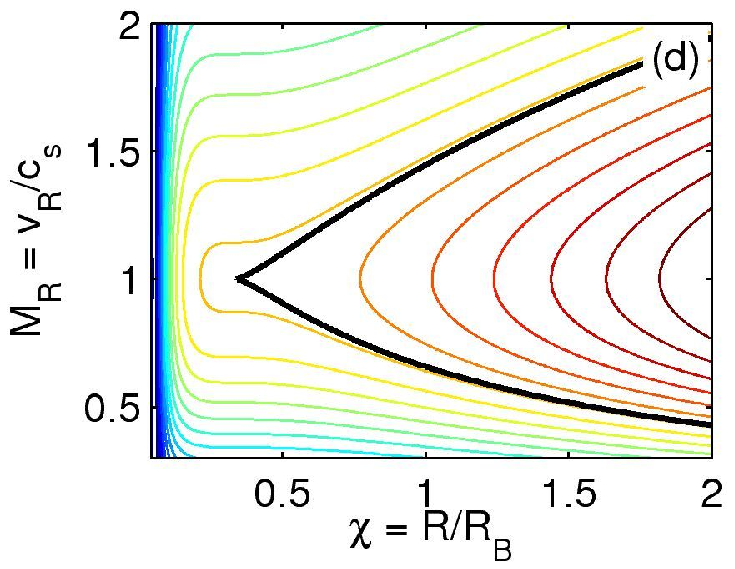} 
\caption{Contours of the dimensionless mass flux, $\lambda$, for a rotating hydrodynamic steady flow with $\gamma = 5/3$ and dimensionless Bernoulli constant $\tilde{\mathcal{B}}=1$.  The scaled angular momentum in each panel is (a) $\omega=0$, (b) $\omega=0.2$, (c) $\omega=0.3$, and (d) $\omega=1/3$.  The critical transonic contours are shown in bold. \label{cylwindrot:lambdacontours}}
\end{figure}

\begin{figure}
\centering
\epsscale{0.85}
\plotone{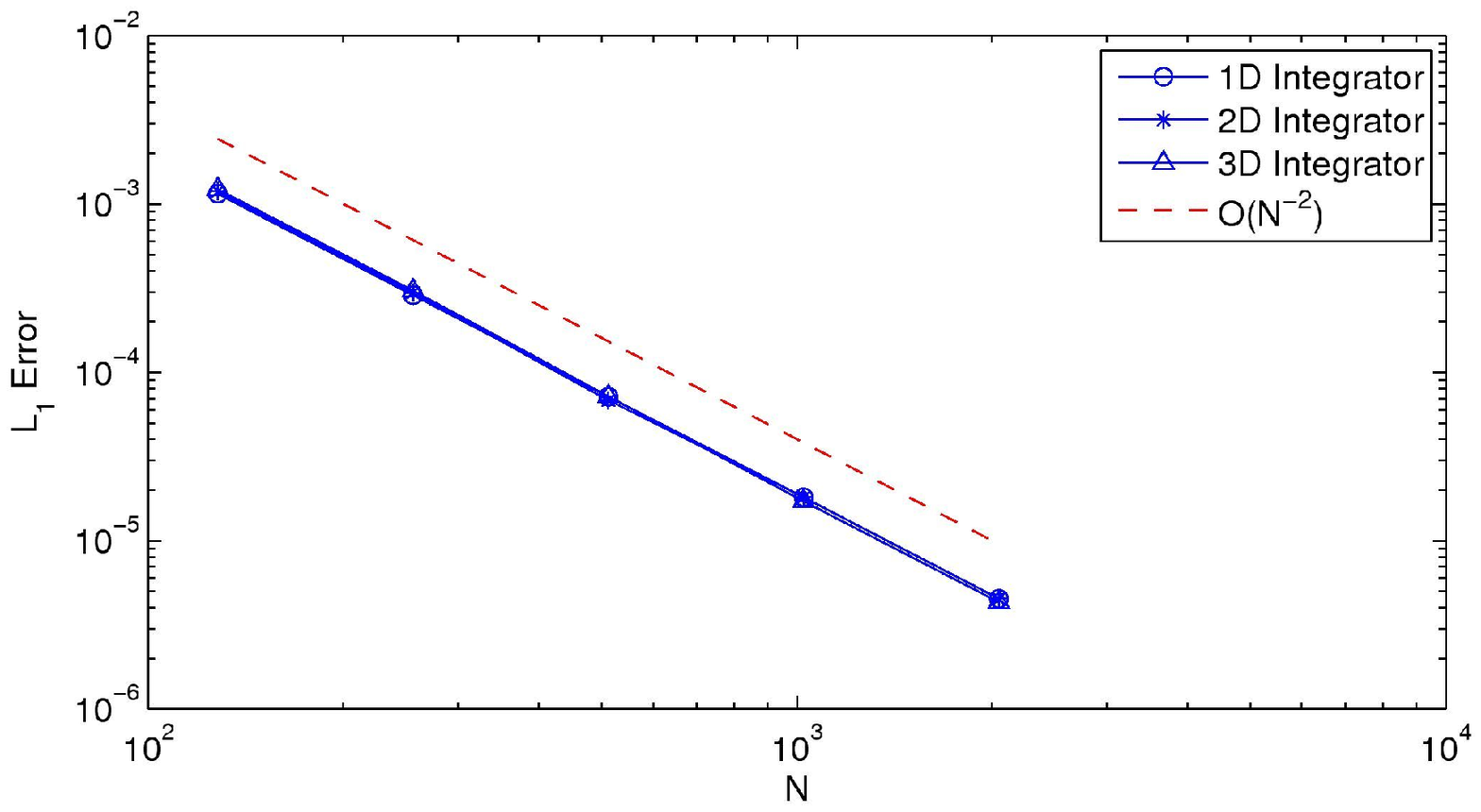}
\caption{Convergence of the RMS error in the $L_1$-norm for various levels of discretization of the rotating hydrodynamic wind test in 1D, 2D and 3D.  For reference, we have plotted a line of slope $-2$ (dashed) to show that the convergence is second-order in $1/N$. \label{cylwindrot:l1error}}
\end{figure}

\subsection{Field Loop Advection}

In this problem, we investigate the advection of a weak field loop in 2D and 3D cylindrical coordinates, analogous to the Cartesian test appearing in \citetalias{gar05,gar08}.  The main difference with our test is that we advect the field loop in the $\phi$-direction only as opposed to a more general advection oblique to the grid.  We use the computational domain $(R,\phi) \in [1,2] \times [-2/3,2/3]$, which has the same total area as the Cartesian version of the test.  We use periodic boundary conditions in $\phi$ and fixed boundary conditions in $R$.  We use uniform initial density $\rho_0=1$ and pressure $P_0=1$ with a solid-body rotation profile of $v_\phi = \Omega_0 R$, where we set $\Omega_0=4/3$ so that the field loop is advected once across the grid by $t=1$.  The initial $z$-component of the magnetic field is $0$, and the $R$- and $\phi$-components are set using the $z$-component of the magnetic vector potential
\begin{equation}
A_z \equiv \left\{ \begin{array}{ll}
A_0 (a_0-r) & \text{for } r \le a_0, \\
0 & \text{for } r > a_0
\end{array} \right.,
\end{equation}
where $a_0$ is the radius of the field loop and $r \equiv \sqrt{R^2 + R_0^2 - 2 R R_0 \cos(\phi-\phi_0)}$ is the distance from the center of the loop $(R_0,\phi_0)$.  We use $A_0 = 10^{-3}$ and $a_0 = 0.3$ so that inside the field loop $P/B = 10^6$ and the field loop should be advected passively.  

Figure~\ref{cylfieldloop:emag_a3} shows the magnetic energy density $B^2/2$ and magnetic field lines at times $t=0$ and $t=2$ for the 2D problem.  The field lines are the contours of $A_z$ which can be readily computed since the field is planar and the CTU+CT algorithm preserves the $\deldot \vct{B} = 0$ condition.  Note that the circular shape of the field lines is nicely preserved.  

Figure~\ref{cylfieldloop:timefit_emag} shows the time evolution of the volume-averaged magnetic energy density.  The dissipation is well-described by a power law of the form $\langle B^2/2 \rangle = A(1-(t/\tau)^\alpha)$ (for $t \ll \tau$), with $A=7.02 \times 10^{-8}$, $\tau=1.46 \times 10^4$, and $\alpha=0.342$, and with a residual error of $0.0580$.  Note that the overall dissipation in this problem is less than that of the Cartesian version since the advection is only in one direction; \citetalias{gar05} found $\tau=1.06 \times 10^4$ and $\alpha = 0.291$ for a similar fit.

\begin{figure}
\centering
\epsscale{0.85}
\plottwo{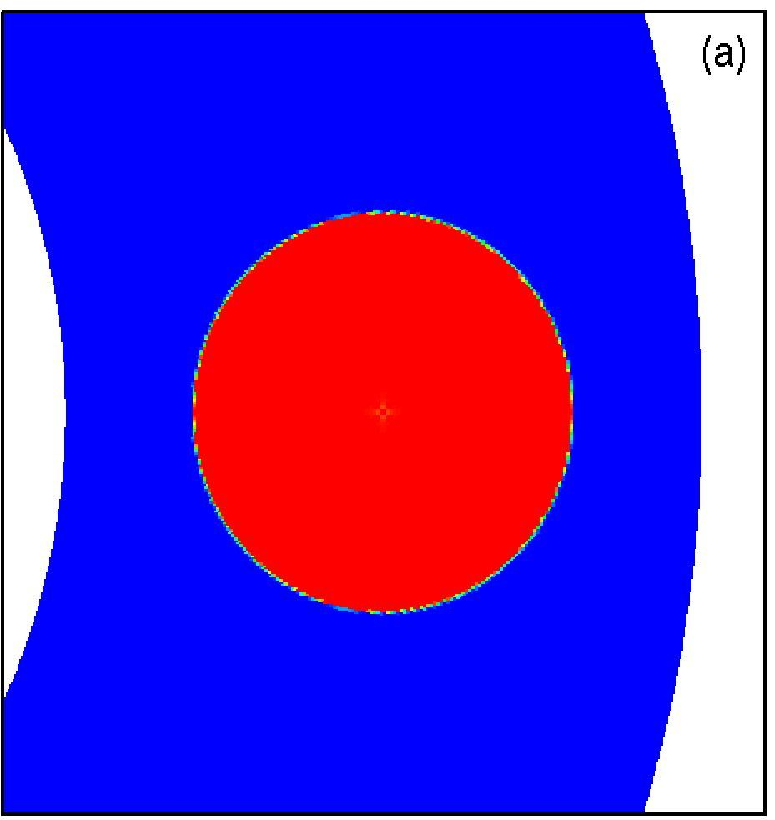}{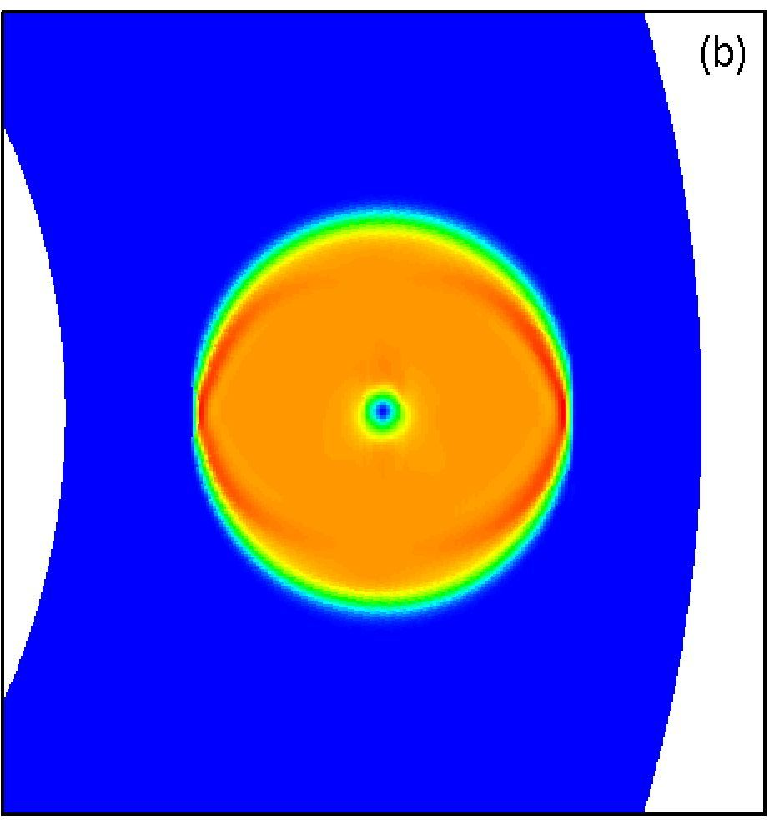} \\
\plottwo{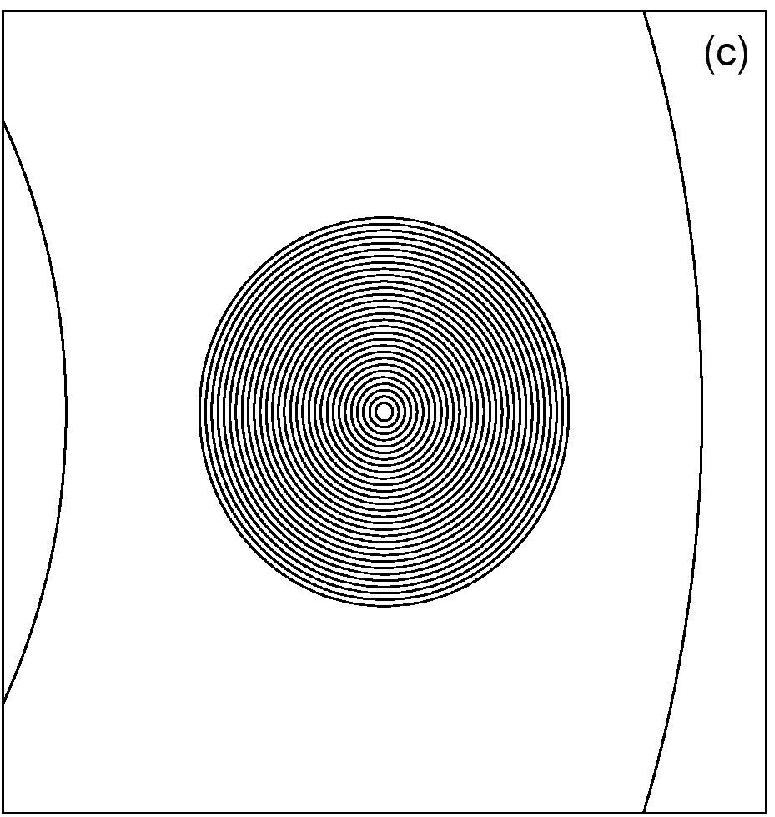}{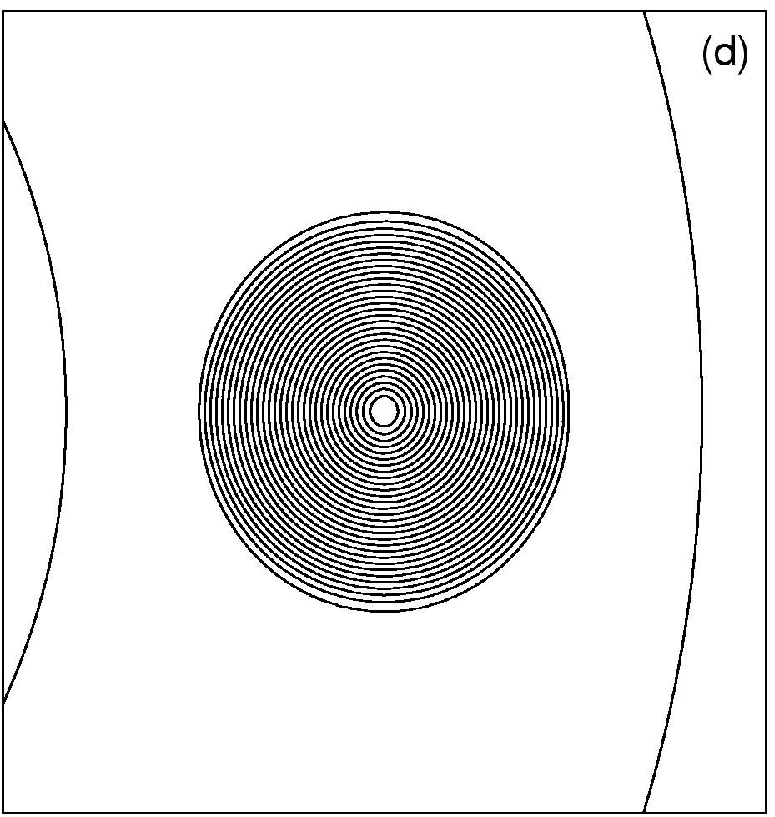}
\caption{For the 2D field loop advection test, we show the magnetic energy density $B^2/2$ at times $t=0$ and $t=2$ in panels (a) and (b), respectively.  Panels (c) and (d) contain magnetic field lines at $t=0$ and $t=2$, respectively. \label{cylfieldloop:emag_a3}}
\end{figure}

\begin{figure}
\centering
\epsscale{0.8}
\plotone{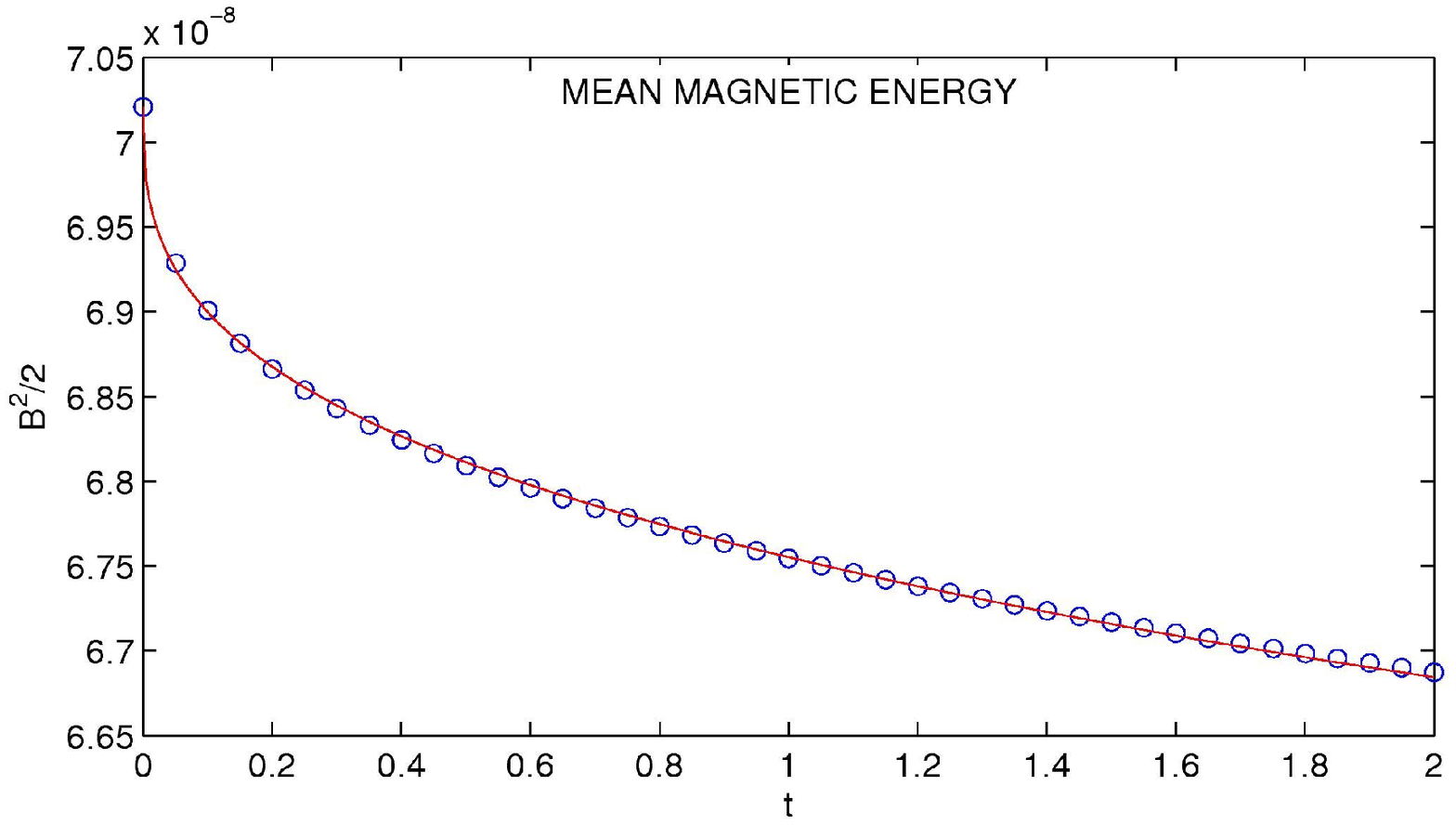}
\caption{Time evolution showing dissipation of the volume-averaged magnetic energy density $B^2/2$ for the 2D field loop advection test.  The solid line represents a power law fit (for $t \ll \tau$) to the data points with residual $0.0580$. \label{cylfieldloop:timefit_emag}}
\end{figure}

\subsection{Blast Wave in a Strong Magnetic Field}
In this problem, we investigate 2D and 3D MHD shocks in a strongly magnetized medium with low plasma-$\beta$, denoted $\beta_p \equiv 2P/B$.  We run two problems, one with $B_0=1$ and $\beta_p=0.2$ using the parameter set of \citetalias{gar08}, and another with $B_0=10$ and $\beta_p=0.02$ using the parameter set of \citet{lon00}.  In each case, we compare the outputs of the cylindrical and Cartesian versions of \athnosp.  

The moderate $B$-field, $\beta_p=0.2$ case uses HLLD fluxes and the same setup as the hydrodynamic blast described in \S\ref{cylblast_B0}, but with a uniform background magnetic field of strength $B_0=1$ oriented at a $45^\circ$ angle to the positive $\hat{x}$- or $\hat{R}$-axis.  For the 2D case, contour plots of the density, pressure, specific kinetic energy, and magnetic energy are shown in Figure~\ref{cylblast_B1:2d_contour} based on the cylindrical (left column) or Cartesian (right column) versions of \athnosp.  Figure~\ref{cylblast_B1:2d_lineout} shows a plot of these variables along a horizontal line through the center of the blast.  Note that in the Cartesian version, the background field is uniformly inclined to the grid, but in the cylindrical version, the angle the background field makes with the grid changes as a function of $\phi$; nonetheless, a high degree of symmetry is observed in the solution.  We have also tested a 3D analogue of this problem on a $200^3$ grid with $z \in [-0.5,0.5]$.  Again, plots of selected variables along a horizontal line through the center of the blast (in the $z=0$ plane) in Figure~\ref{cylblast_B1:3d_lineout} show agreement between the Cartesian and cylindrical versions of \athnosp.

The strong $B$-field, $\beta_p=0.02$ case uses the domain $(x,y) \in [-0.5,0.5] \times [-0.5,0.5]$ for the Cartesian version, and for the cylindrical version, uses the domain $(R,\phi) \in [1,2] \times [2/3,2/3]$, giving a roughly similar domain size in each case.  The initial conditions consist of a circular region of hot gas with radius $R_0=0.125$ and pressure $P=100$ in an ambient medium of uniform pressure $P_0=1$ and density $\rho_0=1$.  There is uniform background magnetic field of strength $B_0=10$ oriented parallel to the $x$-axis.  We use a computational grid of $200^2$ cells, third-order reconstruction and the HLLD fluxes with upwind-only integration for the L/R states.  The density, pressure, specific kinetic energy, and magnetic energy along a horizontal line through the center of the blast at time $t=0.02$ are shown in Figure~\ref{cylblast_B10:2d_lineout}, with comparison to the Cartesian version of \athnosp.  We have also conducted a 3D analogue of this test on a $200^3$ grid with $z \in [-0.5,0.5]$.  Figure~\ref{cylblast_B10:3d_contour} shows a comparison of contour plots with the Cartesian version of \athnosp, and Figure~\ref{cylblast_B10:3d_lineout} shows a comparison along a horizontal line through the center of the blast (in the $z=0$ plane).

\begin{figure}
\centering
\epsscale{0.6}
\plottwo{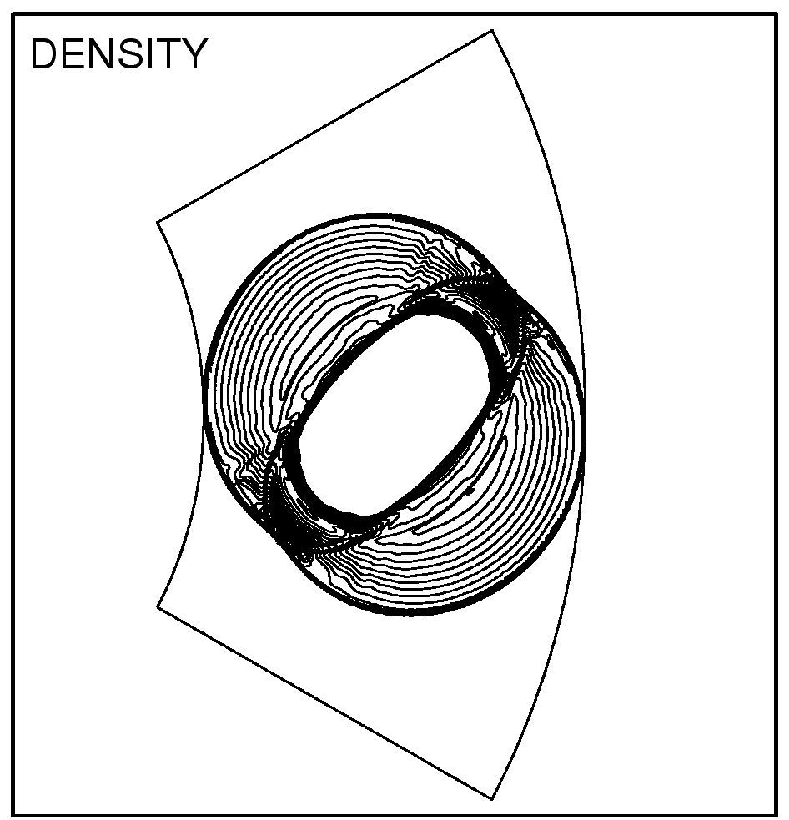}{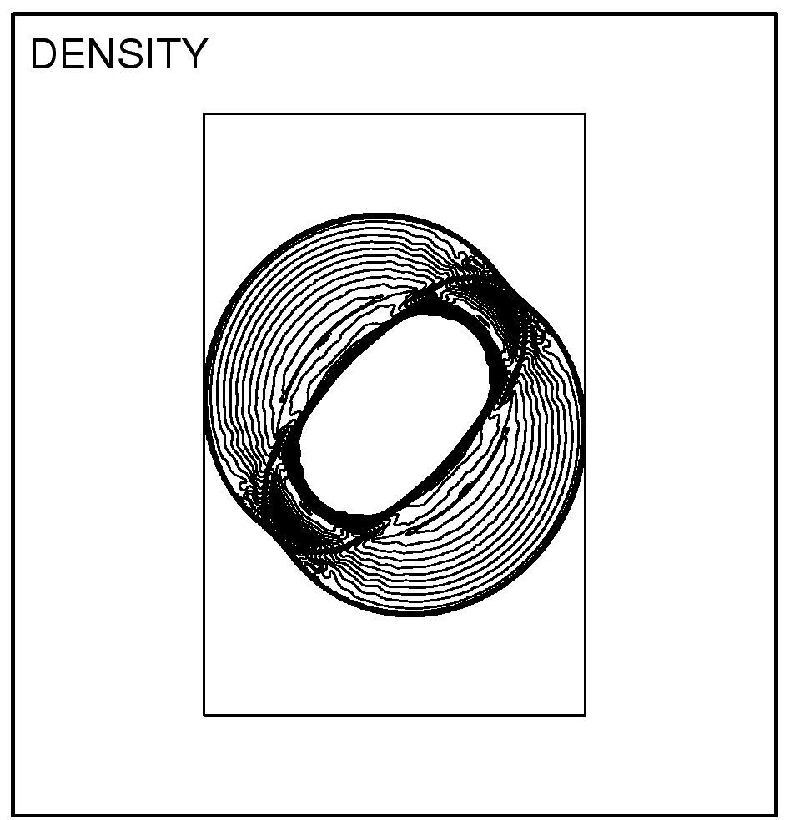} \\
\plottwo{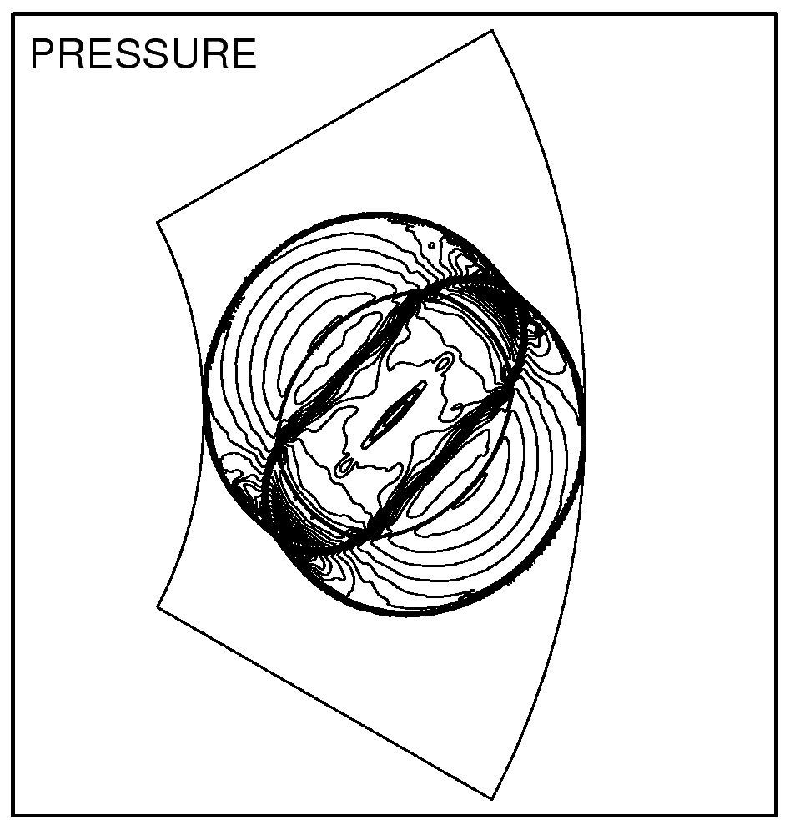}{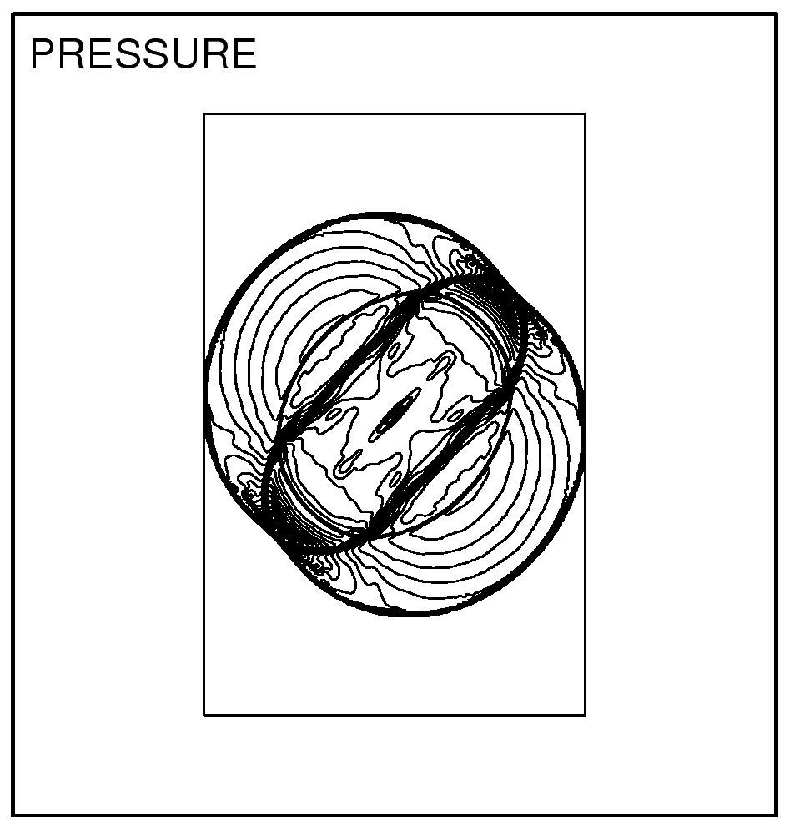} \\
\plottwo{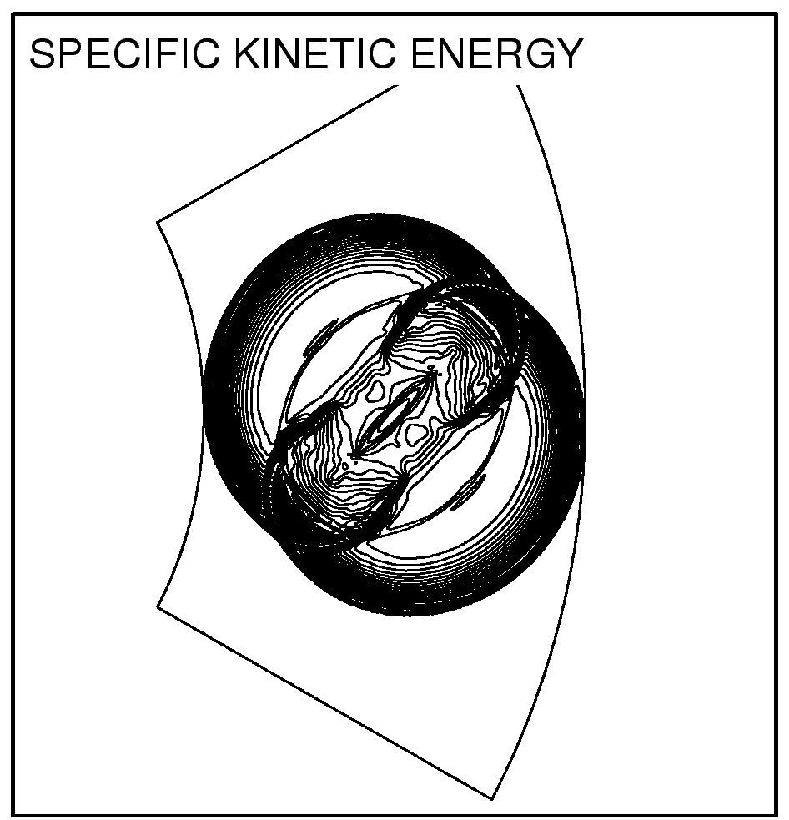}{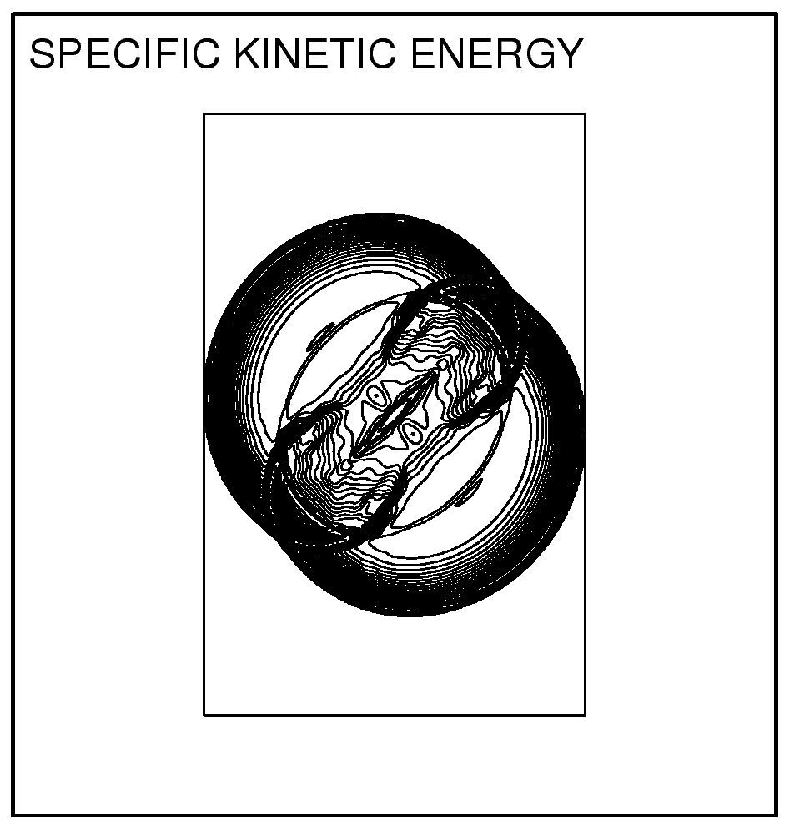} \\
\plottwo{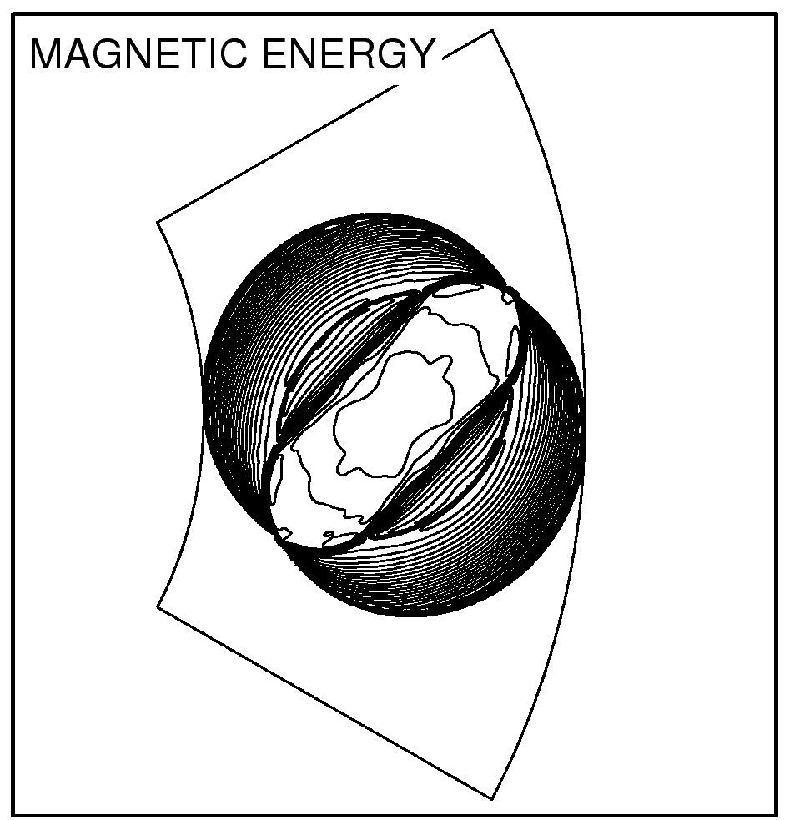}{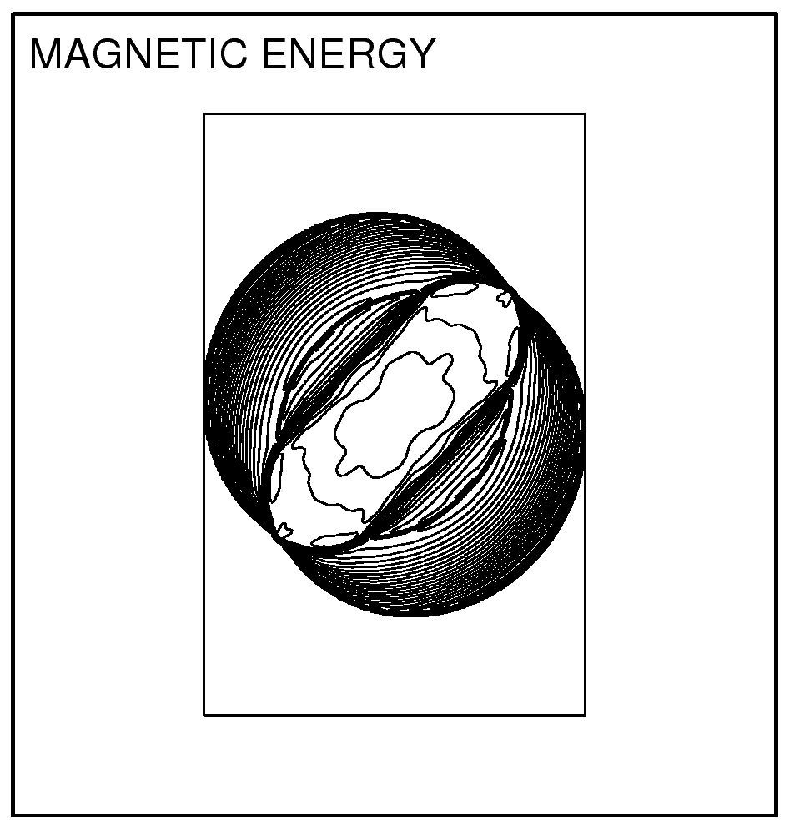}
\caption{Contours of selected variables of the evolved state (at time $t=0.2$) for the 2D MHD blast wave test with $B_0=1$ and $\beta_p=0.2$ using $200 \times 300$ grid cells, third-order reconstruction, HLLD fluxes, and using the cylindrical (left column) or Cartesian (right column) versions of Athena.  Thirty equally spaced contours between the minimum and maximum are drawn in each plot. \label{cylblast_B1:2d_contour}}
\end{figure}

\begin{figure}
\centering
\epsscale{0.75}
\plotone{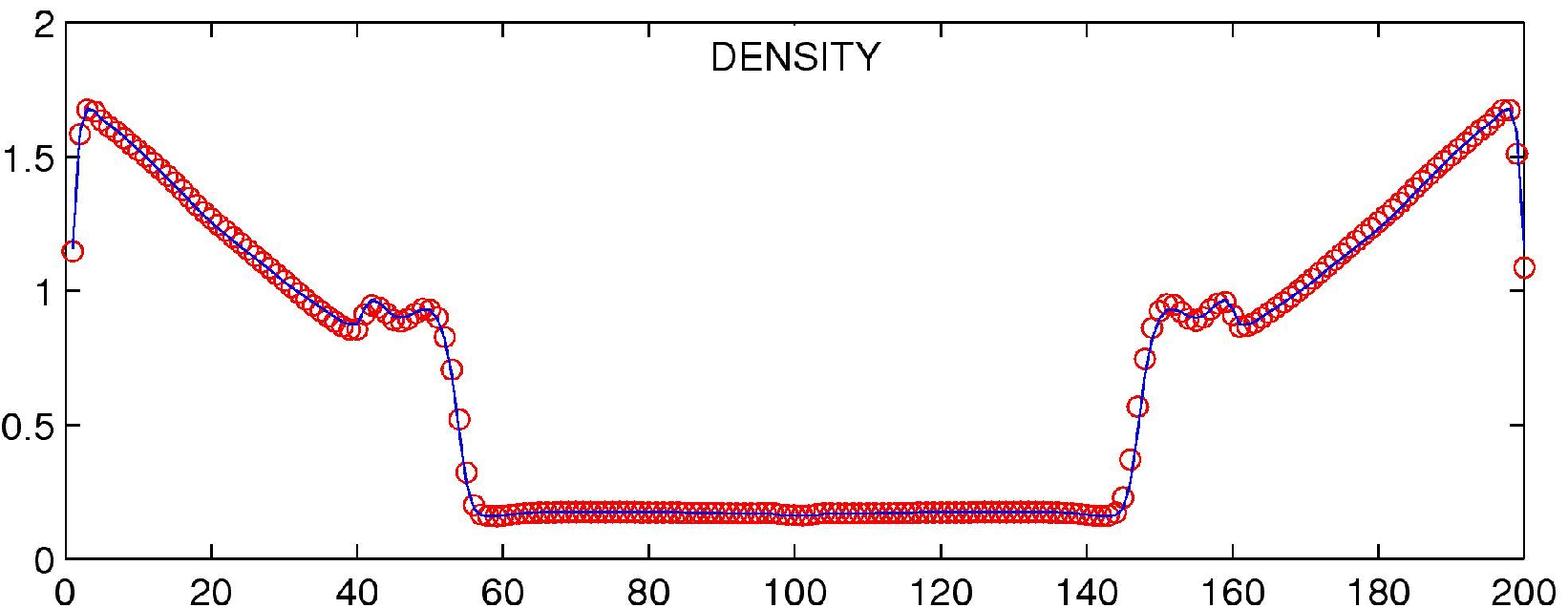} \\
\plotone{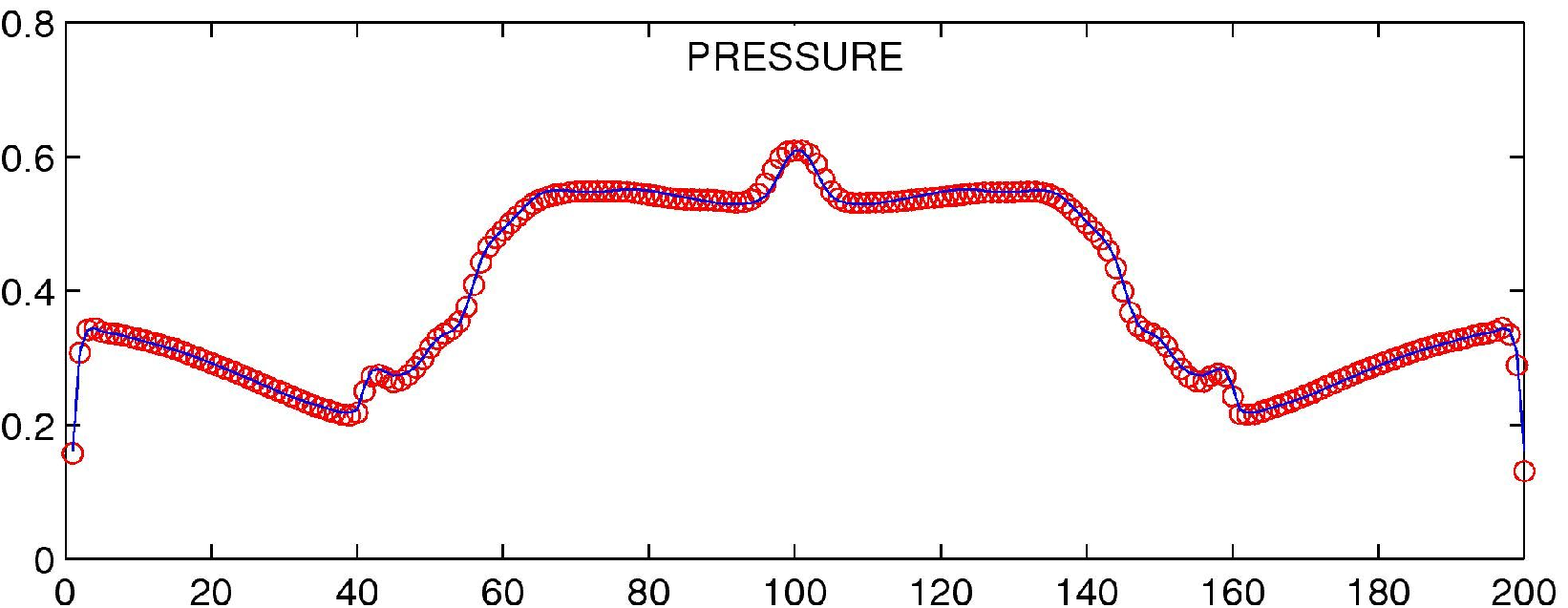} \\
\plotone{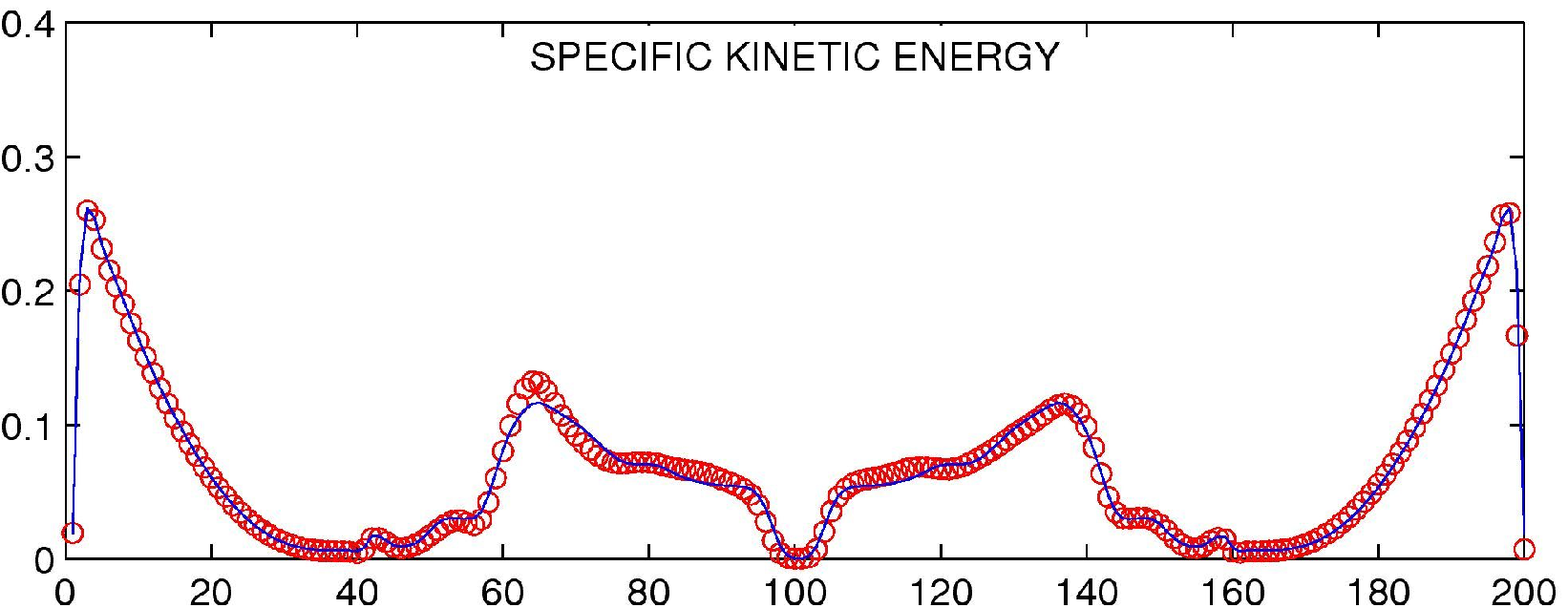} \\
\plotone{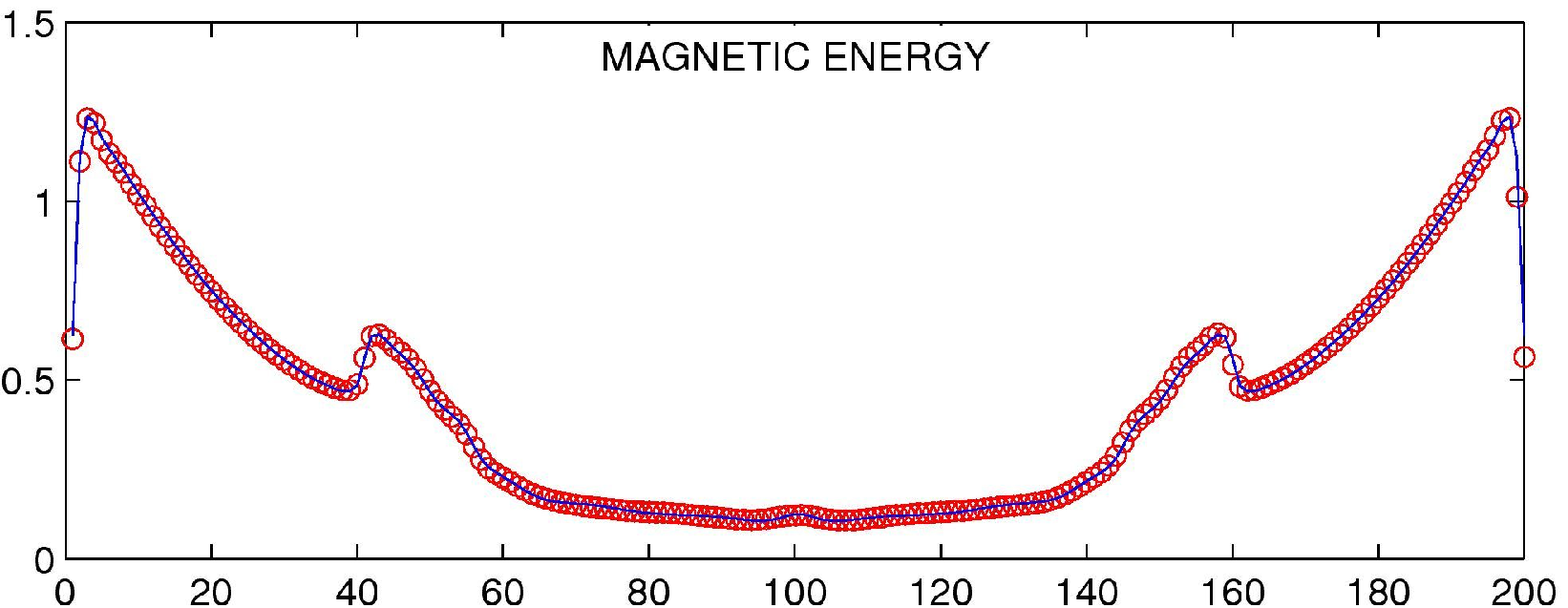}
\caption{Plots of selected variables along a horizontal line through the center of the blast at time \mbox{$t=0.2$} for the 2D MHD blast wave test with $B_0=1$ and $\beta_p=0.2$ using the cylindrical (circles) or Cartesian (solid line) versions of \athnosp. \label{cylblast_B1:2d_lineout}}
\end{figure}

\begin{figure}
\centering
\epsscale{0.75}
\plotone{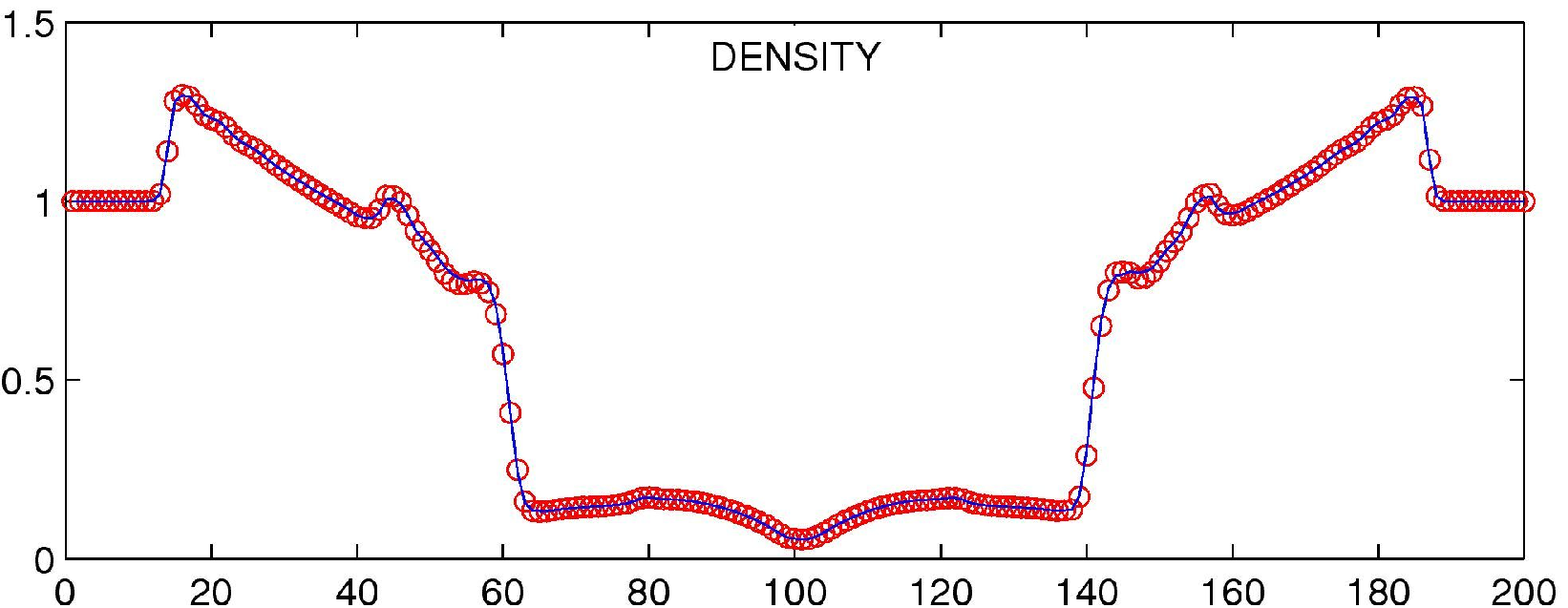} \\
\plotone{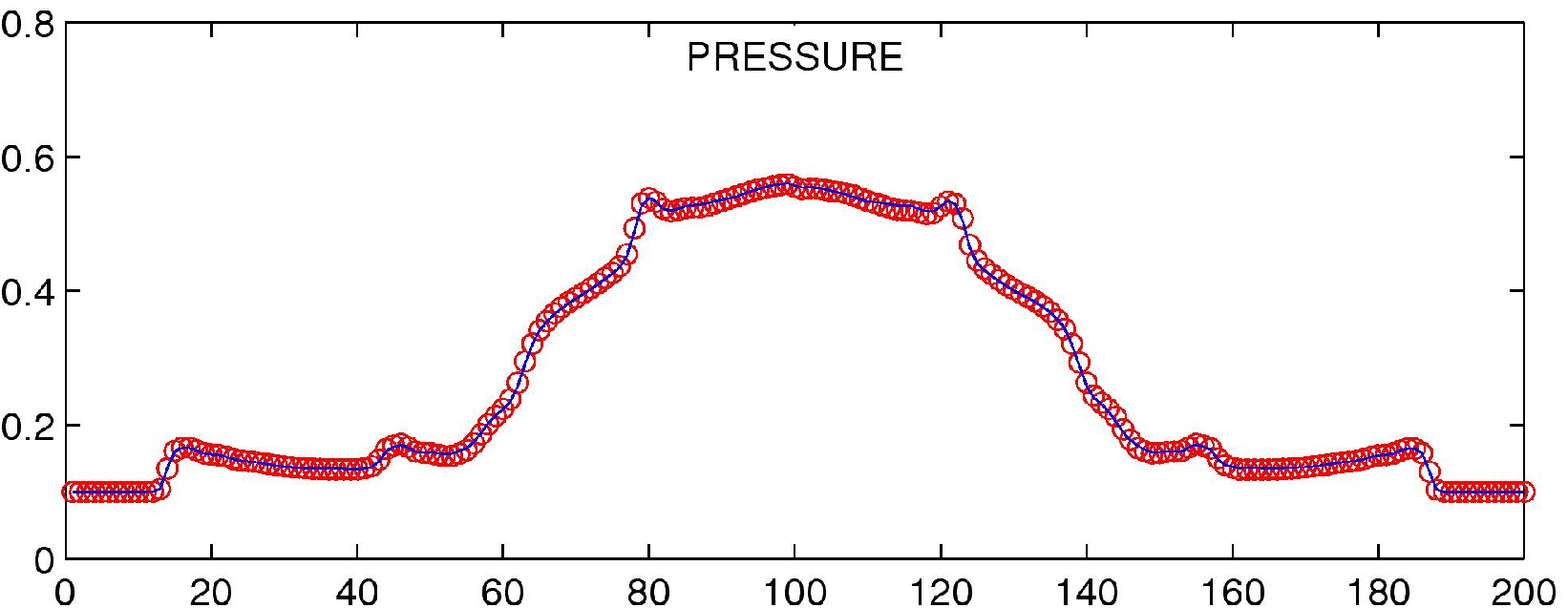} \\
\plotone{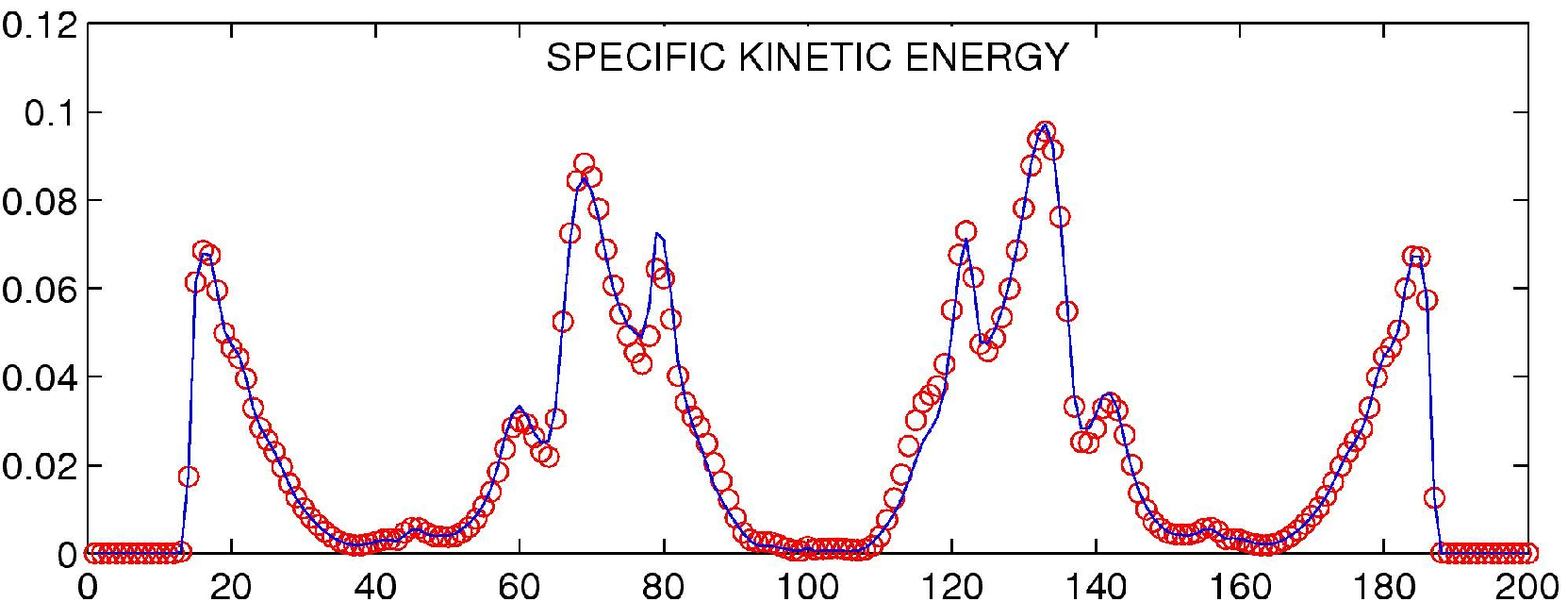} \\
\plotone{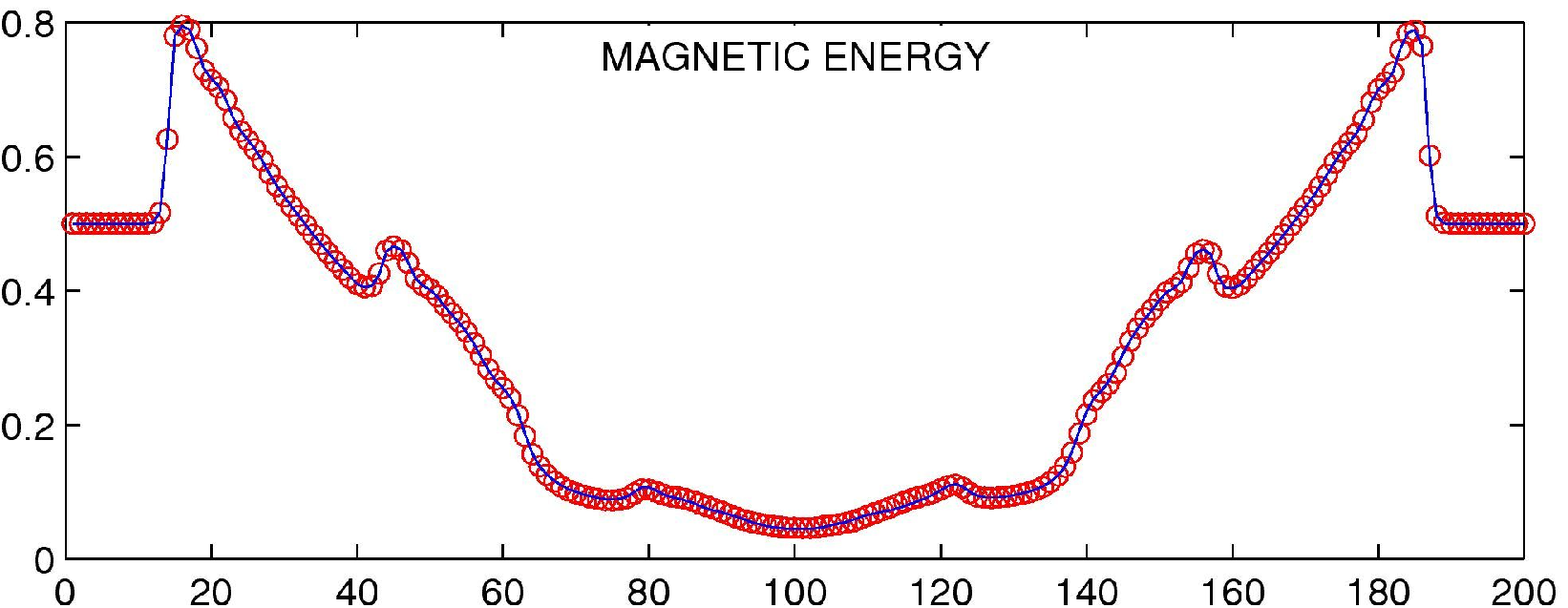}
\caption{Plots of selected variables along a horizontal line through the center of the blast at time \mbox{$t=0.2$} for the 3D MHD blast wave test with $B_0=1$ and $\beta_p=0.2$ using the cylindrical (circles) or Cartesian (solid line) versions of \athnosp. \label{cylblast_B1:3d_lineout}}
\end{figure}

\begin{figure}
\centering
\epsscale{0.75}
\plotone{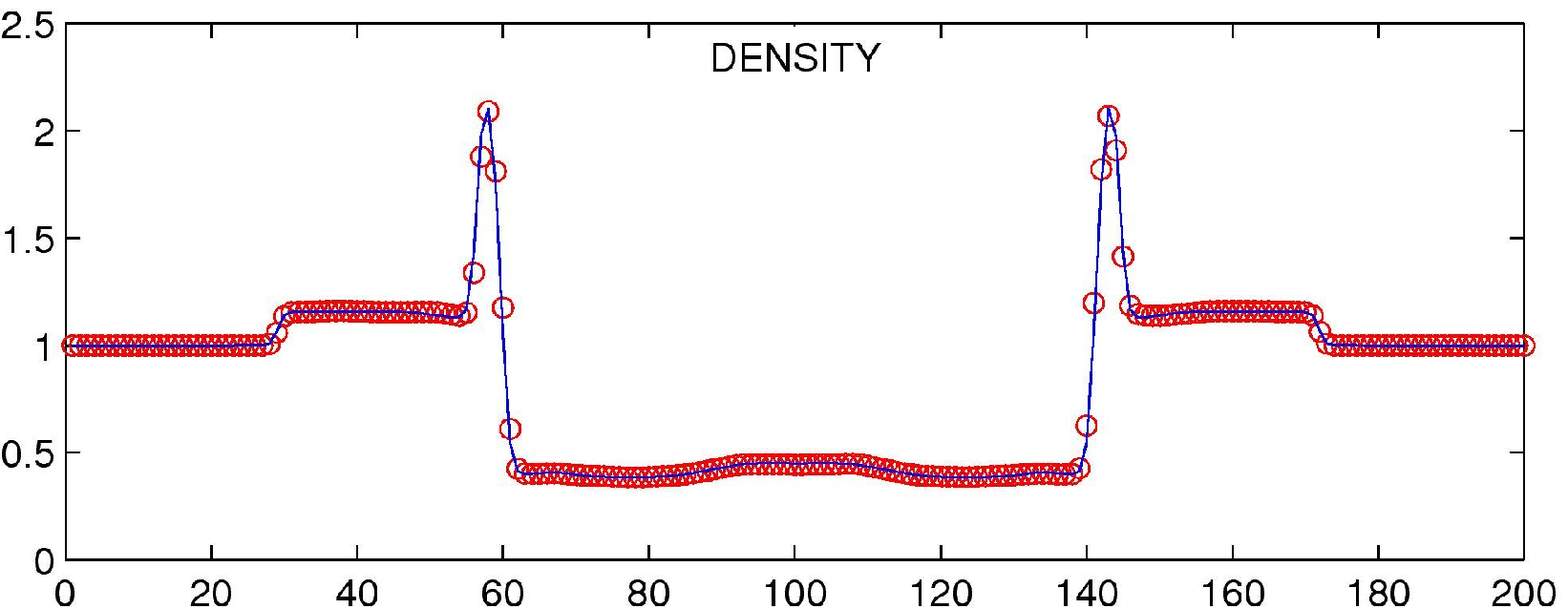} \\
\plotone{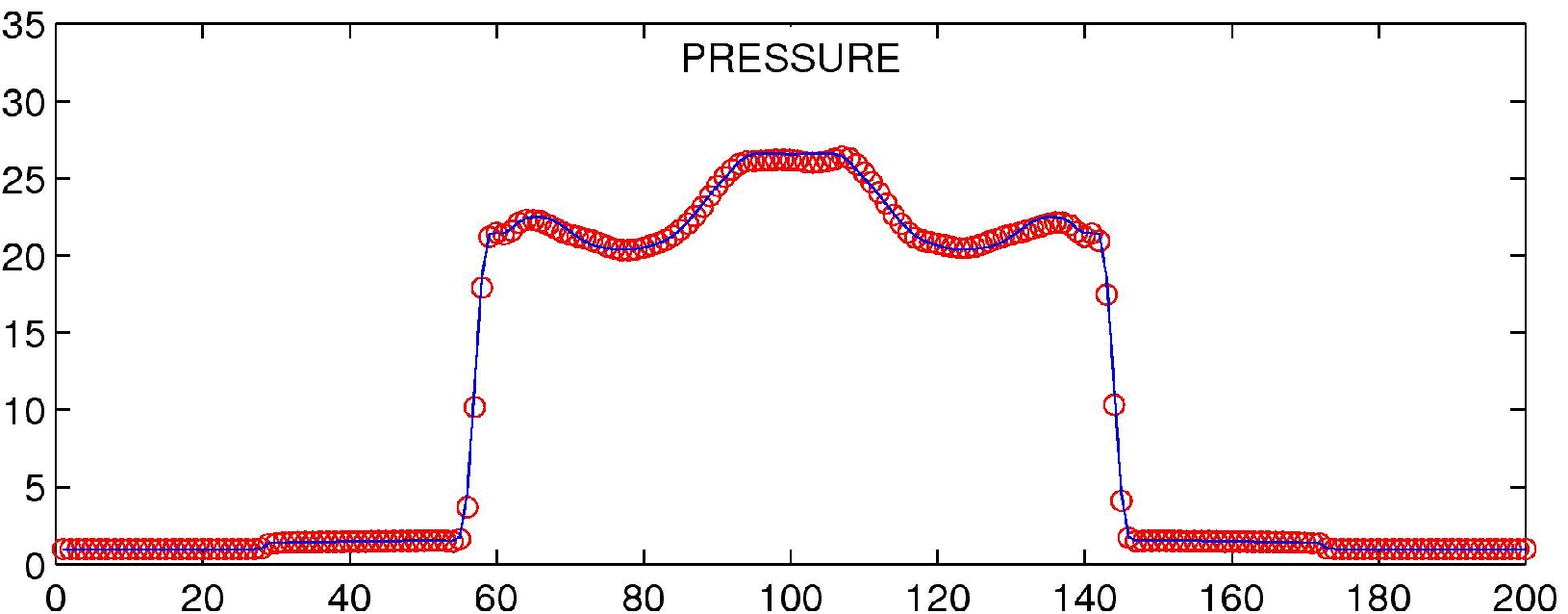} \\
\plotone{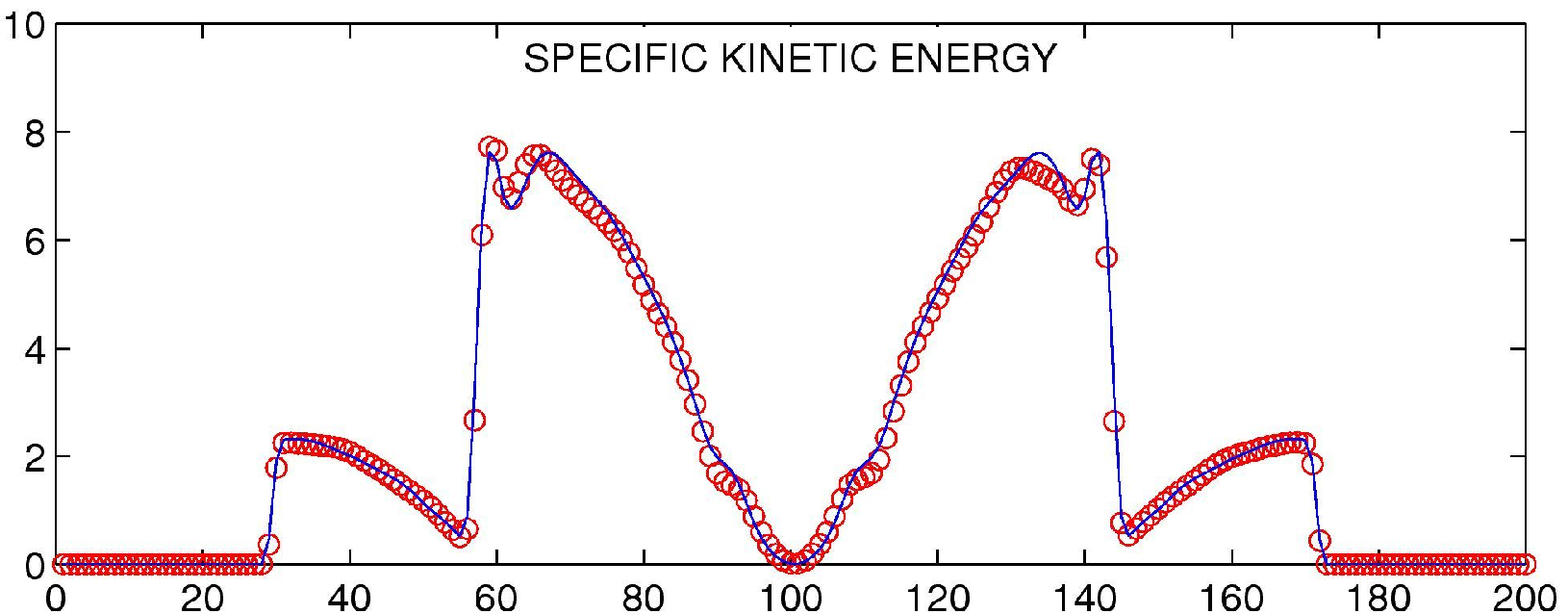} \\
\plotone{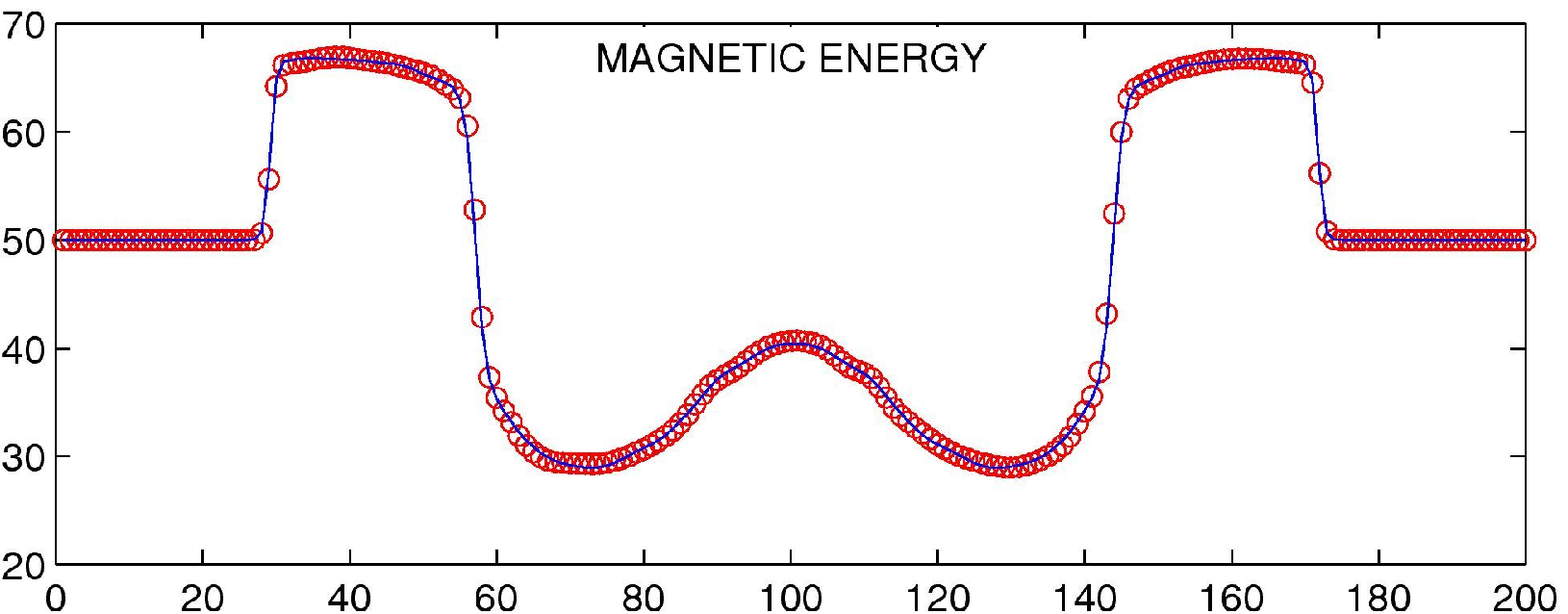}
\caption{Plots of selected variables along a horizontal line through the center of the blast at time \mbox{$t=0.02$} for the 2D MHD blast wave test with $B_0=10$ and $\beta_p=0.02$ using the cylindrical (circles) or Cartesian (solid line) versions of \athnosp. \label{cylblast_B10:2d_lineout}}
\end{figure}

\begin{figure}
\centering
\epsscale{0.6}
\plottwo{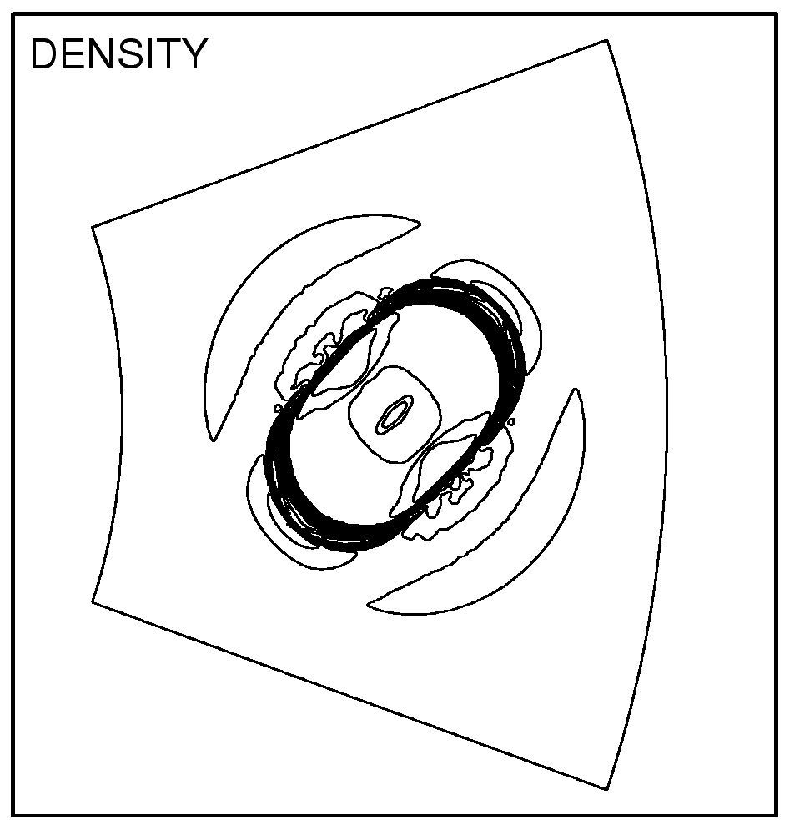}{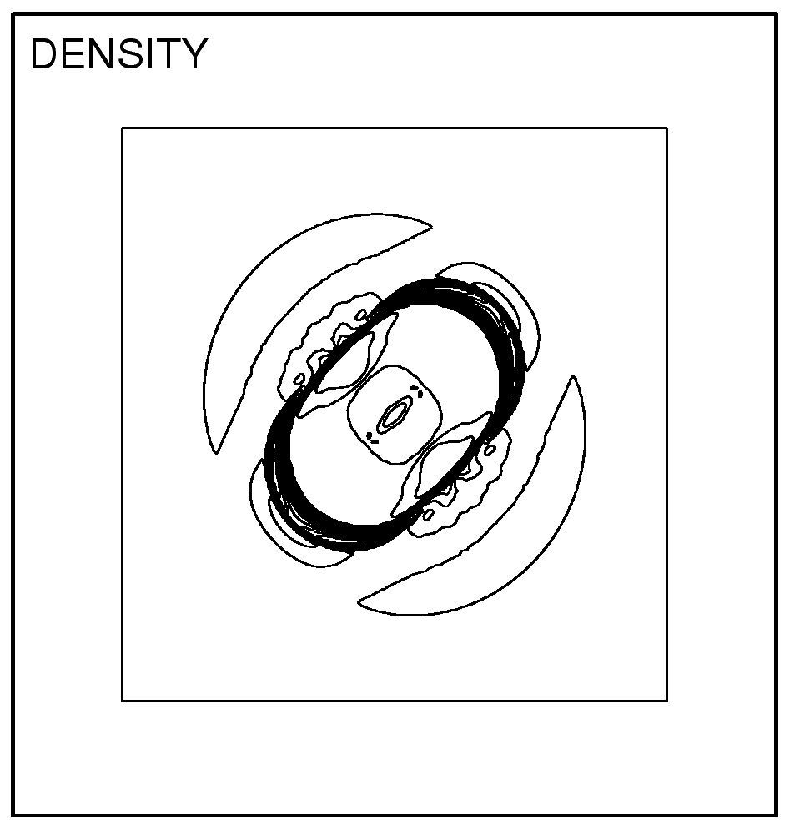} \\
\plottwo{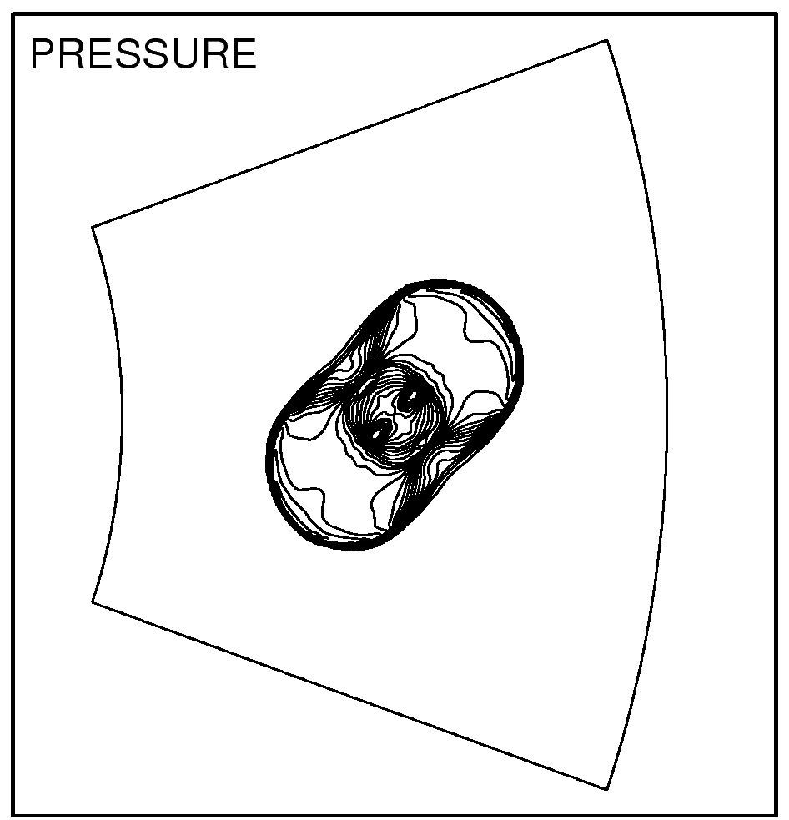}{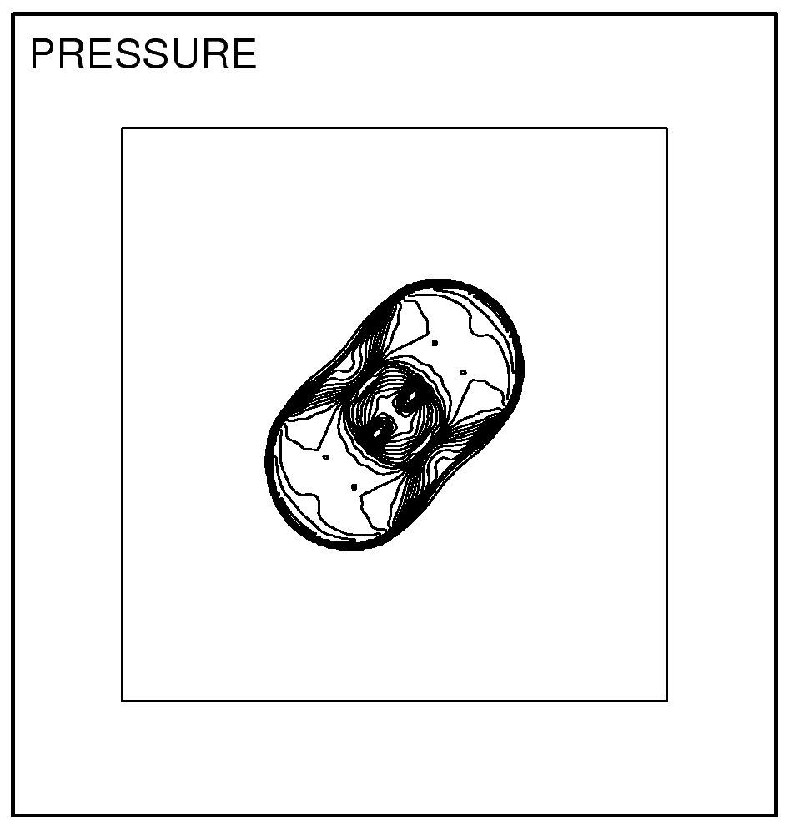} \\
\plottwo{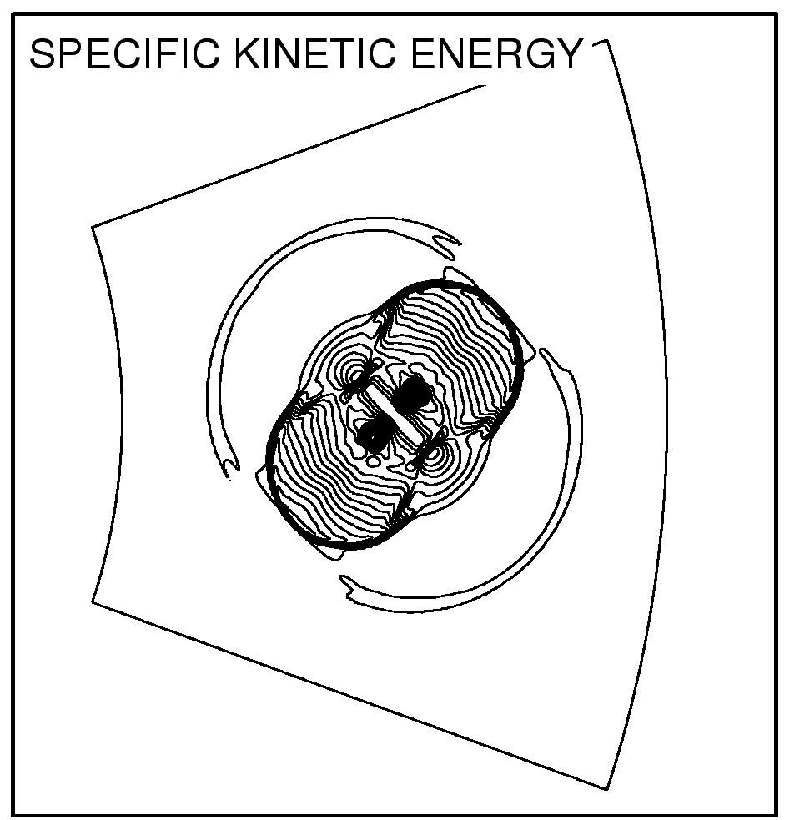}{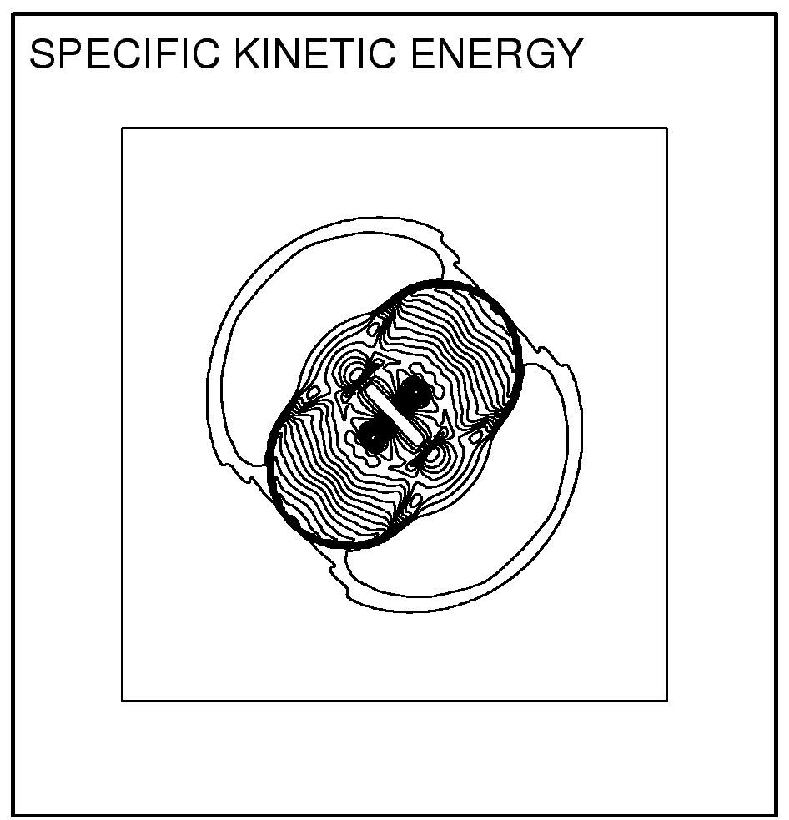} \\
\plottwo{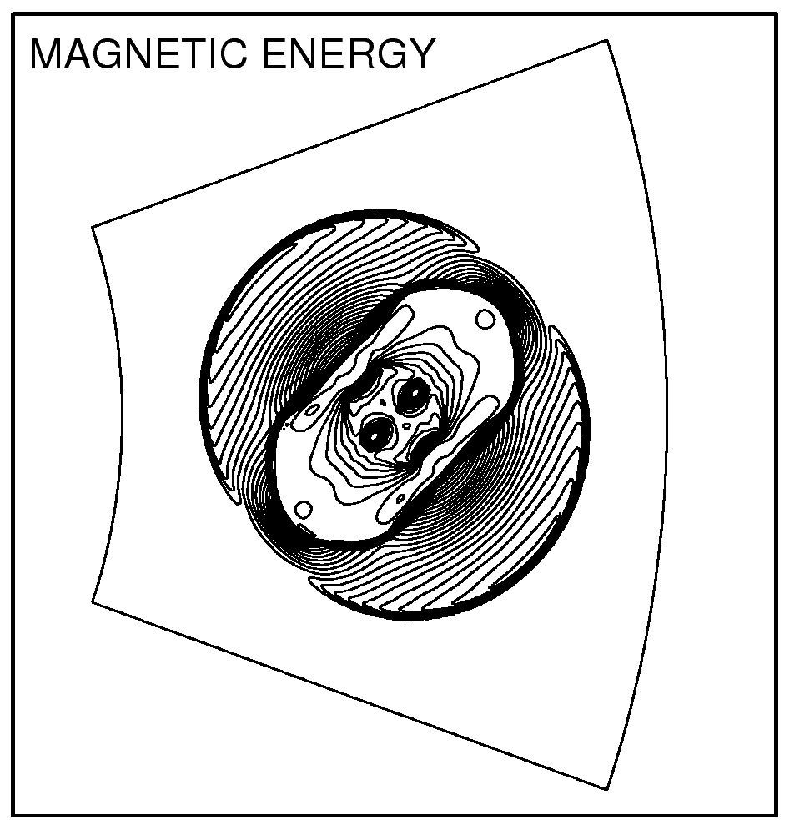}{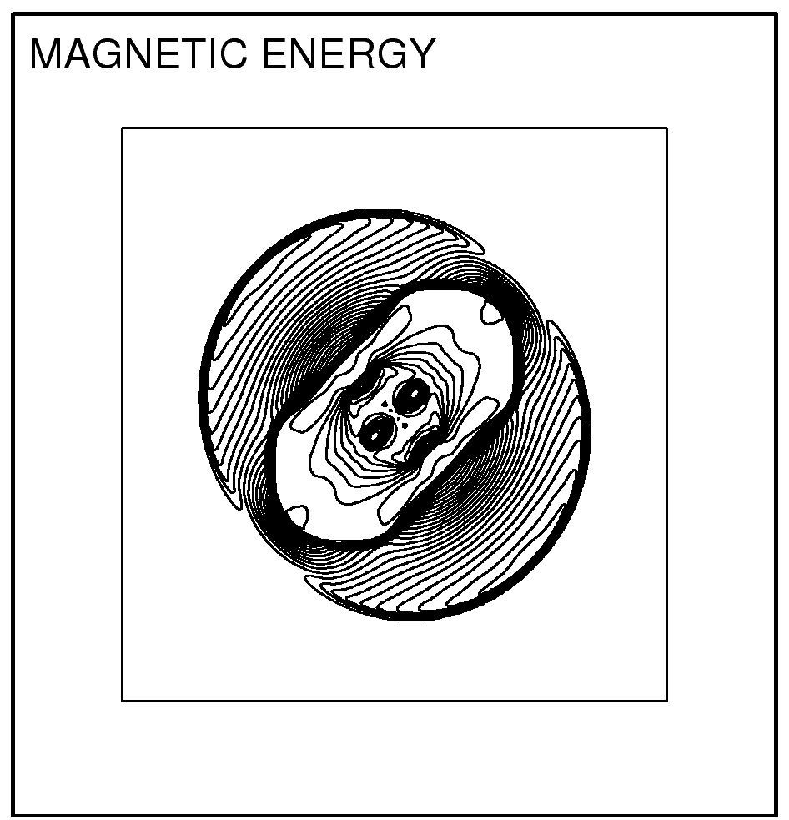}
\caption{Contours of selected variables at time $t=0.02$ for the 3D MHD blast wave test with $B_0=10$ and $\beta_p=0.02$ using $200^3$ grid cells and the cylindrical (top row) or Cartesian (bottom row) versions of \athnosp.  Thirty equally spaced contours between the minimum and maximum are drawn in each plot. \label{cylblast_B10:3d_contour}}
\end{figure}

\begin{figure}
\centering
\epsscale{0.75}
\plotone{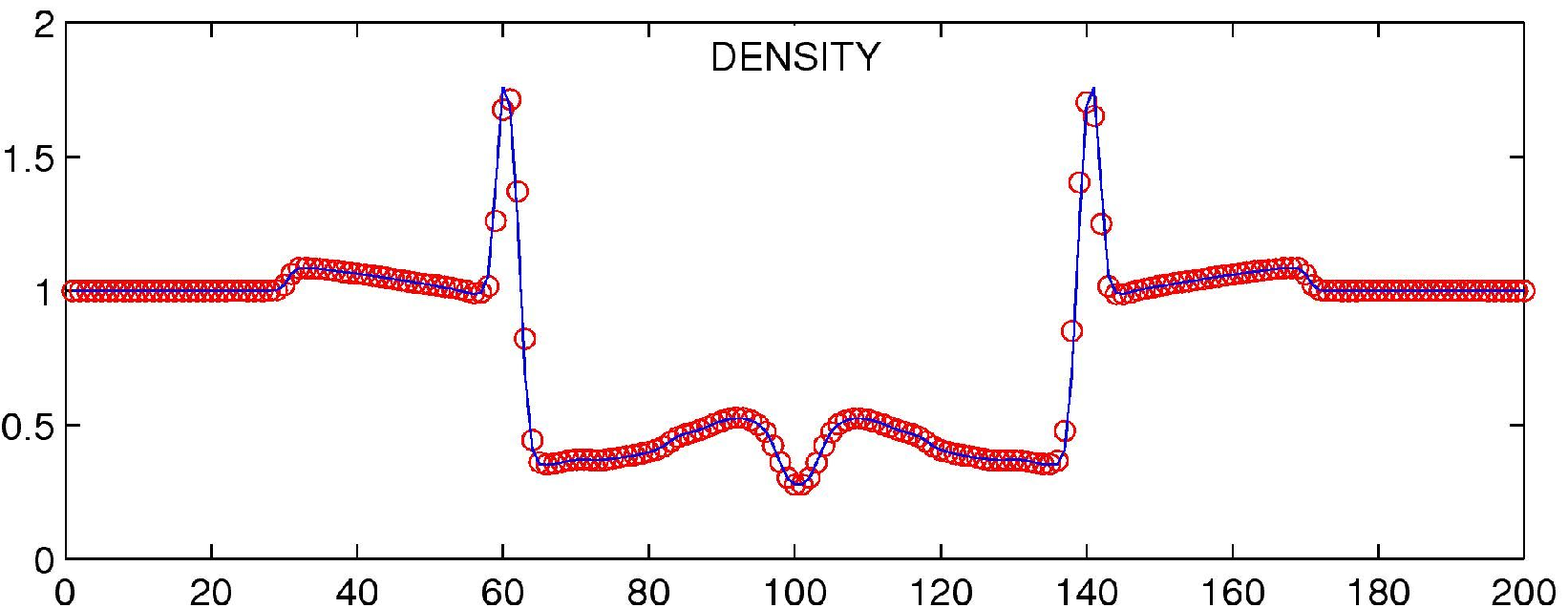} \\
\plotone{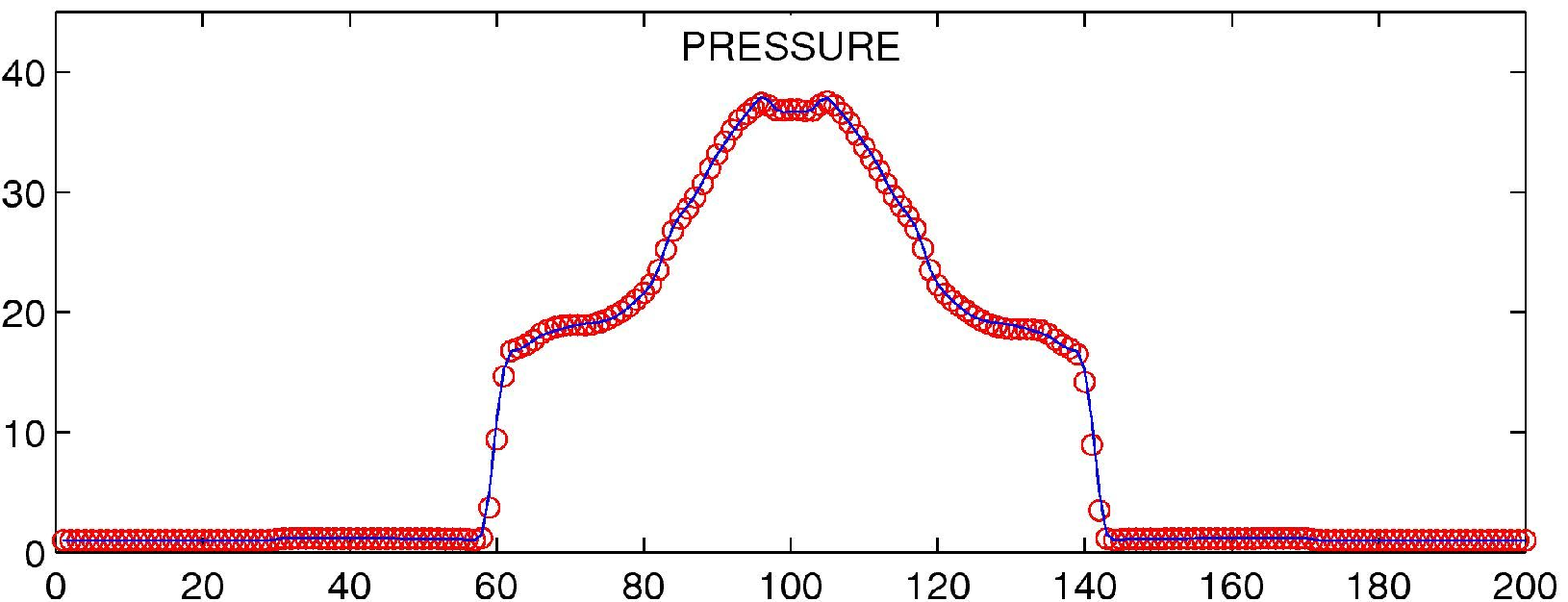} \\
\plotone{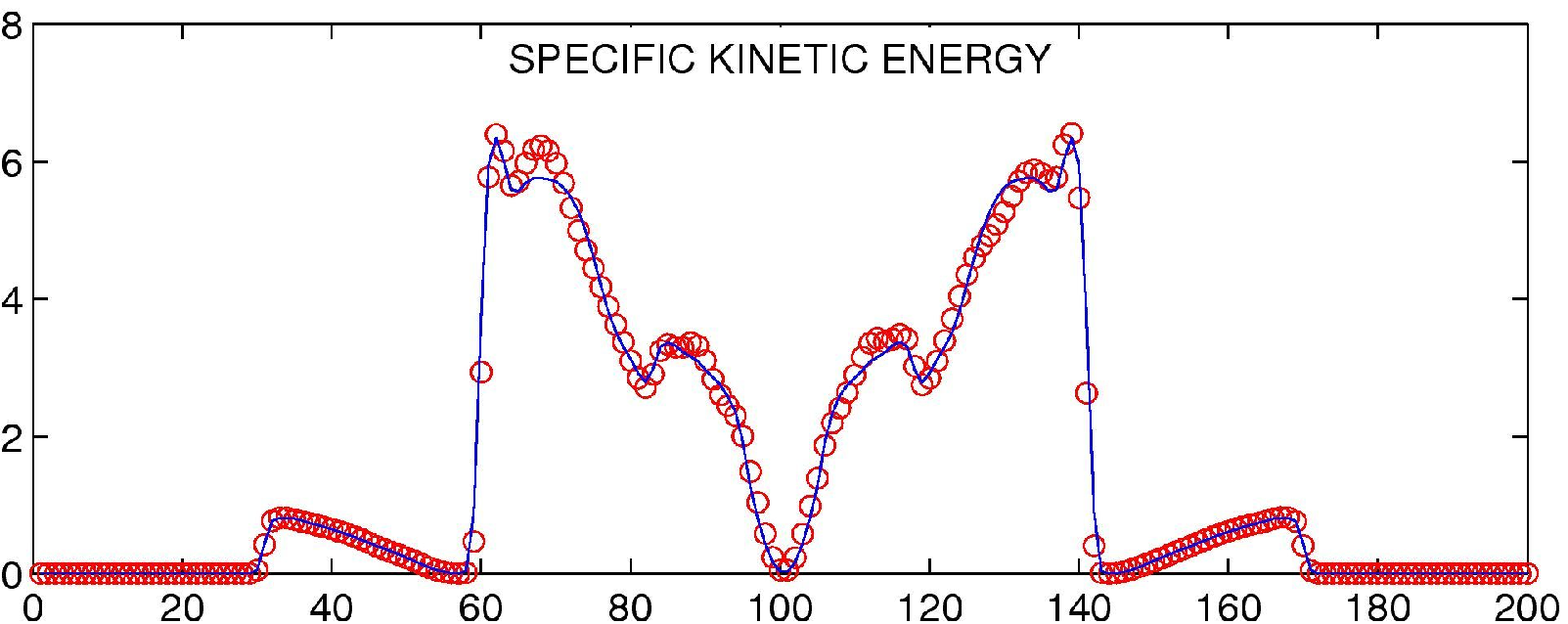} \\
\plotone{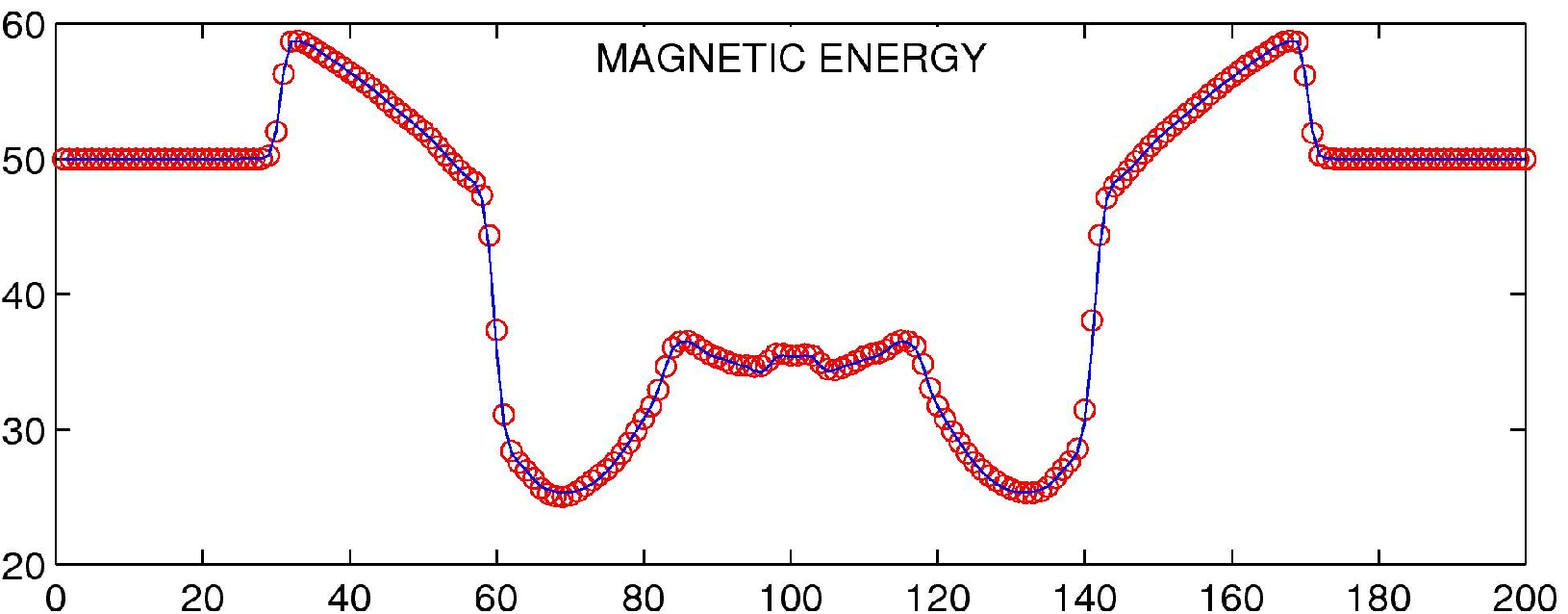}
\caption{Plots of selected variables along a horizontal line through the center of the blast at time \mbox{$t=0.02$} for the 3D MHD blast wave test with $B_0=10$ and $\beta_p=0.02$ using the cylindrical (circles) or Cartesian (solid line) versions of \athnosp. \label{cylblast_B10:3d_lineout}}
\end{figure}

{\subsection{Weber-Davis Wind} \label{cylwindrotb}}
In this problem, we investigate a cylindrical version of the Weber-Davis wind solution as described in \citet{sak85}.  We assume a steady, axisymmetric, 2D MHD flow with planar magnetic field, and a gravitational potential $\Phi_g = -GM/R$.  The constants of motion are $K = P\rho^{-\gamma}$, $\dot{M} = R\rho v_R$, $f = R B_R$, $\beta = B_R/(\rho v_R)$, as well as
\begin{mathletters}
\begin{eqnarray}
\Omega &=& \rinv \left( v_\phi - \frac{B_\phi}{\beta \rho} \right), \\
J &=& R (v_\phi - \beta B_\phi ) = R^2 \Omega + R u_\phi (1 - \beta^2 \rho),  \label{cylwindrotb:J} \\
\mathcal{B} &=& \onehalf (u_R^2 + u_\phi^2) + h + \Phi_g - \onehalf (\Omega R)^2.
\end{eqnarray}
\end{mathletters}
Here, $\vct{u} \equiv (v_R, \,v_\phi-R\Omega,\,0)$ is the velocity in a frame rotating at angular velocity $\Omega$, and in this frame $\vct{B} = \beta \rho \vct{u}$, so that the fluid feels no force from the magnetic field.  Note that the Bernoulli parameter in the rotating frame includes a centrifugal potential contribution.  The Alfv\'en Mach number in the rotating frame is given by $\mathcal{M}_A \equiv u/c_A = 1/\sqrt{\beta^2 \rho}$.  Let $R_A$ and $\rho_A$ denote the radius and density, respectively, at the Alfv\'en Mach point, i.e. where $\mathcal{M}_A = 1$.  Then $\beta = 1/\sqrt{\rho_A}$ and $J=R_A^2 \Omega$.

Letting $x \equiv R/R_A$ and $y \equiv \rho/\rho_A = \rho \beta^2 = \mathcal{M}_A^{-2}$ denote the scaled radius and density, respectively, the Bernoulli parameter is
\begin{equation}
\mathcal{B} = \frac{GM}{R_A} \left[ \frac{\eta}{2 x^2 y^2} + \frac{\omega}{2} \left( \left( \frac{1/x - x}{1-y} \right)^2 - x^2 \right) + \frac{\theta}{\gamma - 1} y^{\gamma-1} - \frac{1}{x} \right], \label{cylwindrotb:scaledbernoulli}
\end{equation}
where we have defined the scaled parameters
\begin{mathletters}
\begin{eqnarray}
\eta &\equiv& \frac{\dot{M}^2}{R_A \rho_A^2 GM}, \\
\theta &\equiv& \frac{\gamma K \rho_A^{\gamma-1} R_A}{GM}, \\
\omega &\equiv& \frac{R_A^3 \Omega^2}{GM}.
\end{eqnarray}
\end{mathletters}
Letting $\tilde{\mathcal{B}} \equiv \mathcal{B}/(GM/R_A)$, we have
\begin{mathletters}
\begin{eqnarray}
x \frac{\partial \tilde{\mathcal{B}}}{\partial x} &=& -\frac{\eta}{x^2 y^2} + \omega \left( \frac{(x^2 - 1/x^2)}{(1-y)^2} - x^2 \right) + \frac{1}{x}, \\
y \frac{\partial \tilde{\mathcal{B}}}{\partial y} &=& -\frac{\eta}{x^2 y^2} + \omega y \frac{(1/x - x)^2}{(1-y)^3} + \theta y^{\gamma-1}.
\end{eqnarray}
\end{mathletters}
To find wind solutions, we solve the Bernoulli equation~(\ref{cylwindrotb:scaledbernoulli}) under the constraint that the solution be locally flat at the slow- and fast-magnetosonic points, i.e. $\partial \tilde{\mathcal{B}}/\partial x = \partial \tilde{\mathcal{B}}/\partial y = 0$ at $(x_s,\,y_s)$ and $(x_f,\,y_f)$.  We will specify $\theta$ and $\omega$, and let $\eta$, $\tilde{\mathcal{B}}$, $x_s$, $y_s$, $x_f$, and $y_f$ vary.  Note that this becomes a system of six equations in six unknowns, so if a solution exists, it must be unique.  Following \citet{sak85}, we use the parameters $\gamma = 1.2$, $\theta = 1.5$ and $\omega = 0.3$ as for the spherically symmetric solar wind model of Weber and Davis.  The numerical solution is given in Table~\ref{cylwindrotb:windsoln}.

We solve the problem on the domain $x \in [0.4,1.8]$, so that the wind solution passes through all three critical points.  We choose units such that $GM=1$ and fix the initial solution at the inner- and outer-boundaries, evolving it long enough for equilibrium to develop.  Figure~\ref{cylwindrotb:l1error} shows the convergence of the RMS error in the $L_1$-norm of the solution at $t=5.0$ compared to the initial solution.  These data were computed using the Roe fluxes, second-order reconstruction, and the 1D integrator; the results were similar for all combinations of Roe/HLLD fluxes and second-/third-order reconstruction.  Because the 2D and 3D algorithms differ significantly from the 1D version, especially in their treatment of magnetic fields, we present the results of the same test using these integrators on grids which are essentially one-dimensional, but contain a few grid cells in each transverse direction considered.  

Evidently, the algorithm yields second-order convergence for smooth MHD flows, and is able to maintain a steady magnetized solution, also conserving angular momentum as it is exchanged between the fluid and magnetic field.

\begin{figure}
\centering
\epsscale{0.85}
\plotone{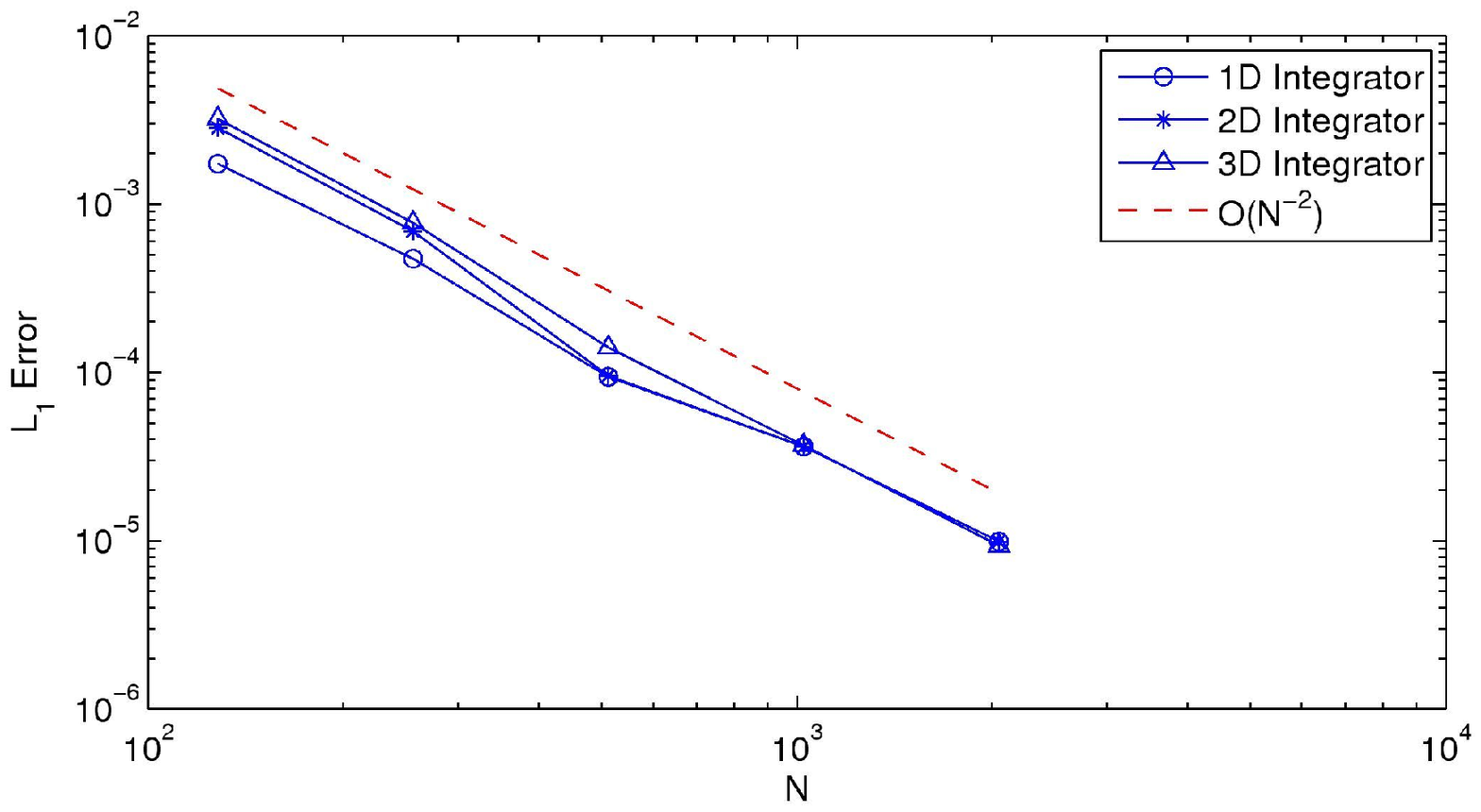}
\caption{Convergence of the RMS error in the $L_1$-norm for various levels of discretization of the Weber-Davis MHD wind test in 1D, 2D and 3D.  For reference, we have plotted a line of slope $-2$ (dashed) to show that the convergence is second-order in $1/N$. \label{cylwindrotb:l1error}}
\end{figure}

{\section{Conclusion} \label{conclusion}}
We have described an adaptation of the \ath astrophysical MHD code for cylindrical coordinates.  The original Cartesian code uses a combination of higher-order Godunov methods (based on the CTU algorithm of \citetalias{col90}) to evolve the mass-density, momenta, and total energy, and constrained transport \citep{eva88} to evolve the magnetic fields.  We have described modifications to the second- and third-order reconstruction schemes, the finite-volume and finite-area formulations of the MHD equations, and the inclusion of geometric source terms.  

Our approach is advantageous in that it does not require modification to the majority of the existing code, in particular to the Riemann solvers and eigensystems.  Furthermore, our approach to implementing cylindrical coordinates could be applied in a straightforward manner to enable other curvilinear coordinate systems, such as spherical coordinates, in the \ath code as well as other higher-order Godunov codes.  

Finally, our code and test suite are publicly available for download on the Web.  The code is currently being used for a variety of applications, including studies of global accretion disks, and we hope it will be of use to many others studying problems in astrophysical fluid dynamics.

\acknowledgments
The authors would like to thank Jim Stone for helpful discussions about the \ath algorithms, and Peter Teuben for his assistance with the code package.  This work was supported by grant AST 0507315 from the National Science Foundation.

\bibliography{apj-jour,apjs_342911}

\begin{thebibliography}{27}
\expandafter\ifx\csname natexlab\endcsname\relax\def\natexlab#1{#1}\fi

\bibitem[{{Blondin} \& {Lufkin}(1993)}]{blo93}
{Blondin}, J.~M., \& {Lufkin}, E.~A. 1993, \apjs, 88, 589

\bibitem[{{Colella}(1990)}]{col90}
{Colella}, P. 1990, J. Comp. Phys., 87, 171

\bibitem[{{Colella} \& {Sekora}(2008)}]{col08}
{Colella}, P., \& {Sekora}, M.~D. 2008, J. Comp. Phys., 227, 7069

\bibitem[{{Colella} \& {Woodward}(1984)}]{col84}
{Colella}, P., \& {Woodward}, P.~R. 1984, J. Comp. Phys., 54, 174

\bibitem[{{Einfeldt} {et~al.}(1991){Einfeldt}, {Roe}, {Munz}, \&
  {Sjogreen}}]{ein91}
{Einfeldt}, B., {Roe}, P.~L., {Munz}, C.~D., \& {Sjogreen}, B. 1991, J. Comp.
  Phys., 92, 273

\bibitem[{{Evans} \& {Hawley}(1988)}]{eva88}
{Evans}, C.~R., \& {Hawley}, J.~F. 1988, \apj, 332, 659

\bibitem[{{Fryxell} {et~al.}(2000){Fryxell}, {Olson}, {Ricker}, {Timmes},
  {Zingale}, {Lamb}, {MacNeice}, {Rosner}, {Truran}, \& {Tufo}}]{fry00}
{Fryxell}, B., {et~al.} 2000, \apjs, 131, 273

\bibitem[{{Gardiner} \& {Stone}(2005)}]{gar05}
{Gardiner}, T.~A., \& {Stone}, J.~M. 2005, J. Comp. Phys., 205, 509

\bibitem[{{Gardiner} \& {Stone}(2008)}]{gar08}
---. 2008, J. Comp. Phys., 227, 4123

\bibitem[{{Harten} {et~al.}(1983){Harten}, {Lax}, \& {van Leer}}]{har83}
{Harten}, A., {Lax}, P.~D., \& {van Leer}, B. 1983, SIAM Review, 25, 35

\bibitem[{{Ji} {et~al.}(2006){Ji}, {Burin}, {Schartman}, \& {Goodman}}]{ji06}
{Ji}, H., {Burin}, M., {Schartman}, E., \& {Goodman}, J. 2006, \nat, 444, 343

\bibitem[{{Leveque}(2002)}]{lev02}
{Leveque}, R.~J. 2002, Finite Volume Methods For Hyperbolic Problems
  (Cambridge: Cambridge University Press)

\bibitem[{{Londrillo} \& {Del Zanna}(2000)}]{lon00}
{Londrillo}, P., \& {Del Zanna}, L. 2000, \apj, 530, 508

\bibitem[{{Mignone} {et~al.}(2007){Mignone}, {Bodo}, {Massaglia}, {Matsakos},
  {Tesileanu}, {Zanni}, \& {Ferrari}}]{mig07}
{Mignone}, A., {Bodo}, G., {Massaglia}, S., {Matsakos}, T., {Tesileanu}, O.,
  {Zanni}, C., \& {Ferrari}, A. 2007, \apjs, 170, 228

\bibitem[{{Mihalas} \& {Mihalas}(1984)}]{mih84}
{Mihalas}, D., \& {Mihalas}, B.~W. 1984, {Foundations of radiation
  hydrodynamics} (New York, Oxford University Press, 1984, 731 p.)

\bibitem[{{Miyoshi} \& {Kusano}(2005)}]{miy05}
{Miyoshi}, T., \& {Kusano}, K. 2005, Journal of Computational Physics, 208, 315

\bibitem[{{Powell} {et~al.}(1999){Powell}, {Roe}, {Linde}, {Gombosi}, \& {de
  Zeeuw}}]{pow99}
{Powell}, K.~G., {Roe}, P.~L., {Linde}, T.~J., {Gombosi}, T.~I., \& {de Zeeuw},
  D.~L. 1999, J. Comp. Phys., 154, 284

\bibitem[{{Roe}(1981)}]{roe81}
{Roe}, P.~L. 1981, Journal of Computational Physics, 43, 357

\bibitem[{{Sakurai}(1985)}]{sak85}
{Sakurai}, T. 1985, \aap, 152, 121

\bibitem[{{Spitzer}(1978)}]{spi78}
{Spitzer}, L. 1978, {Physical processes in the interstellar medium}, ed.
  L.~Spitzer

\bibitem[{{Stone} {et~al.}(2008){Stone}, {Gardiner}, {Teuben}, {Hawley}, \&
  {Simon}}]{sto08}
{Stone}, J.~M., {Gardiner}, T.~A., {Teuben}, P., {Hawley}, J.~F., \& {Simon},
  J.~B. 2008, ArXiv e-prints, 804

\bibitem[{{Stone} \& {Norman}(1992{\natexlab{a}})}]{sto92a}
{Stone}, J.~M., \& {Norman}, M.~L. 1992{\natexlab{a}}, \apjs, 80, 753

\bibitem[{{Stone} \& {Norman}(1992{\natexlab{b}})}]{sto92b}
---. 1992{\natexlab{b}}, \apjs, 80, 791

\bibitem[{{Teyssier}(2002)}]{tey02}
{Teyssier}, R. 2002, \aap, 385, 337

\bibitem[{{Toro}(1999)}]{tor99}
{Toro}, E. 1999, Riemann Solvers and Numerical Methods for Fluid Dynamics: A
  Practical Introduction

\bibitem[{{T{\'o}th}(1996)}]{tot96}
{T{\'o}th}, G. 1996, Astrophysical Letters Communications, 34, 245

\bibitem[{{Ziegler}(2004)}]{zie04}
{Ziegler}, U. 2004, J. Comp. Phys., 196, 393

\end{thebibliography}
\clearpage


\clearpage

\begin{deluxetable}{ll}
\tabletypesize{\scriptsize}
\tablecaption{Weber-Davis Wind Parameters\label{cylwindrotb:windsoln}}
\tablewidth{0pt}
\tablehead{\colhead{Parameter} & \colhead{Value}}
\startdata
$\gamma$ & $1.2$ \\
$\theta$ & $1.5$ \\
$\omega$ & $0.3$ \\
$\eta$ & $2.3609$ \\
$\tilde{\mathcal{B}}$ & $7.8745$ \\
$x_s$ & $0.5243$ \\
$y_s$ & $2.4986$ \\
$x_f$ & $1.6383$ \\
$y_f$ & $0.5374$
\enddata
\end{deluxetable}

\end{document}